\documentclass[aps,prd,preprintnumbers,superscriptaddress,nofootinbib,floatfix,notitlepage,onecolumn,10pt]{revtex4-2}
\usepackage{graphicx,array}
\usepackage{hyperref}
\usepackage{xcolor}
\definecolor{color_git}{HTML}{24292E}
\usepackage{amsmath,amssymb,slashed,latexsym}
\usepackage{bm}
\usepackage{braket}
\usepackage{placeins}
\usepackage{adjustbox}
\usepackage{fontawesome5}
\usepackage{makecell}
\usepackage[caption=false]{subfig}
\usepackage{enumitem}
\usepackage{amsthm}
\usepackage{bbm}
\usepackage{comment}
\usepackage{multirow}
\usepackage{tikz}
\usepackage[compat=1.1.0]{tikz-feynman}

\tikzstyle{block} = [draw, rectangle,
minimum height=3em, minimum width=6em]

\newcommand{\cO}{\mathcal{O}}
\newcommand{\cL}{\mathcal{L}}
\newcommand{\cM}{\mathcal{M}}
\newcommand{\cP}{\mathcal{P}}
\newcommand{\cR}{\mathcal{R}}
\newcommand{\cG}{\mathcal{G}}
\newcommand{\cF}{\mathcal{F}}

\newcommand{\Tr}{{\rm Tr}}

\newcommand{\keV}{\mathrm{keV}}
\newcommand{\MeV}{\mathrm{MeV}}
\newcommand{\GeV}{\mathrm{GeV}}

\newcommand{\gitlink}{\href{https://github.com/maksymovchynnikov/EPXGen}{\textsc{g}it\textsc{h}ub {\large\color{color_git}\faGithub}}}

\def\beq{\begin{equation}}
\def\eeq{\end{equation}}
\def\beqa{\begin{eqnarray}}
\def\eeqa{\end{eqnarray}}

\begin{document}

\preprint{CERN-TH-2026-047}
\title{On Exclusive Coherent Production of Bosons in Electron-Proton Collisions}

\author{Reuven Balkin}
\email{rebalkin@ucsc.edu}
\affiliation{Department of Physics, University of California Santa Cruz and Santa Cruz Institute for Particle Physics, 1156 High St., Santa Cruz, CA 95064, USA}

\author{Ta'el Coren}
\email{tael.coren@campus.technion.ac.il}
\affiliation{Physics Department, Technion - Israel Institute of Technology, Haifa 3200003, Israel}

\author{Alexander Jentsch}
\email{ajentsch@bnl.gov}
\affiliation{Department of Physics,
Brookhaven National Laboratory, Upton, NY 11973, USA}

\author{Hongkai Liu}
\email{hliu6@bnl.gov}
\affiliation{Department of Physics,
Brookhaven National Laboratory, Upton, NY 11973, USA}

\author{Maksym Ovchynnikov}
\email{maksym.ovchynnikov@cern.ch}
\affiliation{Theoretical Physics Department, CERN, 1211 Geneva 23, Switzerland}

\author{Yotam Soreq}
\email{soreqy@physics.technion.ac.il}
\affiliation{Physics Department, Technion - Israel Institute of Technology, Haifa 3200003, Israel}
\affiliation{Theoretical Physics Department, CERN, 1211 Geneva 23, Switzerland}

\author{Sokratis Trifinopoulos}
\email{sokratis.trifinopoulos@cern.ch}
\affiliation{Theoretical Physics Department, CERN, 1211 Geneva 23, Switzerland}
\affiliation{Physik-Institut, Universit\"at Z\"urich, 8057 Z\"urich, Switzerland}

\begin{abstract}
We study the exclusive electroproduction process $e+p\to e'+p'+X$, with $X$ a single-particle final state, in the forward-proton kinematics relevant for the future Electron-Ion Collider~(EIC).
We develop a unified $2\to 3$ framework that provides the full event kinematics and incorporates pseudoscalar and vector mesons, as well as axion-like particles and vector mediators such as dark photons.
It is based on phenomenological amplitudes constrained by existing photo- and electroproduction data and constructed to admit systematic refinement as new measurements become available.
To benchmark the framework, we compare its predictions to flux-factorized descriptions based on the equivalent-photon approximation, demonstrating close agreement for total rates and selected single-differential distributions in the near-real regime, while highlighting the role of finite-$Q^{2}$ correlations for multi-differential observables at larger photon virtualities.
As a case study, we perform a detailed kinematic analysis of the missing-proton-energy signature, illustrating how the full $2\to 3$ treatment informs forward-proton acceptance and signal selection in realistic EIC configurations.
\end{abstract}

\maketitle
\onecolumngrid

\section{Introduction}
\label{sec:intro}

The electron-ion collider~(EIC)~\cite{Accardi:2012qut,AbdulKhalek:2021gbh}, which is planned for construction at Brookhaven National Laboratory, has a rich scientific program that is mostly focused on hadronic physics.
However, the EIC has very good potential to probe new particles in regions that are challenging or inaccessible to other experiments~\cite{Gonderinger:2010yn,Boughezal:2020uwq,Liu:2021lan,Cirigliano:2021img,Davoudiasl:2021mjy,Yan:2021htf,Li:2021uww,Batell:2022ogj,Zhang:2022zuz,Yan:2022npz,Boughezal:2022pmb,Davoudiasl:2023pkq,Balkin:2023gya,Davoudiasl:2024vje,Wang:2024zns,Wen:2024cfu,Gao:2024rgl,Du:2024sjt,Deng:2025hio,Davoudiasl:2025rpn,Bellafronte:2025ubi,Jiang:2025frv,Huang:2025ljp,Bar-Shalom:2026fou,Adhikary:2026rck,UrrutiaQuiroga:2026yee}.
New particles with a GeV mass scale and hadronic couplings to the Standard Model~(SM) are predicted in many Beyond Standard Model~(BSM) extensions~\cite{Basso:2008iv,Okun:1982xi,Holdom:1985ag,Heeck:2014zfa,Essig:2013lka,Beacham:2019nyx,Ilten:2018crw,Kyselov:2024dmi,Dimopoulos:1979pp,Holdom:1982ex,Flynn:1987rs,Rubakov:1997vp,Berezhiani:2000gh,Fukuda:2015ana,Gherghetta:2016fhp,Dimopoulos:2016lvn,Fukuda:2017ywn,Agrawal:2017ksf,Gaillard:2018xgk,Lillard:2018fdt,Hook:2019qoh,Csaki:2019vte,Gherghetta:2020keg,Valenti:2022tsc,Kivel:2022emq,Dunsky:2023ucb,Dolan:2017osp,Alves:2017avw,Marciano:2016yhf,Jaeckel:2015jla,Dobrich:2015jyk,Izaguirre:2016dfi,Knapen:2016moh,Bauer:2018uxu,Mariotti:2017vtv,CidVidal:2018blh,Aloni:2018vki,Aloni:2019ruo,Bauer:2020jbp,Bauer:2021wjo,Sakaki:2020mqb,Florez:2021zoo,Brdar:2020dpr,Bertholet:2021hjl,Co:2022bqq,Trifinopoulos:2022tfx,Ghebretinsaea:2022djg,DallaValleGarcia:2023xhh,Kyselov:2025uez,Afik:2023mhj,Balkin:2021jdr,Blinov:2021say,Balkin:2023gya,Bai:2021gbm,Pybus:2023yex,Bai:2024lpq,Gao:2024rgl,Davoudiasl:2024fiz,Baruch:2025lbw,Ovchynnikov:2025gpx,Boiarska:2019jym,Blackstone:2024ouf,deBlas:2025gyz}.
In this case, their hadronic production shares a lot of similarities with exclusive SM meson production.

In particular, the coherent electroproduction in electron-proton scattering,
\begin{align}
    \label{eq:process}
    e(p_{e})+p(p_{p})
    \to
    e'(p_{e'})+p'(p_{p'})+X(p_{X})\,,
\end{align}
where the forward proton remains intact, and $X$ is a single-particle exclusive final state, is such an example.
Studying this process, and in particular the kinematics of the forward proton, is of special importance for the EIC.
Depending on the identity of $X$, Eq.~\eqref{eq:process} covers a broad class of exclusive final states.
On the SM side, $X$ may be a meson, \emph{e.g.}, a pseudoscalar $X=\pi^0, \eta^{(\prime)}$ or a vector $X=\rho^0,\omega,\phi,J/\psi$.
Alternatively, the states may be new particles that are produced in hadronic interactions, and in particular, those that have mass/kinetic mixing with mesons.
Examples include: hadronically coupled ALPs mixing with pseudoscalar mesons~\cite{Aloni:2018vki,Bauer:2020jbp,Bauer:2021wjo,Ovchynnikov:2025gpx,Balkin:2025enj}, generic CP-even scalars mixing with scalar mesons~\cite{Winkler:2018qyg,Boiarska:2019jym,Foroughi-Abari:2021zbm,Blackstone:2024ouf,Batell:2017kty,Batell:2018fqo,Delaunay:2025lhl,Liu:2025ows}, or a vector/axial-vector particle that couples to hadronic current as a combination of vector and/or axial-vector mesons~\cite{Basso:2008iv,Okun:1982xi,Holdom:1985ag,Heeck:2014zfa,Essig:2013lka,Ilten:2018crw,Baruch:2022esd,Kyselov:2024dmi,Kyselov:2025uez}.

A range of dedicated simulation tools have been developed to generate exclusive electroproduction events in selected channels and kinematic regimes.
Some follow the full one-photon-exchange description~\cite{Ahmed:2024grm,Toll:2013gda}, while alternative treatments approximate photon emission from the electron using the equivalent photon approximation~\cite{Lomnitz:2018juf}.
In either case, the central practical limitation is that the hadronic subprocess, $\gamma^{(*)}p\to p'X$, is typically realized through channel-specific, tabulated or fitted multi-dimensional hadronic tensors (or helicity cross sections), together with auxiliary prescriptions for their kinematic variables dependence.
While this setup can be tuned efficiently in channels where such inputs exist, it does not provide a general, portable recipe for adding new exclusive states:
incorporating a different $X$ commonly requires constructing new high-dimensional inputs from dedicated measurements or bespoke modeling.
This becomes a bottleneck both for SM channels that are not broadly implemented in public generators, such as $\eta$ and $\eta'$ production and scalar mesons like $f_{0}(980)$, and, more fundamentally, for new particles for which the requisite tabulated inputs will not be available in principle.

In this work, we address this limitation by constructing a framework that treats Eq.~\eqref{eq:process} directly at the level of the exact three-body final state phase space.
Our approach is lightweight and modular:
instead of relying on channel-specific multi-dimensional lookup tables for the hadronic tensor, we follow Refs.~\cite{Pichowsky:1996tn,Kaskulov:2010kf,Kaskulov:2011wd,Sakinah:2024cza,Tang:2025qqe} that describe the coherent electroproduction of neutral pions and vector mesons through fully analytic expressions, organized in terms of reggeized trajectories and a small set of phenomenological parameters that can be fitted to existing photo- and electroproduction data.
Additionally, special care is taken to ensure electromagnetic gauge invariance of the hadronic currents -- a requirement that becomes essential at finite photon virtuality and that, as we show, is not satisfied by some of the interaction vertices commonly used in the literature~\cite{Oh:2003aw,Kaskulov:2011wd,Wang:2025rvr}.

By construction, the framework provides fully differential event kinematics and preserves correlations among the invariants that characterize Eq.~\eqref{eq:process}.
This is particularly important for EIC studies of rare processes that depend on multivariable selections, such as electroproduction of vector mesons and searches for new particles~\cite{Davoudiasl:2023pkq,Davoudiasl:2025rpn,Balkin:2025rtc}. The framework, \textsc{EPXGen}, is available on \gitlink~\cite{Ovchynnikov:2026EPXGen}.

Our approach has two advantages.
First, this formulation keeps the dependence on the hadronic input explicit and compact while remaining constrained by existing data.
The assumptions entering the construction are therefore transparent and can be confronted with present constraints, while future EIC electroproduction measurements will allow them to be tested more directly and refined further.
Second, for new particles $X$ mixing with SM mesons, the data-driven calculation can be generalized straightforwardly even in the absence of direct data.
This can be implemented by summing the meson electroproduction amplitudes, weighted by coefficients constructed from basis-independent combinations of the $X$-meson couplings.

We focus on the production of pseudoscalar and vector particles in electron-proton collisions, although the approach may be extended to other particles, such as scalar, axial-vector, and tensor mesons~\cite{Wei:2024lne,Wang:2017plf,Yu:2019wly}, or to different incoming beam configurations, such as muon-proton collisions~\cite{Acosta:2021qpx,Acosta:2022ejc,Davoudiasl:2024fiz}.
For the SM meson channels considered here, the hadronic amplitudes are calibrated to existing photoproduction data, so the remaining uncertainty is associated mainly with the extrapolation to electroproduction and, in some corners of phase space, with the proximity to the baryon-resonance region.
Within the kinematic regime where the framework is constrained by existing data, we expect the residual modeling uncertainty to be at the level of tens of percent; this uncertainty can be reduced further once EIC electroproduction data become available.

The paper is organized as follows.
Sec.~\ref{sec:interactions} is devoted to discussing interactions of pseudoscalar and vector particles.
In Sec.~\ref{sec:matrix-elements}, we summarize the input we use to calculate the matrix elements of the electroproduction processes of Eq.~\eqref{eq:process} for the neutral pseudoscalar and vector mesons, and then generalize them to other pseudoscalar and vector particles.
Sec.~\ref{sec:kinematics-domain} describes the event selection we utilize for the electroproduction events, and analyzes resulting kinematics for various final particles $X$ and beam configurations at the EIC.
In Sec.~\ref{sec:EPA}, we cross-check our framework against the equivalent photon approximation, highlighting the limitations of the latter when used for event analysis.
Finally, conclusions are presented in Sec.~\ref{sec:conclusions}.

\section{Interactions of pseudoscalar and vector particles}
\label{sec:interactions}

Our focus is on the production of hadronically coupled pseudoscalar and vector particles $X$, which may result in a sizable proton energy loss in the process of Eq.~\eqref{eq:process}; see, for example, Figs.~\ref{fig:diagrams-pseudoscalar} and~\ref{fig:diagrams-vector} in Sec.~\ref{sec:matrix-elements}.
In this section, we first discuss the interactions of the light SM mesons that enter the calibrated electroproduction amplitudes used in Sec.~\ref{sec:matrix-elements}.
We then introduce a generic class of new states $X$ whose low-mass phenomenology may be described in terms of mixing with SM mesons of the same spin and parity, and explain two complementary ways of promoting the meson amplitudes to production amplitudes for $X$.
Finally, we specialize this discussion to ALPs and new vector mediators, and present both the corresponding projector coefficients and the explicit $X$--SM couplings.

\subsection{Interactions of SM mesons}
\label{sec:low-mass-couplings}

We begin with the light pseudoscalar and vector mesons
\begin{align}
    P=\pi^0,\eta,\eta'\,,
    \qquad
    V=\rho^0,\omega,\phi\,,
\end{align}
which will later serve both as SM final states and as basis states for the new particles $X$.
Their generators in the $SU(3)_F$ flavor space are
\begin{align}
    \label{eq:TP}
    T_{\pi^{0}}
    =
    \frac{1}{2}\text{diag}(1,-1,0)\,,
    \qquad
    T_{\eta}
    =
    \frac{1}{\sqrt{6}}\text{diag}(1,1,-1)\,,
    \qquad
    T_{\eta'}
    =
    \frac{1}{2\sqrt{3}}\text{diag}(1,1,2)\,,
\end{align}
and
\begin{align}
    \label{eq:TV}
    T_{\rho^{0}}
    =
    \frac{1}{2}\text{diag}(1,-1,0)\,,
    \qquad
    T_{\omega}
    =
    \frac{1}{2}\text{diag}(1,1,0)\,,
    \qquad
    T_{\phi}
    =
    \frac{1}{\sqrt{2}}\text{diag}(0,0,1)\,.
\end{align}
%

\paragraph{Coupling to nucleons.}

Within chiral perturbation theory, the effective interactions of vector mesons $V=\rho^{0},\omega,\phi$ with protons are given by
\begin{align}
    \label{eq:Lagr-Vpp}
    \cL_{Vpp}
    =
    g_{Vpp}\left( \bar{p}\gamma^{\mu}p\,V_{\mu}
    + \frac{\kappa_{V}}{4m_{p}}\bar{p}\,\sigma^{\mu\nu}p\,V_{\mu\nu} \right)\,,
\end{align}
where $g_{Vpp}\,(\kappa_V)$ is the monopole\,(dipole) vector coupling, $m_p$ is the proton mass, and $V_{\mu\nu}=\partial_{\mu}V_{\nu}-\partial_{\nu}V_{\mu}$ is the field-strength tensor of the vector meson $V$.
For convenience, we also provide the corresponding $Vpp$ vertex,
\begin{align}
    \label{eq:Vert-Vpp}
    \bar u(p_{p'})\Gamma_{VNN,\beta}u(p_p)
    =
    i g_{Vpp}\bar u(p_{p'})\bigg[(1+\kappa_V)\gamma_\beta-\frac{\kappa_V}{2m_p}(p_p+p_{p'})_{\beta}\bigg]u(p_p)\,.
\end{align}
Extraction of the $Vpp$ monopole and dipole couplings can be attempted from different processes, including nucleon--nucleon scattering and meson photoproduction data~\cite{Stoks:1996yj,Nagels:1975fb,Nagels:1976xq,Rijken:2006ep,Kaskulov:2010kf,Maessen:1989sx,Machleidt:2000ge}.
Such determinations are, however, subject to sizable model uncertainties~\cite{Yu:2011zu}.
First, they depend sensitively on the choice of hadronic form factors used to parametrize the interaction vertices, as well as on the correlations between the contributions from the $g_{Vpp}$ and $g_{Vpp}\kappa_V$ couplings.
In addition, in Reggeized descriptions, the assumed Regge residue -- including its phase and possible Regge-cut contributions -- constitutes another important source of systematic uncertainty.

For $\rho^0$ and $\omega$, we use the results adopted in Ref.~\cite{Kashevarov:2017vyl}:
\begin{align}
    \label{eq:vector_nucleon_couplings}
    g_{\rho^{0}pp} = 2.7\,,
    \qquad
    \kappa_{\rho} = 4.2\,,
    \qquad
    g_{\omega pp} = 14.2\,,
    \qquad
    \kappa_{\omega} = 0\,.
\end{align}
Unlike the $\rho^0$ and the $\omega$, the $\phi$ couplings cannot be robustly extracted from the photoproduction data.
First, the $\phi\gamma\pi$ vertex is very suppressed, see Tbl.~\ref{tab:Coupl-VPgamma}.
In addition, OZI-rule calculations~\cite{Meissner:1997qt,Palomar:2002hk} predict $|g_{\phi pp}|<0.6$ and $\kappa_{\phi}=1.8$, whereas the CD-Bonn-based calculations supplemented with $SU(3)$-symmetry arguments~\cite{Haidenbauer:1998gr} lead to $g_{\phi pp}\in[-0.16,1.4]$ and $\kappa_{\phi}=\kappa_{\rho}$, with the uncertainty caused by lack of data.
The particular value of this coupling is not important for our studies of the photo- and electroproduction processes, because it is either sub-dominant~\cite{Yu:2016zut,Wang:2025rvr} or degenerate with contributions from other interactions.
Following Ref.~\cite{Yu:2016zut}, we therefore assume
\begin{align}
    g_{\phi pp}=\kappa_\phi=0\,.
\end{align}

For pseudoscalar mesons $P=\pi^0,\eta,\eta'$, the effective interaction with nucleons is given by
\begin{align}
    \cL_{PNN}
    =
    g_{pP}\,\overline{p}\gamma^{\mu}\gamma_{5}p\,\partial_{\mu}P
    +
    g_{nP}\,\overline{n}\gamma^{\mu}\gamma_{5}n\,\partial_{\mu}P\,,
\end{align}
see Appendix~\ref{app:couplings} for details.
The corresponding couplings are
\begin{align}
    \label{eq:Coupl-pP}
    g_{p\pi^0}
    =
    -g_{n\pi^{0}}
    \approx
    6.8\,\GeV^{-1}\,,
    \qquad
    g_{p\eta}
    =
    g_{n\eta}
    \approx
    2.2\,\GeV^{-1}\,,
    \qquad
    g_{p\eta'}
    =
    g_{n\eta'}
    \approx
    0.93\,\GeV^{-1}\,.
\end{align}
%

\paragraph{Coupling between $V$, $\gamma$, and pseudoscalar mesons.}

The Lagrangian of the $V\gamma P$ interaction can be written as
\begin{align}
    \label{eq:Lagr-VYgamma}
    \cL_{V\gamma P}
    =
    -\frac{g_{V\gamma P}}{2}\,P\,F^{\mu\nu}\tilde{V}_{\mu\nu}\,,
\end{align}
where $F_{\mu\nu}$ is the electromagnetic field-strength tensor and $\tilde{V}_{\mu\nu}=\frac{1}{2}\epsilon_{\mu\nu\alpha\beta}V^{\alpha\beta}$ denotes the dual field strength of the vector meson.
The corresponding vertex for an incoming off-shell photon and vector meson with four-momenta $p_{\gamma^*}$ and $p_{V^*}$, and outgoing $P$, reads
\begin{align}
    \label{eq:Vert-VYgamma}
    \Gamma_{V\gamma P}^{\mu\beta}(p_{\gamma^*},p_{V^*})
    =
    i g_{V\gamma P}\,\varepsilon^{\mu\beta\kappa\alpha}\,
    p_{\gamma^*,\kappa}\,p_{V^*,\alpha}\,.
\end{align}
In principle, this and other vertices should also accommodate phenomenological form factors accounting for suppression at large virtualities.
However, as discussed in Sec.~\ref{sec:matrix-elements}, we will adopt a common form factor for the full electroproduction amplitude, which effectively captures the finite-virtuality behavior.
We therefore do not include additional vertex form factors here.
\begin{table}[t]
    \centering
    \begin{tabular}{lcccccc}
    \hline\hline
     & \multicolumn{2}{c}{$\rho^{0}$} & \multicolumn{2}{c}{$\omega$} & \multicolumn{2}{c}{$\phi$} \\
    $P$ & HLS & exp & HLS & exp & HLS & exp \\
    \hline
    $\pi^{0}$
      & $2.4\times 10^{-1}$
      & $(2.2 \pm 0.20)\times 10^{-1}$
      & $7.3\times 10^{-1}$
      & $(7.1 \pm 0.11)\times 10^{-1}$
      & $0$
      & $(4.1 \pm 0.08)\times 10^{-2}$ \\
    $\eta$
      & $6.0\times 10^{-1}$
      & $(4.8 \pm 0.16)\times 10^{-1}$
      & $2.0\times 10^{-1}$
      & $(1.4 \pm 0.08)\times 10^{-1}$
      & $2.8\times 10^{-1}$
      & $(2.1 \pm 0.02)\times 10^{-1}$ \\
    $\eta'$
      & $4.2\times 10^{-1}$
      & $4.2\times 10^{-1}$
      & $1.4\times 10^{-1}$
      & $1.4\times 10^{-1}$
      & $-4.0\times 10^{-1}$
      & $(-2.1 \pm 0.04)\times 10^{-1}$ \\
    \hline\hline
    \end{tabular}
    \caption{The couplings $g_{V\gamma P}$ (in $\GeV^{-1}$) entering the Lagrangian of Eq.~\eqref{eq:Lagr-VYgamma}.
    The ``HLS'' column gives the theoretical result obtained in the Hidden Local Symmetry approach to vector meson dominance, while the ``exp'' column gives the value calibrated on the observed radiative decay widths, where kinematically possible.
    For $g_{\rho^{0}\gamma\eta'}$ and $g_{\omega\gamma\eta'}$, we use the HLS values, which are close to those employed in $\eta'$ production studies~\cite{Nakayama:1999jb,Huang:2012xj}.}
    \label{tab:Coupl-VPgamma}
\end{table}
In Tbl.~\ref{tab:Coupl-VPgamma}, we present the values of the couplings $g_{V\gamma P}$ in the HLS approach under the assumption that the $\phi$ meson is a pure $s\bar s$ state~\cite{Fujiwara:1984mp,Guo:2011ir}, as well as those obtained by fitting to the measured radiative decay widths~\cite{ParticleDataGroup:2024cfk}, which we will use below.
The most important difference between the theoretical and experimental determinations is for $g_{\phi\gamma\pi}$, for which the HLS prediction vanishes.
\paragraph{The $V\gamma f_2$ coupling.}

Next, we need the radiative coupling $\gamma V f_2$, where $f_2\equiv f_2(1270)$ is the lowest-lying spin-2 meson.
We build the Lorentz structure of the $f_2V\gamma$ interaction assuming that
(i)~$f_2$ couples to the hadronic stress-energy tensor, and
(ii)~within the HLS approach to vector meson dominance, the photon exhibits linear mixing with $\rho^0$, $\omega$, and $\phi$; details are provided in Appendix~\ref{app:f2-decays}.
Similar constructions are often adopted in the literature~\cite{Singer:1982tn,Oh:2003aw,Yu:2016zut,Yu:2017vvp,Wang:2025rvr}, although there are important subtleties when applying them to photoproduction and electroproduction.

We start from the interaction
\begin{align}
    \cL_{f_2VV}
    =
    g_{T,V}\,f_{2,\mu\nu}\,T_V^{\mu\nu}\,,
\end{align}
where $T_V^{\mu\nu}$ is the hadronic stress-energy tensor of the free vector meson $V=\rho^0,\omega,\phi$, and $g_{T,V}$ is an interaction coupling.
The couplings $g_{T,V}$ may, in general, be non-universal among the $V$ fields, reflecting the flavor structure of $f_2$ and possible $SU(3)_F$ violation.

Including the electromagnetic interactions within the HLS framework effectively replaces $V_{\mu\nu}\to V_{\mu\nu}+c_{V\gamma}F_{\mu\nu}$ in $T_V^{\mu\nu}$.
This induces the vertex
\begin{align}
    \label{eq:Vert-f2-V-gamma}
    \Gamma^{\mu\nu,\alpha\beta}_{f_2V\gamma}(p_\gamma,p_V)
    &=
    g_{f_{2}V\gamma}\Big[
    (p_{\gamma}\!\cdot\! p_{V})\, g^{\mu\alpha} g^{\nu\beta}
    + p_{\gamma}^{\mu} p_{V}^{\nu}\, g^{\alpha\beta}
    - p_{\gamma}^{\mu} p_{V}^{\alpha}\, g^{\nu\beta}
    - p_{\gamma}^{\beta} p_{V}^{\nu}\, g^{\mu\alpha}
    \nonumber\\
    &\hspace{2.8cm}
    -\frac{1}{2}\, g^{\mu\nu}\Big((p_{\gamma}\!\cdot\! p_{V})\, g^{\alpha\beta}
    - p_{\gamma}^{\beta} p_{V}^{\alpha}\Big)\Big]\,,
\end{align}
with
\begin{align}
    g_{f_2V\gamma}=2g_{T,V}c_{V\gamma}\,,
\end{align}
which has dimension $\GeV^{-1}$.
This vertex is traceless with respect to the $\mu\nu$ indices and vanishes after contraction with $p_{\gamma^*}^{\alpha}$ or $p_{V^*}^{\beta}$, reflecting the properties of the $F_{\mu\nu}V^{\mu\nu}$ interaction.

Notably, it differs from those often used in the literature~\cite{Oh:2003aw,Yu:2016zut,Yu:2019wly,Wang:2025rvr}.
In particular, Refs.~\cite{Yu:2016zut,Yu:2019wly} employ a similar construction but omit the term proportional to $g^{\mu\nu}$.
That term vanishes for an on-shell $f_2$, whose polarization is traceless, but survives for the off-shell $f_2$ relevant to photoproduction.
Ref.~\cite{Oh:2003aw}, on the other hand, constructs a vertex that satisfies Ward identities only after contraction with the polarization tensor of an on-shell $f_2$; for an off-shell $f_2$, contraction with the photon momentum gives a structure proportional to $p_{f_2}^2-m_{f_2}^2\neq 0$.
Using such a vertex in photo- and electroproduction, where $f_2$ is exchanged in the $t$ channel, explicitly violates Ward identities and may lead to unphysical features, such as an incorrect scaling of the electroproduction cross section with the photon momentum transfer.
An expanded discussion is provided in Appendix~\ref{app:f2-decays}.

The couplings $g_{f_2V\gamma}$ may be constrained directly from radiative decays $f_2\to V\gamma$, or indirectly via hadronic decays such as $f_2\to 2V^*\to X$.
Since no radiative decays have been measured~\cite{ParticleDataGroup:2024cfk}, some uncertainty remains.

We fix the value of $g_{f_{2}\rho\gamma}$ to fit the $\rho^0$ photoproduction data. For the remaining couplings, we match the theoretical prediction of the Covariant Oscillator Quark Model~\cite{Ishida:1988uw}, following Refs.~\cite{Yu:2016zut,Yu:2017vvp}. Overall,
\begin{align}
    g_{f_2\rho\gamma}
    &\approx
    0.27\,\GeV^{-1}\,,
    \qquad
    g_{f_2\omega\gamma}
    \approx
    0.15\,\GeV^{-1}\,,
    \qquad
    g_{f_2\phi\gamma}
    \approx
    0.07\,\GeV^{-1}\,.
\end{align}
Notably, the used value of $g_{f_2\rho\gamma}$ lies within the range permitted by the requirement that $g_{f_{2}\rho\gamma}$ not overproduce the $f_2 \to 2\rho^{0*} \to 2\pi^+2\pi^-$ decay relative to the data; see Appendix~\ref{app:f2-decays}.
\paragraph{The $V\gamma\sigma$ coupling.}

Finally, we discuss the coupling $\gamma V\sigma$, where $\sigma\equiv f_0(500)$ is the lowest-lying scalar meson.
We use
\begin{align}
    \label{eq:Lagr-Vsigmagamma}
    \cL_{V\sigma\gamma}
    =
    \frac{e\,g_{V\sigma\gamma}}{2}\,\sigma\,V_{\mu\nu}F^{\mu\nu}\,,
\end{align}
with the corresponding vertex
\begin{align}
    \label{eq:Vert-Vsigmagamma}
    \Gamma_{V\sigma\gamma}^{\mu\nu}(p_{\gamma^*},p_{V^*})
    =
    e g_{V\sigma\gamma}\left[
    (p_{\gamma^*}\!\cdot p_{V^*})\,g^{\mu\nu}
    -p^{\nu}_{\gamma^*}\,p_{V^*}^{\mu}
    \right]\,.
\end{align}
The value of $g_{V\sigma\gamma}$ may be constrained under the assumption that the decays $V\to\pi^0\pi^0\gamma$ are dominated by $\sigma$ exchange.
However, because of the sizable uncertainty in the $\sigma$ mass and width, such estimates are necessarily ambiguous.

Following the photoproduction studies~\cite{Yu:2016zut,Yu:2017vvp}, we take $m_{\sigma}=500\,\MeV$ and use
\begin{equation}
    \label{eq:Coupl-V-sigma-gamma-1}
    g_{\phi\sigma\gamma}\approx -0.279\,\GeV^{-1}\,,
    \qquad
    g_{\omega\sigma\gamma}\approx -0.561\,\GeV^{-1}\,,
\end{equation}
which are inferred from the theoretical predictions for $\phi/\omega\to\sigma\gamma$~\cite{Black:2002ek} and the corresponding measured decay widths~\cite{ParticleDataGroup:2024cfk}.
For the $\rho$ meson, requiring that the $\sigma$ contribution to $\rho^0\to 2\pi^0\gamma$ not exceed the observed width gives $|g_{\rho\sigma\gamma}|=(1.0-1.4)\,\GeV^{-1}$; see Appendix~\ref{app:g-rho-sigma-gamma}.
Interpreting this range as a freedom to be fixed from the $\rho^0$ photoproduction data, we use
\begin{equation}
    \label{eq:Coupl-V-sigma-gamma-2}
    g_{\rho\sigma\gamma}\approx -1.4\,\GeV^{-1}\,,
\end{equation}
with the sign chosen consistently with Eq.~\eqref{eq:Coupl-V-sigma-gamma-1} and to better match the photoproduction data.

\subsection{New states and two descriptions of their production}
\label{sec:BSM}

We now turn to new states $X$ whose phenomenology in the mass range
$m_X\lesssim \text{ few}\,\GeV$ may be described in terms of mixing with SM mesons $h$ carrying the same spin and parity.
At the level of the $SU(3)_F$ flavor representation, we may write their effective generator in the symbolic form
\begin{align}
    \label{eq:TX-generic}
    \widetilde T_X(m_X,f_{X})
    \approx
    \sum_{h\in {\rm basis}(X)} \kappa_{Xh}(m_X,f_{X})\,T_h\,,
\end{align}
where ${\rm basis}(X)$ denotes the light-meson family with the same quantum numbers as $X$, $T_h$ is the generator of the meson $h$ (which, for definiteness, is normalized by $2\,{\rm Tr}[T_{h}T_{h'}] = \delta_{hh'}$), and $\kappa_{Xh}(m_X,f_{X})$ are the corresponding mixing coefficients depending on the $X$'s mass $m_{X}$ and $X$'s coupling $f_{X}$. In many cases, they have a Breit-Wigner-like structure.

In the production analysis below, the amplitudes for the states $X$ will be constructed from the meson electroproduction amplitudes calibrated to existing data.
Conceptually, this may be done with the help of the generator $\widetilde T_X(m_X)$ in two different ways.

The first is a \emph{coupling-based} description.
In this approach, one decomposes the $X$ production matrix element $M_{X}$ into elementary amplitudes of independent subprocesses $i$ (such as different $t$-channel exchanges, bremsstrahlung-like baryonic terms, and so on):
\begin{align}
    \label{eq:MX-coupling-based}
    M_{X}
    =
    \sum_i M_{X,i}\,.
\end{align}
Each contribution depends on a set of effective couplings $g_{X,i}(m_{X})$, which may be obtained from the corresponding meson couplings through
\begin{align}
    \label{eq:gXSM-generic}
    g_{X,i}(m_X,f_{X})
    =
    \sum_{h\in {\rm basis}(X)}
    g_{h,i} \kappa_{Xh}(m_{X},f_{X}) F_{i}(m_{X}) =     \sum_{h\in {\rm basis}(X)} g_{h,i}\, \kappa_{Xh,i}(m_{X},f_{X})\,,
\end{align}
where $F_{i}(m_{X})$ is a phenomenological function accounting for specific sub-process-dependent scaling of the coupling with $m_{X}$ required, e.g., by the QCD sum rules (see below), and
\begin{align}
\label{eq:kappa-process-dependent}
\kappa_{Xh,i}(m_{X},f_{X}) \equiv \kappa_{Xh}(m_{X},f_{X})\, F_{i}(m_{X})\,.
\end{align}
The second is a \emph{projector-based} description.
In this approach, one starts directly from the meson production amplitudes $M_{h,i}$ and constructs the amplitude for $X$ as
\begin{align}
    \label{eq:MX-projector-based}
    M_X
    =
    \sum_{h\in {\rm basis}(X)}
    \sum_{i}  \kappa_{Xh,i}(m_{X},f_{X})
    \left[M_{h,i}\right]_{m_h\to m_X}\,,
\end{align}
where the notation $m_h\to m_X$ indicates that the produced-state mass is replaced by $m_X$ in the external kinematics and phase space.

The projector-based description reduces to the coupling-based one when $ \kappa_{Xh,i}\left[M_{h,i}\right]_{m_h\to m_X} \equiv M_{X,i}$. This is true if $M_{h,i}$ has a clean dependence on the meson $h$ in terms of $g_{h,{\rm SM}}$ and mass $m_{h}$. However, this is not the case once $M_{h,i}$ depends on some phenomenological parameters that cannot be calculated from first principles.

The latter situation applies to the meson production framework we adopt in this study; see Sec.~\ref{sec:matrix-elements}. Namely, for a particular channel $i$ describing the exchange of excitation with a particular spin and parity, it uses a description in terms of the Regge family for this state that also includes absorptive cut contributions from interference with other states, with $h$-dependent coefficients determined by the data.

For this reason, the projector-based description is the more natural one for our implementation, and we follow it below.
Nevertheless, for completeness and generality, in Appendix~\ref{app:couplings-X}, we will also provide the explicit $X$--SM couplings corresponding to the coupling-based viewpoint.

\subsection{Explicit realizations: ALPs and vector mediators}
\label{sec:large-mass-couplings}

We now specialize the general discussion above to two representative classes of new particles: ALPs and new vector mediators.
For $m_X\lesssim 1\,\GeV$, their phenomenology can be described in terms of mixing with the light mesons introduced above.
For larger masses, two additional effects become important:
(i)~suppression of the effective interaction vertices in the limit $m_X\gg\Lambda_{\rm QCD}$, as suggested by quark-counting rules and QCD sum rules~\cite{Lepage:1980fj,Aloni:2018vki,Balkin:2025enj}, and
(ii)~mixing with heavier pseudoscalar or vector excitations.
In the present work, we retain only the lightest mesons explicitly and encode the large-mass behavior through phenomenological scaling factors.
Since the neglected heavier resonances are expected to enhance the production rates, this choice is conservative.

\subsubsection{ALPs}
\label{sec:alp}

At the partonic level, the effective Lagrangian for an ALP $a$ at the QCD scale is~\cite{Georgi:1986df,Bauer:2020jbp}
\begin{align}
    \label{eq:Lagr_alp}
	\mathcal L_a
    \supset
    -c_g\frac{\alpha_s}{8\pi}\frac{a}{f_a}G^{\mu\nu}\tilde G_{\mu\nu}
	+\frac{\partial_\mu a}{f_a}\sum_q \bar q\,c_q\,\gamma^\mu\gamma_5 q
    -\sum_q \bar q\,M_q e^{2i\gamma^5\beta_q \frac{a}{f_a}} q\,,
\end{align}
where $q=(u,d,s)$, $G^{\mu\nu}$ is the gluon field-strength tensor, and $f_a$, $c_g$, $c_q$, and $\beta_q$ encode the UV couplings.
We do not discuss other ALP couplings, such as couplings to photons, electroweak gauge bosons, or leptons, since they do not contribute to the hadronic production considered here.
We also neglect flavor-violating ALP couplings.

The couplings in Eq.~\eqref{eq:Lagr_alp} are not individually physical, since they are not invariant under chiral rotations of the quark fields.
Observables must therefore be organized in terms of the invariant combinations~\cite{Bauer:2021wjo,Bai:2024lpq,Ovchynnikov:2025gpx,Balkin:2025enj}
\begin{align}
    \label{eq:invar_couplings}
    \tilde\beta_q
    \equiv
    \beta_q+c_q\,,
    \qquad
    \tilde c_g
    \equiv
    c_g-2\sum_q c_q\,.
\end{align}
Below, we will present explicit expressions for the gluonic ALP, although the formulas may be generalized straightforwardly to other coupling patterns. The gluonic ALP has the couplings
\begin{align}
    c_g=-2\,,
    \qquad
    \beta_q=c_q=0\,.
\end{align}
We split the discussion of the ALP phenomenology into two mass ranges $m_{a}\lesssim 4\pi f_{\pi}$ and $m_{a}\gtrsim 4\pi f_{\pi}$, using mass $m_{a} = 1.2\,\GeV$ for concreteness.

\paragraph{Mass $m_a < 1.2\,\GeV$.} The interactions of the light ALPs may be understood in terms of the mixing with the pseudoscalar mesons $P = \pi^{0},\eta,\eta'$. The effective ALP generator $\widetilde{T}_{a}$ reads~\cite{Balkin:2025enj}
\begin{align}
    \label{eq:Ta-gluon-light}
    \widetilde{T}_a(m_{a},f_{a})\big|_{m_{a}<1.2\,\GeV}
    = \sum_{P}\kappa_{aP}(m_{a},f_{a})T_{P}\,,
\end{align}
with the following corresponding projector coefficients $\kappa_{aP} = 2\,{\rm Tr}[\widetilde{T}_{a}T_{P}]$:
\begin{align}
    \kappa_{a\pi^{0}}(m_a,f_{a})\big|_{m_{a}<1.2\,\GeV}
    &=
    \frac{f_{\pi}}{f_{a}}\frac{2\delta_I m_\pi^2 m_0^2}{9\left(m_\pi^2-m_a^2\right)}
    \left(
    \frac{1}{m_\eta^2-m_a^2}
    +
    \frac{2}{m_{\eta'}^2-m_a^2}
    \right)\,,
    \nonumber\\\label{eq:projector-light-ALP}
    \kappa_{a\eta}(m_a,f_{a})\big|_{m_{a}<1.2\,\GeV}
    &=
    \frac{f_{\pi}}{f_{a}}\frac{\sqrt{6}\,m_0^2}{9\left(m_\eta^2-m_a^2\right)}\,,\nonumber\\
    \qquad
    \kappa_{a\eta'}(m_a,f_{a})\big|_{m_{a}<1.2\,\GeV}
    &=
    \frac{f_{\pi}}{f_{a}}\frac{4\sqrt{3}\,m_0^2}{9\left(m_{\eta'}^2-m_a^2\right)}\,.
\end{align}
Here, $m_0^2\equiv 3m_\eta^2-3m_\pi^2$ and $\delta_I=(m_d-m_u)/(m_d+m_u)\approx 0.36$~\cite{ParticleDataGroup:2024cfk}.

\paragraph{Mass $m_a>1.2\,\GeV$.} In this domain, the description of the ALP interactions above is no longer sufficient because of the following two effects: large-$m_{a}$ scaling following from quark-counting arguments (QCD sum rules), and mixing with heavy pseudoscalar excitations. Let us discuss them in detail.

According to quark-counting rules, the ALP vertices entering the photo- and electroproduction amplitudes should scale with the largest invariant energy scale of the relevant subprocess, which here is the ALP mass $m_a$. The corresponding power-law behavior is determined by the number of constituent fields~\cite{Lepage:1980fj,Aloni:2018vki,Balkin:2025enj}. Specifically, for a subprocess with a total of $n$ fundamental fields and characteristic energy scale $p$, the matrix element in the limit $p \gg \Lambda_{\text{QCD}}$ behaves as~\cite{Lepage:1980fj} $\mathcal{M} \propto p^{4-n}$.

Let us apply this reasoning to the $app$~\eqref{eq:Coupl-pa} and $V\gamma a$ vertices~\eqref{eq:Coupl-Pgammaa}.
They describe the ALP emission, hence $p^{2} = m^{2}_{a}$.
To get the expected scaling, we count the proton as three fields, mesons as two fields, the photon as two fields (as the $V\gamma P$ coupling originates from the mixing of photons with vector mesons), and ALP as two fields (as in Refs.~\cite{Aloni:2018vki,Balkin:2025enj}).
The number of fields in the $ppa$\,($V\gamma a$) vertex is $8$\,($6$), which gives the required scalings $m_{a}^{-4}$\,($m_{a}^{-2}$).
We note that the field assignments for the ALP and photon used in this counting, \emph{i.e.}, treating each as two fields, are chosen to reproduce these asymptotic scalings, following Refs.~\cite{Aloni:2018vki,Balkin:2025enj}, rather than derived from first principles. The counting should therefore be understood as a phenomenological convention that enforces the correct large-mass behavior, rather than as an independent derivation of it.

However, the low-energy effective projector constants from Eq.~\eqref{eq:projector-light-ALP} do not satisfy the relevant scalings.
Namely, applying them to matrix elements of the processes $a \to p\bar{p}$ and $a\to \gamma V$, we obtain $\cM_{a\to p\bar{p}}\propto m_{a}^{-1}$ and $\cM_{a\to V\gamma}\propto {\rm const}$, where the $m_{a}^{-2}$ powers from the ALP matrix~\eqref{eq:Ta-gluon-light} gets partially canceled by the kinematics structure of the matrix element.

Second, the ALP can mix with heavier pseudoscalar excitations such as~\cite{ParticleDataGroup:2024cfk}
\begin{align}
\label{eq:heavy-pseudoscalar-excitations}
    \eta(1295),\ \pi^0(1300),\ \eta(1405),\ \eta(1475),\ \eta(1760),\ \pi^0(1800),\ \eta(2225),\ \eta(2370)\,,
\end{align}
where sometimes $\eta(1405),\ \eta(1475)$ are interpreted as one meson $\eta(1440)$~\cite{Giacosa:2024epf}.
The widths of most of these mesons
are $(50-200)\,\MeV$.
\emph{I.e.}, they may be simultaneously narrow enough not to completely overlap, but broad enough that resonant enhancement from their mixing with ALPs is significant over a substantial portion of the ALP mass range.
In particular, we expect that the mixing with such species dominates the interactions of the ALPs and vector mediators in the mass range $m_{a}\lesssim 2\,\GeV$ (see Refs.~\cite{Ilten:2018crw,Kyselov:2024dmi,Ovchynnikov:2025gpx}).

The properties of some of the heavy pseudoscalar mesons are not known very well~\cite{ParticleDataGroup:2024cfk,Giacosa:2024epf}.
Ref.~\cite{Ovchynnikov:2025gpx} incorporated mixing with $\eta(1295)$, $\pi^{0}(1300)$, and $\eta(1440)$, but so far, there is no approach systematically including all these resonances.
Moreover, the interaction couplings of these mesons needed to calculate their photoproduction are also poorly known, and data are not yet available.

For these reasons, we leave the investigation of this interesting question for future work, while in this study, we do not include these resonances.
Instead, we use a simplified large-mass interpolation of the ALP interaction vertices accounting for the QCD sum rules, following Refs.~\cite{Aloni:2018vki,Balkin:2025enj}.

First, for a gluonic ALP with $m_a\gtrsim 1.2\,\GeV$, we replace the low-mass ALP generator~\eqref{eq:Ta-gluon-light} with
\begin{align}
    \label{eq:Ta-gluon-heavy}
    \widetilde T_a(m_a,f_{a})\bigg|_{m_{a}>1.2\,\GeV}
    =
    -\frac{f_{\pi}}{f_{a}}\frac{2\alpha_s(m_a)}{\sqrt{6}}\,I\,,
\end{align}
where $I$ is the identity in flavor space, with
\begin{align}
    \nonumber
    \kappa_{a\pi}(m_a,f_{a})\big|_{m_{a}>1.2\,\GeV}
    &=
    0\,,
    \\\nonumber
    \kappa_{a\eta}(m_a,f_{a})\big|_{m_{a}>1.2\,\GeV}
    &=
    -\frac{f_{\pi}}{f_{a}}\frac{2}{3}\alpha_s(m_a)\,, \\
    \label{eq:kappa-aP-high}
    \kappa_{a\eta'}(m_a,f_{a})\big|_{m_{a}>1.2\,\GeV}
    &=
    -\frac{f_{\pi}}{f_{a}}\frac{4\sqrt{2}}{3}\alpha_s(m_a)\,.
\end{align}
This choice is motivated by recovering the interaction structure as predicted by the genuine hadronic interactions of the gluonic ALP $\propto \alpha_{s}G_{\mu\nu}\tilde{G}^{\mu\nu}$. In general, this and other similar choices for the large-mass asymptotics create a discontinuity in the ALP production and decay rates.
Ref.~\cite{Balkin:2025enj} minimized the discontinuity in the ALP decay rates.
For $m_a>1.5\,\GeV$, we evaluate $\alpha_s(m_a)$ using \textsc{RunDec}~\cite{Chetyrkin:2000yt}, while for $1\,\GeV<m_a<1.5\,\GeV$, we use the interpolation $\alpha_s(m_a)=\left(1\,\GeV/m_a\right)^{2.6}\,$.

In addition, Ref.~\cite{Aloni:2018vki} approximated the domain of pseudoscalar excitations by a fit of the scattering data $e^{+}e^{-}\to V^{*}\to VP$, where $V$ are various vector states. This resulted in a phenomenological function $\cF(m_a)$, given by
\begin{align}
\label{eq:cF}
    \cF(m_a)
    =
    \begin{cases}
    1\,,
    & m_a<\beta_{\mathcal F}\,,
    \\[4pt]
    \text{interpolation}\,,
    & \beta_{\mathcal F}\le m_a\le 2\,\GeV\,,
    \\[4pt]
    \left(\dfrac{\beta_{\mathcal F}}{m_a}\right)^4\,,
    & m_a>2\,\GeV\,,
    \end{cases}
\end{align}
where $\beta_{\mathcal F}\approx 1.4\,\GeV$, and interpolation resembles the shape from Fig.~1 of Ref.~\cite{Aloni:2018vki}. In the large-$m_{a}$ domain for the $Va\gamma$ process with Eq.~\eqref{eq:kappa-aP-high}, it also implies the large-$m_{a}$ scaling dictated by the QCD sum rules.

Collecting these results, we may construct channel-dependent projectors $\kappa_{aP,i}(m_{a})$ defined in Eqs.~\eqref{eq:gXSM-generic},~\eqref{eq:kappa-process-dependent}. For the interactions $i = app,aV\gamma$, they are given by
\begin{align}
    \label{eq:kappa-aPZ}
    \kappa_{aP,i}(m_a,f_{a})
    =
       \kappa_{aP}(m_a,f_{a})\times \cF(m_{a})\times \begin{cases}
    1\,,
    & m_a<\beta_{\mathcal F}\,,
    \\[6pt]
    \left(\dfrac{\beta_{\mathcal F}}{m_a}\right)^{n_{i}}\,,
    & m_a\ge \beta_{\mathcal F}\,,
    \end{cases}
\end{align}
where $n_{aV\gamma} = 0$ and $n_{app} = 1$ are required by the QCD sum rules.

In the case of the $app$ interaction, Ref.~\cite{Blinov:2021say} also introduced the timelike proton form factor $F_{A}(m_{a})$, which incorporates the mixing with the axial-vector state $a_{1}^{0}$. It does not lead to a resonant enhancement of the production amplitude, which is due to the Lorentz-structure of the ALP-$a_{1}$ operator~\cite{Ovchynnikov:2025gpx}. We do not include it because it does not account for the truly resonant mixing with heavy pseudoscalar mesons.

Overall, Eq.~\eqref{eq:kappa-aPZ} provides a highly conservative estimate and may substantially underestimate the ALP yield. In the mass range $m_a \lesssim 2.4\,\GeV$, it neglects contributions from heavier excitations, which are expected only to enhance the couplings. Moreover, instead of retaining the naive low-mass behavior $\widetilde{T}_a \propto 1/m_a^2$ of the generator in Eq.~\eqref{eq:Ta-gluon-light}, we impose the much stronger suppression $1/m_a^{4+n_i}$ suggested by QCD sum rules.

\subsubsection{Vector mediators}
\label{sec:vector-mediators}

We also consider new vector states $V'$ coupled to anomaly-free combinations of baryon and lepton currents:
\begin{align}
    \label{eq:Lagr_V'}
    \cL_{V'}
    =
    g_{V'}V^{'\mu}\sum_q x_q \bar q\gamma_\mu q
    +
    g_{V'}V^{'\mu}\sum_\ell
    \left(
    x_\ell \bar\ell\gamma_\mu\ell
    +
    x_{\nu_\ell}\bar\nu_\ell\gamma_\mu P_L\nu_\ell
    \right)\,,
\end{align}
where the coefficients $x_q$ and $x_\ell$ are chosen such that the corresponding current is anomaly-free.
Typical examples include dark photons, for which $x_f=Q_f$, and $B-L$ mediators, for which $x_\ell=x_{\nu_\ell}=-3x_q=1$~\cite{Holdom:1985ag,Basso:2008iv,Heeck:2014zfa,Ilten:2018crw}.

For $m_{V'}\lesssim 1\,\GeV$, the hadronic interactions of $V'$ may be understood using the HLS implementation of vector meson dominance~\cite{Fujiwara:1984mp,Ilten:2018crw,Baruch:2022esd,Kyselov:2024dmi}, according to which $V'$ mixes with families of $\rho^{0},\omega,\phi$ mesons.
The generator of the $V'$ particle is
\begin{align}
    \label{eq:Ttilde-Vprime}
    \widetilde T_{V'}(m_{V'},g_{V'})\big|_{m_{V'}\lesssim 1\,\GeV}
    =
    \sum_{V=\rho^0,\omega,\phi}
    \kappa_{V'V}(m_{V'},g_{V'})\big|_{m_{V'}<1\,\GeV}\,T_V\,.
\end{align}
Explicit form of the coefficients $\kappa_{V'V}(m_{V'})$ is
\begin{align}
    \kappa_{V'V}(m_{V'})\big|_{m_{V'}<1\,\GeV} = g_{V'}\frac{m_{V}^{2}}{m_{V'}^{2}-m_{V}^{2}-i\Gamma_{V}m_{V}}2\, {\rm Tr}[T_{V}\times{\rm diag}(x_{u},x_{d},x_{s})],
\end{align}
or, after using the generators~\eqref{eq:TV},
\begin{align}
 &\kappa_{V'\rho^{0}}(m_{V'},g_{V'})\big|_{m_{V'}<1\,\GeV} = g_{V'}\frac{m_{\rho^{0}}^{2}}{m_{V'}^{2}-m_{\rho^{0}}^{2}-i\Gamma_{\rho^{0}}m_{\rho^{0}}}(x_{u}-x_{d})\,, \\ &\kappa_{V'\omega}(m_{V'},g_{V'})\big|_{m_{V'}<1\,\GeV} = g_{V'}\frac{m_{\omega}^{2}}{m_{V'}^{2}-m_{\omega}^{2}-i\Gamma_{\omega}m_{\omega}}(x_{u}+x_{d})\,, \\ &\kappa_{V'\phi}(m_{V'},g_{V'})\big|_{m_{V'}<1\,\GeV} = -g_{V'}\frac{m_{\phi}^{2}}{m_{V'}^{2}-m_{\phi}^{2}-i\Gamma_{\phi}m_{\phi}}\sqrt{2}x_{s}\,.
\end{align}
For larger masses, the description becomes more complicated.
Similar to the ALPs, the vector mediator may mix with heavier vector excitations, such as $\omega(1420)$, $\rho^0(1450)$, $\omega(1650)$, $\phi(1680)$, $\rho^0(1700)$, and $\phi(2170)$.
Unlike the ALP case, the mixing structure of $V'$ with these excitations may be accurately captured by the EM scattering data~\cite{Ilten:2018crw}, which replaces the Breit-Wigner form with an appropriate line shape; in addition, they are typically much broader and overlap.
However, with the EIC data, one can systematically estimate these rates based on the $\ell^+\ell^-$ spectra as done, for example, at LHCb dark photon searches~\cite{Ilten:2016tkc,LHCb:2017trq} and rescale them to other models based on $SU(3)_F$ symmetry.
Currently, we neglect heavier vector resonances.
As in the pseudoscalar case, this should be viewed as a conservative approximation, since the omitted states are generally expected to enhance the production rate.
\section{Matrix elements of electroproduction}
\label{sec:matrix-elements}

The matrix elements of the electroproduction process of Eq.~\eqref{eq:process} for pseudoscalars and vectors, denoted as $X$, can be parameterized as
\begin{align}
    \label{eq:electroproduction-matrix-element}
    i\cM
    =
    i \,\frac{J_{\mu}^{e}}{(p_{e}-p_{e'})^{2}}\, \cM^{\mu}_{X},
\end{align}
with $J_{\mu}^{e} = e\bar{u}(p_{e'})\gamma_{\mu}u(p_{e})$ the electron electromagnetic current, and $\cM^{\mu}_{X}$ a process-dependent hadronic current describing the subprocess $\gamma^{*}+p \to X+p'$. The object $\cM^{\mu}_{X}$ satisfies the Ward identity,
$(p_{e}-p_{e'})_{\mu}\cM^{\mu}_{X} = 0$.
The Lorentz-invariant kinematical variables are defined as
\begin{align}
    \label{eq:LorentzInv}
    s \equiv (p_e+p_p)^2\,,\quad
    s_1 \equiv (p_{e'}+p_X)^2\,,\quad
    s_{\gamma^{*}p} \equiv (p_X+p_{p'})^2 \,, \quad
    Q^2 \equiv -(p_e-p_{e'})^2\,,\quad
    t_{p} \equiv (p_p-p_{p'})^2\,.
\end{align}
Crucially, the beam-energy configurations at the EIC, the kinematic selection we adopt (a forward proton $p'$ and energetic $X$ in the transverse direction), and the structure of the matrix element with the $Q^{2}$ pole favor the domain
$Q^{2}\ll 1\,\GeV^{2}$, $|t_{p}| < 2\,\GeV^{2}$, and, for most accepted events, $\sqrt{s_{\gamma^{*}p}}\gtrsim 2\,\GeV$; see Sec.~\ref{sec:events} and Fig.~\ref{fig:distributions-sgammap-Q2-tp}. In this range, the electroproduction cross section is modeled in terms of exchanges with various intermediate hadronic states.

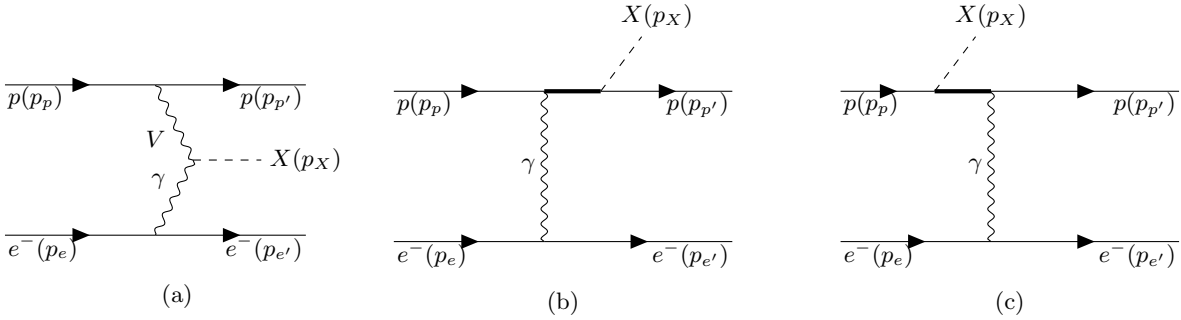
\begin{figure*}[t!]
\centering

\tikzset{
  extlabelL/.style={below=5pt, anchor=west, inner sep=1pt},
  extlabelR/.style={below=5pt, anchor=east, inner sep=1pt},
}

\begin{minipage}{0.31\textwidth}
\vspace{1cm}
\centering
\begin{tikzpicture}
\begin{feynman}
  \vertex (pL) at (-2,  1);
  \vertex (eL) at (-2, -1);
  \vertex (pR) at ( 2,  1);
  \vertex (eR) at ( 2, -1);

  \vertex (Vp) at (0,  1);
  \vertex (Ve) at (0, -1);

  \vertex (F)  at (0.5, 0);
  \vertex (X)  at (2, 0) {$X(p_X)$};

  \diagram*{
    (pL) -- [fermion] (Vp) -- [fermion] (pR),
    (eL) -- [fermion] (Ve) -- [fermion] (eR),
    (Ve) -- [boson, edge label={$\gamma$}] (F),
    (Vp) -- [boson, edge label'={$V$}] (F),
    (F)  -- [scalar] (X)
  };
\end{feynman}

\node[extlabelL] at (pL) {$p(p_p)$};
\node[extlabelL] at (eL) {$e^-(p_e)$};
\node[extlabelR] at (pR) {$p(p_{p'})$};
\node[extlabelR] at (eR) {$e^-(p_{e'})$};

\end{tikzpicture}

\vspace{4pt}
(a)
\end{minipage}
\begin{minipage}{0.31\textwidth}
\centering
\begin{tikzpicture}
\begin{feynman}
  \vertex (pL) at (-2,  1);
  \vertex (eL) at (-2, -1);
  \vertex (pR) at ( 2.5,  1);
  \vertex (eR) at ( 2.5, -1);

  \vertex (Vp) at (0,  1);
  \vertex (Ve) at (0, -1);

  \vertex (pm) at (0.75, 1);
  \vertex (a)  at (1.5, 2) {$X(p_X)$};

  \diagram*{
    (pL) -- [fermion] (Vp) -- [plain, ultra thick] (pm) -- [fermion] (pR),
    (eL) -- [fermion] (Ve) -- [fermion] (eR),
    (Vp) -- [photon, edge label'={$\gamma$}] (Ve),
    (pm) -- [scalar, dashed] (a)
  };
\end{feynman}

\node[extlabelL] at (pL) {$p(p_p)$};
\node[extlabelL] at (eL) {$e^-(p_e)$};
\node[extlabelR] at (pR) {$p(p_{p'})$};
\node[extlabelR] at (eR) {$e^-(p_{e'})$};

\end{tikzpicture}

\vspace{4pt}
(b)
\end{minipage}
\hfill
\begin{minipage}{0.31\textwidth}
\centering
\begin{tikzpicture}
\begin{feynman}
  \vertex (pL) at (-2,  1);
  \vertex (eL) at (-2, -1);
  \vertex (pR) at ( 2.5,  1);
  \vertex (eR) at ( 2.5, -1);

  \vertex (Vp) at (0,  1);
  \vertex (Ve) at (0, -1);

  \vertex (pm) at (-0.75, 1);
  \vertex (a)  at (0., 2) {$X(p_X)$};

  \diagram*{
    (pL) -- [fermion] (pm) -- [plain, ultra thick] (Vp) -- [fermion] (pR),
    (eL) -- [fermion] (Ve) -- [fermion] (eR),
    (Vp) -- [photon, edge label'={$\gamma$}] (Ve),
    (pm) -- [scalar, dashed] (a)
  };
\end{feynman}

\node[extlabelL] at (pL) {$p(p_p)$};
\node[extlabelL] at (eL) {$e^-(p_e)$};
\node[extlabelR] at (pR) {$p(p_{p'})$};
\node[extlabelR] at (eR) {$e^-(p_{e'})$};

\end{tikzpicture}

\vspace{4pt}
(c)
\end{minipage}
\hfill

\caption{Feynman diagrams for the electroproduction process $e\,p \to e\,p X$ with $X=(\pi^0\,,\eta\,,\eta'\,,a)$:
(a) $t$-channel fusion production of $X$, where $V$ is a vector meson $V = \rho,\omega$;
(b) $s$-channel bremsstrahlung-like process with an intermediate baryonic resonance, where $X$ is emitted by the outgoing proton leg (retarded emission);
(c) $u$-channel bremsstrahlung-like process with $X$ emitted from the incoming proton leg (advanced emission). In this work, we adopt the description of Ref.~\cite{Kashevarov:2017rmk}, where the production is described via the $V$-exchange process (a), with absorptive cut contributions from a Pomeron $\mathbb{P}$ and the tensor meson $f_{2}$.}
\label{fig:diagrams-pseudoscalar}
\end{figure*}

The first broad class is given by $t$-channel exchange processes, in which the proton legs do not emit the final-state particle $X$.
These include non-diffractive contributions, such as exchanges by mesons with different spins depending on the quantum numbers of $X$, see Fig.~\ref{fig:diagrams-pseudoscalar}\,(a) and Fig.~\ref{fig:diagrams-vector}\,(b), as well as diffractive contributions, typically described by effective Pomeron exchange $\mathbb{P}$, which controls the energy dependence at large $\sqrt{s_{\gamma^{*} p}}$, see Fig.~\ref{fig:diagrams-vector}\,(a).
For $t$-channel exchange, the intermediate states are off-shell independently of the value of $s_{\gamma^{*}p}$.

The second broad class is given by bremsstrahlung-like $s$- and $u$-channel topologies involving intermediate baryon resonances with spins
$S = 1/2, 3/2, 5/2, \dots$, as illustrated in Fig.~\ref{fig:diagrams-pseudoscalar}\,(b) and (c).
Such excitations may go on-shell if $\sqrt{s_{\gamma^{*} p}}\lesssim 2\,\GeV$~\cite{ParticleDataGroup:2024cfk}.
In general, describing this region requires an explicit treatment of individual baryonic resonances.

Before specifying the channel-dependent amplitudes, let us emphasize an important conceptual point.
In the kinematic domain relevant for this work, there is, in general, no unique microscopic decomposition of exclusive photo- or electroproduction into a fixed set of intermediate hadronic mechanisms.
Different studies often describe the same final state $X$ using different effective sets of explicitly included exchange mechanisms or production topologies~\cite{Chiang:2002vq,Nakayama:2005ts,Sibirtsev:2010cj,Kaskulov:2008ej,Kaskulov:2010kf,Kaskulov:2011wd,Kashevarov:2017vyl,Kashevarov:2017rmk,Yu:2019wly,Zhang:2021esc,Friman:1995qm,Pichowsky:1996tn,Donnachie:1999yb,Ryu:2012tw,Obukhovsky:2009th,Sibirtsev:2003qh,Wei:2019imo,Oh:2003aw,Wang:2025rvr,Xu:2021mju,Wei:2024lne,Wang:2017plf,Tang:2024pky,Tang:2025qqe}.
The corresponding pole contributions may then be promoted to Reggeized amplitudes, and, in some cases, supplemented by absorptive Regge-cut terms that effectively account for additional non-pole strength not included as separate explicit exchanges.
Such choices are typically motivated by phenomenological dominance in a given kinematic regime, simplicity of the parametrization, and the availability of data with which the corresponding description may be calibrated.
As a result, distinct representations of the hadronic current can yield very similar predictions for integrated rates and for broad classes of differential distributions across overlapping regions of phase space.

Accordingly, the intermediate states appearing in our amplitudes should not be interpreted as defining a unique microscopic production mechanism.
Rather, they provide a phenomenological representation of the hadronic current appropriate for the observables and kinematic regime under consideration.
An explicit example, discussed below in Sec.~\ref{sec:production-pseudoscalar}, is neutral pion production: some descriptions include contributions from nucleon resonances together with vector-meson exchange~\cite{Kaskulov:2008ej,Kaskulov:2010kf,Kaskulov:2011wd,Zhang:2021esc}, whereas others employ a forward Regge-based description with vector exchanges only, supplemented by absorptive cuts from additional trajectories~\cite{Yu:2011zu,Kashevarov:2017vyl}.
Future electroproduction data from the EIC may help discriminate between such effectively degenerate descriptions and thereby guide the development of a more unified treatment across different final states.

Within any chosen phenomenological representation, the relative signs and magnitudes of the individual contributions matter through interference and can affect differential distributions more strongly than inclusive or broadly integrated rates.
However, these parameters are fixed collectively to production data and are strongly correlated, so varying one coupling or sign independently would not provide a meaningful uncertainty estimate; we instead characterize their combined impact through the empirical agreement of the full amplitude with the measured cross sections.

The selected kinematics motivates three simplifications in our study.
The first is that we model the electroproduction current $\cM^{\mu}_{X}$ in Eq.~\eqref{eq:electroproduction-matrix-element} using the corresponding photoproduction amplitudes for the SM mesons
$\pi^{0},\eta,\eta',\rho^{0},\omega,\phi$, which we denote by
$\left[\cM^{\mu}_{X}\right]_{\rm photopr.}$.
The extrapolation to electroproduction is performed in two steps:
(i) the incoming on-shell photon momentum $p_{\gamma}$ is replaced by the off-shell virtual-photon momentum
$p_{\gamma^{*}} \equiv p_{e}-p_{e'}$,
and
(ii) a finite-$Q^{2}$ form factor $F_{X}(Q^{2})$ is included to account effectively for the compositeness of the hadronic electromagnetic vertices,
\begin{align}
    \label{eq:photo-to-electro}
    \cM^{\mu}_{X}
    \approx
    F_{X}(Q^{2})\,
    \left[\cM^{\mu}_{X}\right]_{\rm photopr.}\Big|_{p_{\gamma}\to p_{\gamma^{*}}}\, .
\end{align}
This construction is similar in spirit to the approaches adopted in
Refs.~\cite{Kaskulov:2008ej,Kaskulov:2010kf,Lomnitz:2018juf}.

For the $\rho^{0}$ and $\phi$ mesons, we use form factors fitted to HERA data~\cite{H1:2009cml}.
Those fits are based on the approximation that the total electroproduction cross section factorizes relative to the photoproduction one at fixed $\sqrt{s_{\gamma^{*}p}}=75\,\GeV$,
\begin{align}
    \label{eq:factorization-HERA}
    \sigma_{\gamma^{*}+p\to p'+V}\Big|_{\sqrt{s_{\gamma^{*}p}}=75\,\GeV}
    \approx
    |F_{V}(Q^{2})|^{2}\times
    \sigma_{\gamma+p\to p'+V}\Big|_{\sqrt{s_{\gamma p}}=75\,\GeV}\,.
\end{align}
The resulting fitted form factors for $\rho^0$ and $\phi$ are
\begin{align}
    \label{eq:photo-to-electro-vectors}
    F_{V}(Q^{2})
    =
    \left(\frac{1}{1+Q^2/m_{V}^2}\right)^{n},
    \quad
    n
    =
    \begin{cases}
    1.05+3.5\times 10^{-3}\,\GeV^{-2}(m_{\rho}^{2}+Q^{2}),
    \quad V = \rho^{0}\,,\\
    1.07+3.7\times 10^{-3}\,\GeV^{-2}(m_{\phi}^{2}+Q^{2}),
    \quad V = \phi\,.
    \end{cases}
\end{align}
For $\omega$, for which the available electroproduction data are more limited, we assume the same functional form as for $\rho^{0}$.
For the pseudoscalar mesons and ALPs, we use the analogous ansatz
\begin{equation}
    F_{P}(Q^{2})
    =
    \left(\frac{1}{1+Q^{2}/m_{\rho}^{2}}\right)^{1+3.5\times 10^{-3}\,\GeV^{-2}(m_{P}^{2}+Q^{2})}\,,
    \label{eq:photo-to-electro-pseudoscalars}
\end{equation}
which mirrors the observed scaling pattern of the light vector mesons.

The use of the empirical form factors in Eq.~\eqref{eq:photo-to-electro} is necessarily approximate and may be systematically improved once electroproduction data from the EIC become available.
First, they are extracted at fixed $\sqrt{s_{\gamma^{*} p}} = 75\,\GeV$, which lies well above the $s_{\gamma^*p}$ domain most relevant for the EIC.
Second, Eq.~\eqref{eq:factorization-HERA} encodes the $Q^{2}$ dependence only at the level of the integrated cross section and therefore does not capture correlations between $Q^{2}$ and the other kinematic variables.
In our treatment based on Eq.~\eqref{eq:photo-to-electro}, the exact $2\to 3$ kinematics is retained, and the finite photon virtuality is kept explicitly in the operator for the subprocess $\gamma^{*}+p\to p'+X$.
As a result, the form factors in Eq.~\eqref{eq:photo-to-electro-vectors} should be viewed only as a practical phenomenological ansatz rather than as a unique prediction for the full multidifferential electroproduction amplitude.
These form factors may be refined once EIC electroproduction data become available; we leave this to future work.

The HERA $\rho^0$ and $\phi$ electroproduction data partially constrain the finite-$Q^2$ dependence at the level of integrated cross sections~\cite{H1:2009cml}.
For the $\omega$ and pseudoscalar channels, we are not aware of measurements that directly constrain the corresponding form-factor ans\"atze in the low-$Q^2$, few-GeV-energy region relevant here.

To assess the associated uncertainty, we vary the exponent of the baseline form factor over $0\leq n\leq2$, spanning no finite-$Q^2$ suppression through dipole-like suppression.
In the region $Q^2\lesssim0.1\,\GeV^2$, which dominates the fiducial yield, this changes $|F_X(Q^2)|^2$ by at most approximately $40\%$ relative to the baseline and by substantially less at smaller $Q^2$.
We therefore adopt approximately $40\%$ as a conservative estimate of the form-factor uncertainty on the predicted fiducial rates.

The second simplification is that we do not include individual baryonic excitations explicitly in the electroproduction amplitudes.
This is motivated by several considerations.
Our event selection mostly favors the region $\sqrt{s_{\gamma^{*}p}}>2\,\GeV$, where baryonic excitations do not go on-shell.
In addition, the number of resonances is large, and their widths are not narrow, so their contributions overlap strongly and do not generically produce isolated resonant enhancements.
Finally, in electroproduction, we integrate over a continuous range of $s_{\gamma^{*} p}$, which further dilutes the impact of individual resonances.
In practice, we include such effects only collectively, through phenomenological Reggeized descriptions where appropriate.

The third simplification is that we restrict ourselves to unpolarized $e$ and $p$ beams.
First, for the observables emphasized in this work -- total rates, fiducial yields, and broad multidifferential kinematic distributions -- this is expected to provide an adequate baseline description.
Polarization-dependent effects typically enter through spin asymmetries and azimuthal modulations, which are usually subleading corrections to the unpolarized cross section rather than order-one distortions of the event kinematics.
Moreover, such observables are commonly constructed from comparisons between different regions in the azimuthal angle $\phi$, and are therefore not expected to be strongly affected by acceptance asymmetries in the approximately azimuthally symmetric central ePIC detector, aside from possible complications associated with selected beamline instrumentation.
Finally, while most of the EIC data will be taken with polarized beams, the unpolarized sample can always be obtained by averaging over data taken with different beam-polarization configurations.
A dedicated treatment of polarization-dependent observables would nevertheless be valuable, but goes beyond the scope of the present work.

For the SM meson channels considered in this work, the hadronic amplitudes are calibrated to existing photoproduction data.
The residual modeling uncertainty is therefore best characterized empirically, by the level at which the framework reproduces these data, rather than by propagating uncertainties on individual model ingredients (Regge residues, absorptive-cut parameters, effective couplings), which are not independent once the fit is performed.
In the kinematic domain where the fitted description applies, namely forward kinematics, away from on-shell baryonic resonances, and $\sqrt{s_{\gamma^{*}p}} \gtrsim 2\,\mathrm{GeV}$, the empirical accuracy of the photoproduction component is typically within $30\%$.
An exception is neutral pion production in the lowest-energy EIC configuration, where the accepted events probe $\sqrt{s_{\gamma^{*}p}} \lesssim 2\,\mathrm{GeV}$.

An additional source of uncertainty arises from the extrapolation from photoproduction to electroproduction, encoded in our treatment of the finite-$Q^{2}$ dependence of the amplitudes.
The quoted $30\%$ empirical accuracy does not include this separate finite-$Q^{2}$ uncertainty, which is quantified above for the low-$Q^{2}$ region dominating the fiducial yields.

Let us now turn to the new states $X$. We construct the electroproduction cross-section using the recipe of Eq.~\eqref{eq:MX-projector-based} and the results discussed in Secs.~\ref{sec:alp},~\ref{sec:vector-mediators}. The main source of the uncertainty, especially in the mass domain $m_{X}\gtrsim 1\,\GeV$, is a model dependence associated with the treatment of mixing with excited hadronic states.
Therefore, the quoted $\sim 30\%$ accuracy should be understood as applying only to the SM meson channels in the validated kinematic regime, and not to the heavier BSM scenarios considered below. For the latter, our estimates are very conservative, with the true production rates potentially larger by an order of magnitude or more. Because the spectrum and couplings of the omitted heavier excitations are model dependent, no model-independent upper envelope can be assigned.

%

\subsection{Pseudoscalar mesons}
\label{sec:production-pseudoscalar}

The photoproduction of pseudoscalar mesons $P=\pi^{0},\eta,\eta'$ has been studied in numerous works~\cite{Chiang:2002vq,Nakayama:2005ts,Sibirtsev:2010cj,Kaskulov:2008ej,Kaskulov:2010kf,Kaskulov:2011wd,Kashevarov:2017vyl,Kashevarov:2017rmk,Yu:2019wly,Zhang:2021esc}.
A common description for $\sqrt{s_{\gamma p}}\simeq\text{few GeV}$ combines $t_{p}$-channel exchanges by a photon, vector $V$, and axial-vector mesons (diagram (a) in Fig.~\ref{fig:diagrams-pseudoscalar}), and/or $s/u$-channel exchanges by baryonic resonances $B$ (diagrams (b), (c)).
The photon-photon fusion topology produces the proton at very small momentum transfer, $|t_{p}|\ll 1\,\GeV^2$, corresponding to a very soft $X$.
It is therefore outside our event selection discussed in Sec.~\ref{sec:events}.
The axial-vector contributions to unpolarized differential rates are typically strongly suppressed. Unless one considers polarization-sensitive observables (e.g., Fig.~1 of~\cite{Kaskulov:2011wd}), they may be safely neglected.

Finally, for sufficiently small $|t_{p}|\lesssim 2\,\GeV^2$ and $\sqrt{s_{\gamma p}}> (2-3)\,\GeV$, the net impact of explicit $s/u$-channel baryonic-resonance exchanges becomes largely degenerate with an effective forward-angle description based on $t$-channel natural-parity exchange.
This is reflected by the fact that models including both $V$ and $B$ mechanisms~\cite{Kaskulov:2011wd,Zhang:2021esc} and, alternatively, Regge-based approaches dominated by $t$-channel $V$ exchange~\cite{Yu:2011zu,Kashevarov:2017vyl} can provide comparably good descriptions of $\pi^{0}/\eta$ photoproduction data in the forward region.
In particular, Refs.~\cite{Yu:2011zu,Kashevarov:2017vyl} start from Reggeized $\rho$ and $\omega$ exchange and supplement it with additional non-pole strength via Regge cuts (e.g. $\rho\mathbb{P}$, $\rho f_{2}$, $\omega\mathbb{P}$, $\omega f_{2}$).
These cut contributions are branch-point (continuum) terms associated with two-Reggeon exchange/absorptive unitarity corrections.
They should therefore not be interpreted as introducing additional single-particle exchanges with independent vertex structures.

For the $\pi^{0},\eta$ production, we will follow Ref.~\cite{Kashevarov:2017vyl} and, specifically, adopt their ``Solution~I'', which accurately fits the photoproduction data in the domain $\sqrt{s_{\gamma p}}\gtrsim 3\,\GeV$.
This provides a baseline Regge(-cut) description in which the forward-angle cross section at high energy is governed by natural-parity $t_{p}$-channel exchange: Reggeized $\rho$ and $\omega$ pole terms, supplemented by absorptive cut corrections driven by $Y=\mathbb{P},f_{2}$, in both polarized and unpolarized beams.
In this construction, the $Y$ states enter through the cut trajectories and phases defining the $YV$ cuts, implemented by modifying the Regge propagator factor multiplying the same $\lambda_{V}g^{v,t}_{V}$ residues (cf.~Eq.~(23) of~\cite{Kashevarov:2017vyl}), rather than by introducing separate dedicated $f_{2}$ or Pomeron amplitudes with their own vertices.
This is particularly natural for the Pomeron, which in Regge phenomenology is treated as an effective vacuum exchange with quantum numbers $J^{PC} = 0^{++}$.
Unnatural-parity axial-vector exchanges are neglected at the level of unpolarized rates and are only expected to matter for selected polarization observables through interference.

For the lightest pseudoscalar channel, $X=\pi^{0}$, the accepted electroproduction sample can probe the low-intermediate region $\sqrt{s_{\gamma^{*}p}}\sim 2$--$3\,\GeV$. In this range, the description of Ref.~\cite{Kashevarov:2017vyl} is expected to be less robust because of the proximity to the domain of baryonic resonances. To assess the sensitivity of our results to the modeling choice, in Appendix~\ref{app:pi0-production-kaskulov} we compare our baseline treatment to an alternative description of the high-$s_{\gamma p}$ data based on Refs.~\cite{Kaskulov:2008ej,Kaskulov:2010kf,Kaskulov:2011wd}. As seen from Fig.~1 of Ref.~\cite{Kaskulov:2011wd}, the alternative approach approximately fits the photoproduction for the domain $\sqrt{s_{\gamma p}}>3.5\,\GeV$. However, it overestimates the differential cross-section $d\sigma_{\gamma p\to \pi^{0}p}$ at low invariant masses, with the discrepancy being a factor of $\simeq 3$ at the lowest invariant mass $\sqrt{s_{\gamma p}}\approx 2.9\,\GeV$ for which the approach of Ref.~\cite{Kashevarov:2017vyl} matches the data well (see Fig.~\ref{fig:comparison-kaskulov-vs-kashevarov} of the Appendix).

After imposing our event selection for the C1 configuration from Tbl.~\ref{tab:energy-configurations}, this alternative approach gives a fiducial $\pi^{0}$ cross section larger by a factor of about $4-5$ than the baseline result of Ref.~\cite{Kashevarov:2017vyl}. In light of what was written above, we do not interpret the difference between the two predictions as a realistic estimate of the uncertainty.
Rather, it should be viewed as a very conservative upper bound on the uncertainty associated with the low-energy extrapolation.

For $\eta'$ photoproduction, existing observations are concentrated in the near-threshold and low-intermediate energy region, with high-statistics unpolarized differential cross sections available up to $\sqrt{s_{\gamma p}}\simeq 2.84\,\GeV$~\cite{CLAS:2009wde,Kashevarov:2017rmk,CBELSATAPS:2009ntt}.
Phenomenological descriptions in this domain typically rely on the $s,u$-baryon exchange/isobar amplitudes with $s$-channel resonance contributions plus $t$-channel vector meson exchange~\cite{Chiang:2002vq,Tiator:2018heh,Zhang:2021esc}.
At higher energies, GlueX~\cite{GlueX:2019adl} reported the first measurement of the $\eta'$ photon-beam asymmetry $\Sigma$ in the range $\sqrt{s_{\gamma p}}\simeq (4.03-4.17)\,\GeV$, measured as a function of $-t$ using a linearly polarized tagged photon beam.
This observable strongly constrains the relative weight of natural- vs.\ unnatural-parity exchange in the forward (Regge) regime, but it does not by itself fix the absolute normalization of the production rate in the absence of corresponding high-energy $d\sigma/dt$ data.

Since our selection prefers $\sqrt{s_{\gamma p}}>3\,\GeV$ largely independent of the EIC beam configuration, we therefore adopt the same baseline description for $\eta'$ production as for $\eta$, and treat the resulting $\eta'$ yields as model-dependent extrapolations.
The fit parameters entering the $\eta'$ calculation can be revisited and tuned once additional high-energy cross-section data and polarization observables become available.

Within the approach of Ref.~\cite{Kashevarov:2017vyl}, the hadronic current for pseudoscalar photoproduction is written as
\begin{align}
    \label{eq:matrix-element-pseudoscalar-meson}
    \cM^{\mu}_{P}(s_{\gamma^{*}p},t_p)
    =
    \sum_{V=\rho,\omega}\cM^{\mu}_{P,V}(s_{\gamma^{*}p},t_p)\,,
\end{align}
where
\begin{align}
   \cM^{\mu}_{P,V}(s_{\gamma^{*}p},t_p)= \bar u(p_{p'})
    \,\Gamma_{VNN,\beta}(p_{p'},p_{p})\,
    u(p_{p})\;
    \Gamma_{V\gamma P}^{\mu\beta}(p_{\gamma^{*}},p_{p'}-p_p)\;
    \tilde{\cR}^{P}_{V}(s_{\gamma^{*}p},t_{p}) \, ,
\end{align}
where $\Gamma_{VNN,\beta}$ and $\Gamma^{\mu\beta}_{V\gamma P}$ stand for the $VNN$ and $V\gamma P$ interactions from Eqs.~\eqref{eq:Vert-Vpp} and \eqref{eq:Vert-VYgamma}, respectively.

The $V=\rho,\omega$ Regge factors entering Eq.~\eqref{eq:matrix-element-pseudoscalar-meson} include the Regge-pole contribution $\cR_{V}(s_{\gamma^{*}p},t_{p})$ and absorptive Regge-cut corrections $\cR_{YV}(s_{\gamma^{*}p},t_{p})$,
\begin{align}
    \label{eq:Rtilde}
    \tilde{\cR}^{P}_{V}(s_{\gamma^{*}p},t_{p})
    =
    \cR_{V}(s_{\gamma^{*}p},t_{p})
    +
    \sum_{Y=\mathbb{P},f_2}
    \cR^{P}_{YV}(s_{\gamma^{*}p},t_{p})\,.
\end{align}
In this way, the pole and cut contributions enter at the amplitude level and therefore contribute to the cross-section both individually and through their interference.
Explicitly,
\begin{align}
    \label{eq:ReggeVector}
    \cR_{V}(s_{\gamma^{*}p},t_{p})
    &=
    -\left(\frac{s_{\gamma^{*}p}}{s_0}\right)^{\alpha_V(t_{p})-1}
    \alpha_{V,1}\,
    \Gamma\!\big(1-\alpha_V(t_{p})\big)\,
    \frac{1-\exp[-i\pi\alpha_V(t_{p})]}{2}\,,
    \\
    \label{eq:ReggeCut}
    \cR^{P}_{YV}(s_{\gamma^{*}p},t_{p})
    &=
    c^{P}_{YV}\,
    \exp\!\left[-\frac{i\pi}{2}\alpha_{YV}(t_{p})\right]\,
    \left(\frac{s_{\gamma^{*}p}}{s_0}\right)^{\alpha_{YV}(t_{p})-1}\,
    \exp\!\big(d^{P}_{YV}\,t_{p}\big)\,.
\end{align}
Here, $s_{0} = 1\,\GeV^{2}$, $\Gamma(x)$ is the gamma function, and $\alpha_{Y}(t_{p})$, $\alpha_{YV}(t_{p})$ are Regge trajectories assumed to be linear in $t_p$,
\begin{align}
    \label{eq:linear-trajectories}
    \alpha_{Y}(t_{p})
    =
    \alpha_{Y,0}+\alpha_{Y,1}\,t_{p}\,,
    \quad \quad
    \alpha_{YV}(t_{p})
    =
    \alpha_{YV,0}+\alpha_{YV,1}\,t_{p}\,,
\end{align}
where the intercepts $\alpha_{Y/YV,0}$ and slopes $\alpha_{Y/YV,1}$ are taken from Ref.~\cite{Kashevarov:2017vyl} and are collected in Tbl.~\ref{tab:regge-params-solutionI}. 
Likewise, the values of $c_{YV}^P$ and $d_{YV}^P$ in Eq.~\eqref{eq:ReggeCut} are also given in Tbl.~\ref{tab:regge-params-solutionI} for $\pi^0$ and $\eta$ production; for the $\eta'$ case, in the absence of data, we take the same values as for the $\eta$ production.
\begin{table*}[t]
\centering
\begin{ruledtabular}
\begin{tabular}{lcc cc cc}
\multicolumn{1}{c}{Exchange / cut}
& \multicolumn{2}{c}{Trajectory $\alpha = \alpha_{0}+\alpha_{1}t_{p}$}
& \multicolumn{2}{c}{$\gamma p\to\pi^0 p$}
& \multicolumn{2}{c}{$\gamma p\to\eta p$}
\\
\multicolumn{1}{c}{}
& \multicolumn{1}{c}{$\alpha_{Y,0}$}
& \multicolumn{1}{c}{$\alpha_{Y,1}$ [$\GeV^{-2}$]}
& \multicolumn{1}{c}{$c^{\pi^0}_{YV}$}
& \multicolumn{1}{c}{$d^{\pi^0}_{YV}$ [$\GeV^{-2}$]}
& \multicolumn{1}{c}{$c^{\eta}_{YV}$}
& \multicolumn{1}{c}{$d^{\eta}_{YV}$ [$\GeV^{-2}$]}
\\ \hline
$\rho$                         & 0.477 & 0.885 & \multicolumn{2}{c}{---} & \multicolumn{2}{c}{---} \\
$\omega$                       & 0.434 & 0.923 & \multicolumn{2}{c}{---} & \multicolumn{2}{c}{---} \\
$\mathbb{P}\rho$               & 0.557 & 0.195 & 0.52    & 1.07  & $-2.27$ & 5.5  \\
$\mathbb{P}\omega$             & 0.514 & 0.197 & $-0.06$ & 0.37  & 0.016   & 5.5  \\
$f_2\rho$                      & 0.148 & 0.425 & 0.72    & 0.62  & 5.89    & 2.36 \\
$f_2\omega$                    & 0.106 & 0.436 & 2.98    & 5.02  & $-5.96$ & 2.36 \\
\end{tabular}
\end{ruledtabular}
\caption{Regge trajectories and Regge-cut parameters used for ``Solution~I'' of Ref.~\cite{Kashevarov:2017vyl} and entering the reggeized propagators of $\omega$ and $\rho$ exchange in Eq.~\eqref{eq:Rtilde}. All trajectories are linear, $\alpha_{Y}(t_p)=\alpha_{Y,0}+\alpha_{Y,1} t_p$, with $t_p$ in $\GeV^2$, separately for the production of $\pi^{0}$ and $\eta$ mesons. In this study, for the photoproduction of $\eta'$ mesons,  we take the same values as for the $\eta$ case.}
\label{tab:regge-params-solutionI}
\end{table*}
%
\subsection{Vector mesons}
\label{sec:production-vector-mesons}

\begin{figure*}[t!]
\centering
\tikzfeynmanset{
  h-exchange/.style={draw=black, thick, densely dash dot},
  pomeron/.style={draw=black, thick, double, double distance=1.2pt},
}
\tikzset{
  extlabelL/.style={below=5pt, anchor=west, inner sep=1pt},
  extlabelR/.style={below=5pt, anchor=east, inner sep=1pt},
}

\begin{minipage}[b]{0.24\textwidth}\centering
\vspace{0pt}
\begin{tikzpicture}[baseline={(base)},scale=0.88,transform shape]
\begin{feynman}
  \vertex (base) at (0,0);

  \vertex (pL) at (-1.8,  0.9);
  \vertex (eL) at (-1.8, -0.9);
  \vertex (pR) at ( 1.9,  0.9);
  \vertex (eR) at ( 1.9, -0.9);

  \vertex (Vp) at (0,  0.9);
  \vertex (Ve) at (0, -0.9);
  \vertex (F)  at (0.5, 0);
  \vertex (Vout) at (1.9, 0) {$V(p_V)$};

  \diagram*{
    (pL) -- [fermion] (Vp) -- [fermion] (pR),
    (eL) -- [fermion] (Ve) -- [fermion] (eR),
    (Ve) -- [photon, edge label={$\gamma^*$}] (F),
    (Vp) -- [pomeron, edge label'={$\mathbb{P}$}] (F),
    (F)  -- [boson] (Vout)
  };
\end{feynman}

\node[extlabelL] at (pL) {$p(p_p)$};
\node[extlabelL] at (eL) {$e^-(p_e)$};
\node[extlabelR] at (pR) {$p(p_{p'})$};
\node[extlabelR] at (eR) {$e^-(p_{e'})$};

\end{tikzpicture}\vspace{3pt}\par (a)
\end{minipage}\hfill
\begin{minipage}[b]{0.24\textwidth}\centering
\vspace{0pt}
\begin{tikzpicture}[baseline={(base)},scale=0.88,transform shape]
\begin{feynman}
  \vertex (base) at (0,0);

  \vertex (pL) at (-1.8,  0.9);
  \vertex (eL) at (-1.8, -0.9);
  \vertex (pR) at ( 1.9,  0.9);
  \vertex (eR) at ( 1.9, -0.9);

  \vertex (Vp) at (0,  0.9);
  \vertex (Ve) at (0, -0.9);
  \vertex (F)  at (0.5, 0);
  \vertex (Vout) at (1.9, 0) {$V(p_V)$};

  \diagram*{
    (pL) -- [fermion] (Vp) -- [fermion] (pR),
    (eL) -- [fermion] (Ve) -- [fermion] (eR),
    (Ve) -- [photon, edge label={$\gamma^*$}] (F),
    (Vp) -- [plain, h-exchange, edge label'={$h$}] (F),
    (F)  -- [boson] (Vout)
  };
\end{feynman}

\node[extlabelL] at (pL) {$p(p_p)$};
\node[extlabelL] at (eL) {$e^-(p_e)$};
\node[extlabelR] at (pR) {$p(p_{p'})$};
\node[extlabelR] at (eR) {$e^-(p_{e'})$};

\end{tikzpicture}\vspace{3pt}\par (b)
\end{minipage}\hfill
\begin{minipage}[b]{0.24\textwidth}\centering
\vspace{0pt}
\begin{tikzpicture}[baseline={(base)},scale=0.88,transform shape]
\begin{feynman}
  \vertex (base) at (0,0);

  \vertex (pL) at (-1.8,  0.9);
  \vertex (eL) at (-1.8, -0.9);
  \vertex (pR) at ( 2.0,  0.9);
  \vertex (eR) at ( 2.0, -0.9);

  \vertex (Vp) at (0,  0.9);
  \vertex (Ve) at (0, -0.9);
  \vertex (pm)   at (0.65, 0.9);
  \vertex (Vmes) at (1.25, 1.75) {$V(p_V)$};

  \diagram*{
    (pL) -- [fermion] (Vp) -- [plain, ultra thick, edge label={$B^*$}] (pm) -- [fermion] (pR),
    (eL) -- [fermion] (Ve) -- [fermion] (eR),
    (Vp) -- [photon, edge label'={$\gamma^*$}] (Ve),
    (pm) -- [boson] (Vmes)
  };
\end{feynman}

\node[extlabelL] at (pL) {$p(p_p)$};
\node[extlabelL] at (eL) {$e^-(p_e)$};
\node[extlabelR] at (pR) {$p(p_{p'})$};
\node[extlabelR] at (eR) {$e^-(p_{e'})$};

\end{tikzpicture}\vspace{3pt}\par (c)
\end{minipage}\hfill
\begin{minipage}[b]{0.24\textwidth}\centering
\vspace{0pt}
\begin{tikzpicture}[baseline={(base)},scale=0.88,transform shape]
\begin{feynman}
  \vertex (base) at (0,0);

  \vertex (pL) at (-1.8,  0.9);
  \vertex (eL) at (-1.8, -0.9);
  \vertex (pR) at ( 2.0,  0.9);
  \vertex (eR) at ( 2.0, -0.9);

  \vertex (Vp) at (0,  0.9);
  \vertex (Ve) at (0, -0.9);
  \vertex (pm)   at (-0.65, 0.9);
  \vertex (Vmes) at (0.0,   1.75) {$V(p_V)$};

  \diagram*{
    (pL) -- [fermion] (pm) -- [plain, ultra thick, edge label'={$B^*$}] (Vp) -- [fermion] (pR),
    (eL) -- [fermion] (Ve) -- [fermion] (eR),
    (Vp) -- [photon, edge label'={$\gamma^*$}] (Ve),
    (pm) -- [boson] (Vmes)
  };
\end{feynman}

\node[extlabelL] at (pL) {$p(p_p)$};
\node[extlabelL] at (eL) {$e^-(p_e)$};
\node[extlabelR] at (pR) {$p(p_{p'})$};
\node[extlabelR] at (eR) {$e^-(p_{e'})$};

\end{tikzpicture}\vspace{3pt}\par (d)
\end{minipage}

\caption{Feynman diagrams for exclusive electroproduction of vector mesons $V$.
(a)~Diffractive contribution, modeled by effective Pomeron exchange: the proton emits a Pomeron $\mathbb{P}$ which fuses with the virtual photon $\gamma^*$ to produce $V$.
(b)~Non-diffractive contribution, modeled as an effective $t$-channel exchange of a generic hadron $h$ (with $h$ standing for the set of non-diffractive exchanges included in the hadronic current, e.g., $\pi^0$, $\sigma$, $f_2$, etc.); the exchanged $h$ fuses with the virtual photon to produce $V$.
(c),(d)~Non-diffractive bremsstrahlung-type topologies with an intermediate baryonic excitation $B^*$ in the $s$- and $u$-channel of the $\gamma^{*}p$ pair, corresponding to emission of $V$ from the outgoing (c) or incoming (d) proton leg.}
\label{fig:diagrams-vector}
\end{figure*}
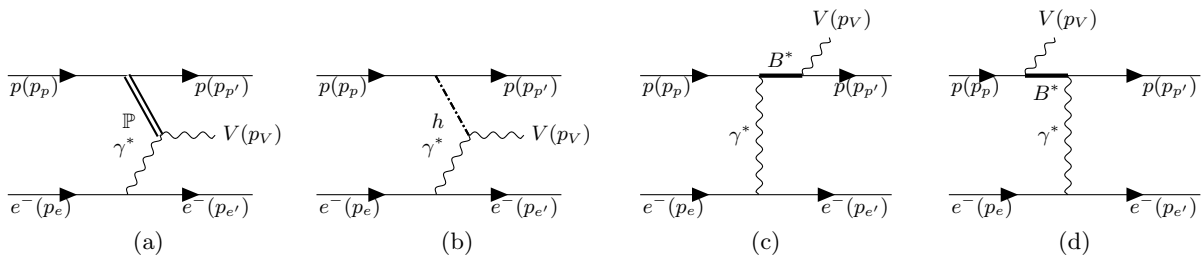

Depending on the invariant mass of the photon-proton pair, $s_{\gamma p}$, different mechanisms may dominate the photoproduction of vector mesons.
It has been investigated in various studies covering diffractive and non-diffractive regimes and their interplay,  e.g.~\cite{Friman:1995qm,Pichowsky:1996tn,Donnachie:1999yb,Ryu:2012tw,Sibirtsev:2002at,Obukhovsky:2009th,Sibirtsev:2003qh,Wei:2019imo,Oh:2003aw,Wang:2025rvr,Xu:2021mju,Wei:2024lne,Wang:2017plf,Yu:2019wly,CBELSATAPS:2015ftl,Zhang:2021esc,Tang:2024pky,Tang:2025qqe}.
Conceptually, one distinguishes two contributions, see Fig.~\ref{fig:diagrams-vector}:
(i)~a diffractive component, dominating the domain $\sqrt{s_{\gamma p}}\gg 1\,\GeV$, well described by an effective Pomeron exchange
$\mathbb{P}$~\cite{Donnachie:1994zb,Sibirtsev:2003qh,Tang:2025qqe,Donnachie:1999yb,Ryu:2012tw} (the diagram (a)), and
(ii)~a non-diffractive component, being the dominant one in the domain $\sqrt{s_{\gamma p}}\simeq \text{few }\GeV$, and modeled through a combination of $t$-channel exchanges by scalar, pseudoscalar, axial-vector, and tensor states (the diagram (b)) and bremsstrahlung-like processes via intermediate baryonic resonances~\cite{Friman:1995qm,Pichowsky:1996tn,Sibirtsev:2003qh,Ryu:2012tw,CBELSATAPS:2015ftl,Obukhovsky:2009th,Oh:2003aw,Wei:2019imo,Wang:2025rvr,Xu:2021mju} (see the diagrams (c) and (d)).

The axial-vector and $\eta^{(\prime)}$ pieces are typically sub-dominant compared to the other contributions~\cite{Wang:2025rvr,Sibirtsev:2003qh,Yu:2016zut}.
The baryonic excitations are uniquely responsible for a flat behavior $d\sigma_{\omega}/dt_{p}$~\cite{Sibirtsev:2002at,CBELSATAPS:2015ftl} for large $|t_{p}|$.
Whether this baryonic component is relevant within our kinematic window $|t_{p}|\lesssim 2\,\GeV^{2}$ depends on the produced meson.
For $\rho^{0}$ production, the baryonic resonances are important in this range~\cite{Wang:2025rvr,Oh:2003aw}, whereas for $\omega$ and $\phi$ production, the nucleon-resonance contributions may be neglected~\cite{Sibirtsev:2002at,CBELSATAPS:2015ftl}.
For $\omega$, two ingredients are particularly important:
the large $t$-channel contribution from $\pi$ exchange, driven by the sizable $\omega\pi\gamma$ coupling (see Tbl.~\ref{tab:Coupl-VPgamma}), and the isospin structure of the intermediate baryonic channel~\cite{Ajaka:2006bn}.
Since the $\omega,p$ isospins are $I_{\omega}=0$ and $I_{p}=1/2$, the $\omega p$ system has only $I=1/2$ and may couple only to the sector of nucleon excitations $N^{*}$.
In contrast, for $\rho$, we have $I_{\rho}=1$, and the $\rho p$ system may have total isospin $I=1/2$ or $I=3/2$, therefore receiving contributions from both $N^{*}$ and $\Delta^{*}$ resonances.
For $\phi$ photoproduction, the direct nucleon couplings are OZI-suppressed (recall the discussion around Eq.~\eqref{eq:Lagr-Vpp}), and over most of the forward kinematic region, the data are well described by the diffractive Pomeron contribution, with non-diffractive pieces providing only subleading corrections~\cite{Tang:2025qqe,Yu:2016zut}.
For the photoproduction of $\phi$ and $\omega$, we largely follow Refs.~\cite{Yu:2016zut,Yu:2017vvp}.
These works fit the contributions in terms of Reggeized $t$-channel non-diffractive exchanges supplemented by the Pomeron to CBELSA/TAPS, SLAC/LBL, LEPS, and CLAS data for differential cross-sections $d\sigma_{\gamma p\to Vp}/dt_{p}$ for a broad range of $\sqrt{s_{\gamma p}}\gtrsim 2\,\GeV$.
Within that framework, very good agreement with experiment has been obtained for $|t_{p}|\lesssim 2\,\GeV^2$, which is our target kinematic space.

The $\rho^{0}$ channel has been considered in many studies, including Refs.~\cite{Oh:2003aw,Tang:2025qqe,Wang:2025rvr}; however, to the best of our knowledge, no analysis combines the various contributions to the amplitude in a fully self-consistent manner.
In particular, Ref.~\cite{Oh:2003aw} employs an $f_{2}\gamma V$ vertex that violates gauge invariance for off-shell $f_{2}$ (recall Sec.~\ref{sec:interactions}), which leads to an unphysical scaling $d\sigma/dt_{p}$ in the domain of small $t_{p}$, which is essential for successful fit of the photoproduction.
In addition, the approach there does not include reggeizing spin-0 and spin-2 exchanges.
Ref.~\cite{Wang:2025rvr}, on the other hand, implements Reggeized trajectories, but its fits rely on the same $f_{2}\gamma V$ vertex. Finally, Ref.~\cite{Tang:2025qqe} restricts the discussion to the domain of high $\sqrt{s_{\gamma p}}\gg1\,\GeV$, which may be described solely by the Pomeron exchange.

For a unified treatment of all light vector mesons, we extend the approach of Refs.~\cite{Yu:2016zut,Yu:2017vvp} to the $\rho^{0}$ channel.
There, we incorporate the gauge-invariant treatment of the $f_{2}V\gamma$ interaction. However, as we discussed above, the photoproduction data in the domain $|t_{p}|\lesssim 2\,\GeV^{2}$ cannot be explained without adding the baryonic contributions~\cite{Oh:2003aw,Wang:2025rvr}. These contributions are not included in the fit of Refs.~\cite{Yu:2016zut,Yu:2017vvp}, and we have to incorporate them separately.

The descriptions from Refs.~\cite{Oh:2003aw,Wang:2025rvr} add towers of spin 1/2, 3/2, 5/2 resonances and are very sophisticated. Namely, they include many fitting ingredients, such as independent $s/u$-channel form factors modulating the suppression of each of the matrix elements, and also ambiguous gauge-invariance-restoring currents constructed by the recipe of Ref.~\cite{Davidson:2001rk} (see also Appendix~\ref{app:pi0-production-kaskulov} for related discussions). In practice, we find that this construction may be cumulatively described by just simple $s/u$ proton exchange diagrams supplemented by a single form factor, see details below.
We find that our approach reproduces the GlueX~\cite{GlueX:2023fcq}, CLAS~\cite{CLAS:2001zxv}, and SLAC~\cite{Ballam:1972eq} data for the differential cross-section $d\sigma_{\gamma p\to \rho^{0}p}/dt_{p}$ to better than $\sim 30\%$ in the target region $|t_{p}|\lesssim 2\,\GeV^{2}$ and $\sqrt{s_{\gamma p}}\lesssim 5\,\GeV$.

Namely, we write the hadronic current for $\gamma^{*}p\to Vp$ as
\begin{align}
    \label{eq:matrix-element-vector-meson}
    \cM_V^\mu
    = \cM_{V,\pi}^\mu + \cM_{V,\sigma}^\mu + \cM_{V,f_2}^\mu + \cM_{V,\mathbb{P}}^\mu + \cM_{V,B}^{\mu} \, .
\end{align}
Here, the terms $\pi$, $\sigma$, $f_2$, and $\mathbb{P}$ are pion, scalar, tensor, and Pomeron exchange, respectively, and the last term denotes the contribution from a family of baryon resonances, including spins $1/2, 3/2, 5/2$.
We label the incoming vector meson four-momentum and its outgoing polarization vector by $p_V$ and $\epsilon_V^\mu(p_V)$.
\begin{table}[t]
\centering
\begin{tabular}{lccc}
\hline\hline
$X$ & $\alpha_{0,X}$ & $\alpha_{1,X},\,[\GeV^{-2}]$ & $\mathcal{P}_{X}(t_{p})$ \\
\hline
$\pi$        & $-0.013$ & $0.7$  & $\exp[-i\pi\alpha_\pi(t_{p})]$ \\
$\sigma$     & $-0.18$  & $0.7$  & $(1+\exp[-i\pi\alpha_\sigma(t_{p})])/2$ \\
$f_2$        & $0.55$   & $0.9$  & $(1+\exp[-i\pi\alpha_{f_2}(t_{p})])/2$ \\
$\mathbb{P}$ & $1.08$   & $0.25$ & $\exp[-i\pi(\alpha_{\mathbb{P}}(t_p)-1)/2]$ \\
\hline\hline
\end{tabular}
\caption{Parameters of the Regge trajectories~\eqref{eq:TrajectoriesSymbolic} and the phases $\mathcal{P}_{X}(t_{p})$ entering the Reggeized form-factors~\eqref{eq:ReggePropagatorGeneric},~\eqref{eq:ReggePomeron} used for the vector meson electroproduction, following Refs.~\cite{Yu:2016zut,Yu:2017vvp}.}
\label{tab:ReggeTrajectories}
\end{table}
To account for the forward-angle regime, for all the channels except for $\cM_{V,B}$, we employ the description of the exchanges in terms of Reggeized trajectories.
For a meson $h\in\{\pi^0,\sigma,f_2\}$ of spin $J_{h}\in\{0,2\}$ and for Pomerons $\mathbb{P}$, we use the Reggeized propagators~\cite{Yu:2016zut,Yu:2017vvp}
\begin{align}
    \label{eq:ReggePropagatorGeneric}
    &\cR_{h}(s_{\gamma^{*}p},t_p)
    =
    -\alpha_{1,h}\Gamma(J_{h}-\alpha_{h}(t_{p}))\,
    \left(\frac{s_{\gamma^{*}p}}{s_0}\right)^{\alpha_h(t_p)-J_{h}}\,
    \cP_h(t_p)\, , \\ &\cR_{\mathbb{P}}(s_{\gamma^{*}p},t_p)
    =
    \mathcal{P}_{\mathbb{P}}(t_{p})\,
    \Big(\frac{s_{\gamma^{*}p}}{4s_0}\Big)^{\alpha_{\mathbb{P}}(t_p)-1}\,.
    \label{eq:ReggePomeron}
\end{align}
$\alpha_{X}(t_{p})$ is the Regge trajectory parameterized as
\begin{align}
    \label{eq:TrajectoriesSymbolic}
    \alpha_X(t_{p})
    =
    \alpha_{0,X}+\alpha_{1,X}\,t_{p}\,.
\end{align}
Finally, $\mathcal{P}_{X}(t_{p})$ are complex phases.
The parameters of the trajectories $\alpha_{0,X}$ and $\alpha_{1,X}$, as well as the $\mathcal{P}_{X}(t_{p})$ are summarized in Tbl.~\ref{tab:ReggeTrajectories}.

In $\omega$ photo/electroproduction, the pion-exchange contribution modeled in this way leads to an overly large cross-section, and a phenomenologically important refinement adopted in Ref.~\cite{Yu:2017vvp} is to include absorptive Regge-cut corrections that dress the pion exchange:
\begin{align}
    \label{eq:pi-cuts}
    \cR^{(\omega)}_{\pi}(s_{\gamma^{*}p},t_p)
    =
    \cR_{\pi}(s_{\gamma^{*}p},t_p)
    +
    \sum_{\varphi\in\{f_2,\mathbb{P}\}}
    C_{\pi\varphi}\,e^{d_{\pi\varphi}t_p}\,
    e^{-i\frac{\pi}{2}\alpha_{\pi\varphi}(t_p)}
    \left(\frac{s_{\gamma^{*}p}}{s_0}\right)^{\alpha_{\pi\varphi}(t_p)-1} \,,
\end{align}
where $\varphi$ labels the trajectory participating in the absorptive cut with the pion, \emph{i.e.}, the $\pi$--$f_2$ and $\pi$--$\mathbb{P}$ cuts.
The cut trajectories are taken to be linear,
\begin{align}
    \label{eq:cut-trajectories}
    \alpha_{\pi\varphi}(t_p)
    =
    \alpha_{0,\pi}+\alpha_{0,\varphi}-1
    +\frac{\alpha_{1,\pi}\,\alpha_{1,\varphi}}
    {\alpha_{1,\pi}+\alpha_{1,\varphi}}\,t_p\,,
\end{align}
The parameters $C_{\pi\varphi}$ and $d_{\pi\varphi}$ are phenomenological constants with mass dimension of $-2$ that set, respectively, the strength and the $t$-range of each cut contribution, and are fixed from fits to $\omega$ photoproduction data:
\begin{align}
    \label{eq:cut-params-omega}
    C_{\pi f_2}=41\,\GeV^{-2}\,,\quad
    d_{\pi f_2}=2.2\,\GeV^{-2}\,,\qquad
    C_{\pi\mathbb{P}}=-2.5\,\GeV^{-2}\,,\quad
    d_{\pi\mathbb{P}}=2.0\,\GeV^{-2}\,.
\end{align}
In this construction, the $\pi$--$\sigma$ cut is typically neglected due to its small intercept.
In our $\rho^0$ extension, we do not include absorptive cut dressing of the pion term because the $\pi \rho \gamma$ coupling is much smaller.

Next, we introduce the different contributions of the vector mesons electroproduction current in Eq.~\eqref{eq:matrix-element-vector-meson}.
Using the $\pi^{0}pp$ and $V\gamma \pi^{0}$ vertices from Eqs.~\eqref{eq:Vert-Vpp},~\eqref{eq:Vert-VYgamma} and the coupling constants $g_{\pi^{0}p}$, $g_{\gamma \pi^{0}V}$ defined and discussed in Eq.~\eqref{eq:Coupl-ppi0} and Tbl.~\ref{tab:Coupl-VPgamma}, the Reggeized pseudoscalar-exchange amplitude reads
\begin{align}
    \label{eq:Mpseudoscalar}
    \cM_{V,\pi^{0}}^\mu
    =
    i\,g_{\gamma \pi^{0}V}\,g_{p\pi^{0}}\,
    \epsilon^{\mu\nu\alpha\beta}\,
    \epsilon_{V,\nu}^*(p_V)\,p_{\gamma^{*},\alpha}\,p_{V,\beta}\;
    \bar u(p')\gamma_5 u(p)\; \cR_{\pi}(s_{\gamma^{*}p},t_p) \, .
\end{align}
By using Eq.~\eqref{eq:Vert-Vsigmagamma} for the vertex $\Gamma_{V\sigma \gamma}^{\mu\nu}$ and contracting it with the polarization vector $\epsilon_{\nu}(p_{V})$, and the $\sigma pp$ Lagrangian $\cL_{\sigma pp}=g_{\sigma pp}\sigma\bar{p}p$ with the standard value $g_{\sigma NN}=14.6$~\cite{Erkol:2005jz,Erkol:2006eq}, we find the scalar-exchange amplitude
\begin{equation}
    \label{eq:Msigma}
    \cM_{V,\sigma}^\mu
    =
    e\,g_{V\sigma\gamma}\,g_{\sigma pp}\;
    \left[ (p_{\gamma^{*}}\!\cdot p_V)\,\epsilon_V^{*\mu}(p_V)
    -(\epsilon_V^*(p_V)\!\cdot p_{\gamma^{*}})\,p_V^\mu \right]\;
    \bar{u}(p')u(p)\; \cR_\sigma(s_{\gamma^{*}p},t_p),
\end{equation}
where the channel-dependent radiative coupling $g_{V\sigma\gamma}$ is given by Eqs.~\eqref{eq:Coupl-V-sigma-gamma-1},~\eqref{eq:Coupl-V-sigma-gamma-2}.

The tensor-exchange current is constructed from the $f_{2}NN$ vertex, the spin-2 projector, and the $\gamma^* f_2 V$ vertex (given by Eq.~\eqref{eq:Vert-f2-V-gamma}).
We find
\begin{align}
    \label{eq:Mf2}
    \cM_{V,f_2}^\mu
    =
    \bar u(p_{p'})\,\Gamma^{\lambda\sigma}_{f_2NN}(p_{p},p_{p'})\,u(p_{p})\;
    \Pi_{\lambda\sigma;\beta\rho}(p_{p}-p_{p'})\;
    \Gamma^{\beta\rho,\mu\nu}_{\gamma^* f_2 V}(p_{\gamma^{*}},p_V)\epsilon_{\nu}(p_{V})\;
    \cR_{f_2}(s_{\gamma^{*}p},t_p)\,.
\end{align}
The $f_2NN$ vertex is given by~\cite{Yu:2016zut}
\begin{align}
    \label{eq:f2NN-vertex}
    \Gamma^{\lambda\sigma}_{f_2NN}(p_{p'},p_{p})
    =
    \frac{2g^{(1)}_{f_2NN}}{m_N}\,\Big(P^\lambda\gamma^\sigma+P^\sigma\gamma^\lambda\Big)
    +\frac{4g^{(2)}_{f_2NN}}{m_N^2}\,P^\lambda P^\sigma\,,
\end{align}
with $P^\mu\equiv (p_{p'}+p_{p})^{\mu}/2$, $g^{(1)}_{f_2NN}=6.45$ and $g^{(2)}_{f_2NN}=0$~\cite{Yu:2016zut}.
Lastly, $\Pi_{\mu\nu;\alpha\beta}(p)$ is the spin-2 field sum over polarizations
\begin{align}
    \label{eq:f2-polarization-sum-rule}
    \Pi_{\mu\nu;\alpha\beta}(p)
    =
    \frac{1}{2}\Big(\bar g_{\mu\alpha}(p)\,\bar g_{\nu\beta}(p)
    +\bar g_{\mu\beta}(p)\,\bar g_{\nu\alpha}(p)\Big)
    -\frac{1}{3}\,\bar g_{\mu\nu}(p)\,\bar g_{\alpha\beta}(p) \, ,
\end{align}
with $\bar g_{\mu\nu}(p)= -g_{\mu\nu}+ p_{\mu} p_{\nu}/m_{f_2}^2$.

The diffractive (natural-parity) Pomeron contribution is taken as in
Refs.~\cite{Yu:2016zut,Yu:2017vvp}:
\begin{align}
    \label{eq:MPomeron}
    \cM_{V,\mathbb{P}}^\mu
    &=
    i\,12  e \,\frac{\beta_{\mathbb{P}p}\,\beta_{\mathbb{P}V}}{f_V}\,\frac{m_V^2}{m_V^2-t_p}\,
    \Big(\frac{2\mu_0^2}{2\mu_0^2+m_V^2-t_p}\Big)\, F_1(t_p)\,
    \cR_{\mathbb{P}}(s_{\gamma^{*}p},t_p)
    \nonumber\\
    &\hspace{2cm}\times
    \bar u(p')\Big(\slashed{p}_{\gamma^{*}}\,\epsilon_V^{*\mu}(p_V)-\gamma^\mu\,\epsilon_V^*(p_V)\!\cdot p_{\gamma^{*}}\Big)u(p)\,,
\end{align}
with $\mu_0^2=1.1\,\GeV^2$ and $F_1(t_p)=\frac{4m_N^2-2.8t_p}{(4m_N^2-t_p)(1-t_p/0.71)^2}$ being the nucleon isoscalar form-factor~\cite{Donnachie:1999yb}.
The factors $\beta_{\mathbb{P}p}$ and $\beta_{\mathbb{P}V}$ denote effective Pomeron-proton and Pomeron-V couplings.
They correspond to the dominant quark inside the states $p,V$, and are $\beta_{\mathbb{P}p}=\beta_{\mathbb{P}\rho}=\beta_{\mathbb{P}\omega} \approx 2.1\,\GeV^{-1}$ (the dominant quarks are $u,d$), $\beta_{\mathbb{P}\phi} = 1.6\, \GeV^{-1}$ (the dominant quark is $s$).
Finally, $(f_\rho,f_\omega,f_\phi)=(5.2,15.6,-13.4)$ are the
dimensionless VMD $\gamma$--$V$ couplings~\cite{Yu:2016zut,Yu:2017vvp}.

Finally, let us comment on the contribution from baryon resonances, $\cM_{V,B}^\mu$, which is represented by the diagrams (c), (d) in Fig.~\ref{fig:diagrams-vector}.
As we discussed earlier, these resonances are important for the $\rho^{0}$ photoproduction in the domain of large $|t_{p}|\gtrsim 1\,\GeV^{2}$. Being summed, they give a flat scaling of $d\sigma_{\gamma p\to \rho p}/dt_{p}$~\cite{Oh:2003aw,Wang:2025rvr} at intermediate $\sqrt{s_{\gamma p}}\simeq \text{few}\,\GeV$.

To properly describe this contribution, we should consider families of the excitations with distinct quantum numbers, such as spin, and within each family, treat every resonance independently. This results in a very complicated set of amplitudes with a non-trivial construction ensuring gauge invariance.
However, a crucial observation is that for $\sqrt{s_{\gamma p}}\gtrsim 2\,\GeV$, it may be possible to reproduce the cumulative contribution of all baryonic resonances with a simple \emph{Born} matrix element.
Namely, we consider
\begin{align}
    \label{eq:Melement-vector-production-nucleon}
    \cM^{\mu}_{V,B}
    =
    iG_{N}^{V}(s_{\gamma^{*}p})\bar{u}(p_{p'})\bigg[ &\left(g_{VNN}\gamma^{\nu}\epsilon_{\nu}(p_{V})+\frac{\kappa_{V}}
    {4m_{N}}\sigma^{\alpha\beta}V_{\alpha\beta}(p_{V})\right)\frac{\slashed{p}_{p}+\slashed{p}_{\gamma^{*}}+m_{p}}{s_{\gamma^{*}p}-m_{p}^{2}}\gamma^{\mu} \nonumber \\
    &+ \gamma^{\mu}\frac{\slashed{p}_{p}-\slashed{p}_{\gamma^{*}}+m_{p}}{u-m_{p}^{2}}\left(g_{VNN}\gamma^{\nu}\epsilon_{\nu}(p_{V})+\frac{\kappa_{V}}{4m_{N}}\sigma^{\alpha\beta}V_{\alpha\beta}(p_{V})\right) \bigg]u_{p}(p_{p})\,,
\end{align}
where the couplings $g_{VNN}$ and $\kappa_{V}$ are introduced in Eq.~\eqref{eq:vector_nucleon_couplings}, and $V_{\alpha\beta}(p_{V}) = p_{V,\alpha}\epsilon_{\beta}(p_{V})-p_{V,\beta}\epsilon_{\alpha}(p_{V})$.
The function $G^{N}_{V}(s_{\gamma p})$ is effectively capturing  two effects:
describing the cumulative effect of towers of all baryonic states at $\sqrt{s_{\gamma p}}\simeq (1-2)\,\GeV$, as well as the suppression of the photoproduction cross-section in the limit $\sqrt{s_{\gamma p}}\gg 1\,\GeV$, which originates from matching to the QCD sum rules.
The crucial point that makes our approximation plausible is that the matrix element in Eq.~\eqref{eq:Melement-vector-production-nucleon} leads to an approximately flat contribution to $d\sigma_{\gamma p \to Vp}/dt_{p}$ for a broad range of $t_{p}$ and grows only close to the boundary $|t_{p}| = |t_{p,\text{max}}(s_{\gamma p})|$.
This is similar to the cumulative contribution from baryonic spin $1/2,3/2,5/2$ resonances as obtained in Fig.~2 of~\cite{Wang:2025rvr}.

Let us discuss $G_{V}^{N}(s_{\gamma p})$.
The expected high-energy scaling of the photoproduction cross-section from the QCD sum rules is~\cite{Brodsky:1973kr}
\begin{align}
    s_{\gamma p}^{n-2}\frac{d\sigma_{\gamma p\to Vp}}{dt_{p}}
    \simeq
    \text{const} \, ,
\end{align}
where $n = 9$ is the number of constituent fields, counting each proton as three fields, a photon as one field, and a vector meson as two fields.
By using the Born cross-section from Eq.~\eqref{eq:Melement-vector-production-nucleon} with $G^{N}_{V} = 1$, we find that $d\sigma/dt_{p} \propto s_{\gamma p}^{-2}$.
Therefore, the high-energy asymptotics is $|G^{V}_{N}(s_{\gamma p})|^{2} \propto s_{\gamma p}^{-5}$.
To find the behavior at smaller $s_{\gamma p}$ near the threshold, we used the $d\sigma_{\gamma p\to \rho^{0}p}/dt_{p}$ data in the range $|t_{p}|\gtrsim 1\,\GeV^{2}$ (where the nucleon piece is expected to dominate) from GlueX~\cite{GlueX:2023fcq}, CLAS~\cite{CLAS:2001zxv}, and SLAC~\cite{Ballam:1972eq}.
The resulting function is
\begin{align}
    G_{V}^{N}(s_{\gamma p})
    \approx
    \left\{
    \begin{matrix}
    3.9
    & s_{\gamma p} \le 5.5\,\GeV^2, \\[2mm]
    \left(\dfrac{6.9\,\GeV^2}{s_{\gamma p}}\right)^{6.0}
    &  5.5\,\GeV^2 < s_{\gamma p} \le 6.2\,\GeV^2, \\[2mm]
    \left(\dfrac{8.3\,\GeV^2}{s_{\gamma p}}\right)^{2.2} &
    s_{\gamma p}>6.2\,\GeV^{2}
    \end{matrix}\right. \,.
\end{align}
\begin{figure}
    \centering
    \includegraphics[width=0.33\linewidth]{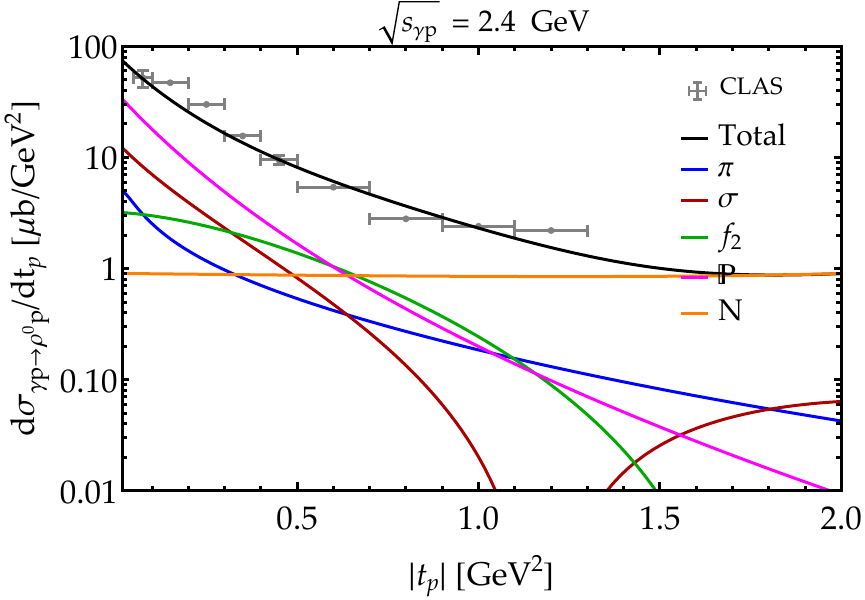}~\includegraphics[width=0.33\linewidth]{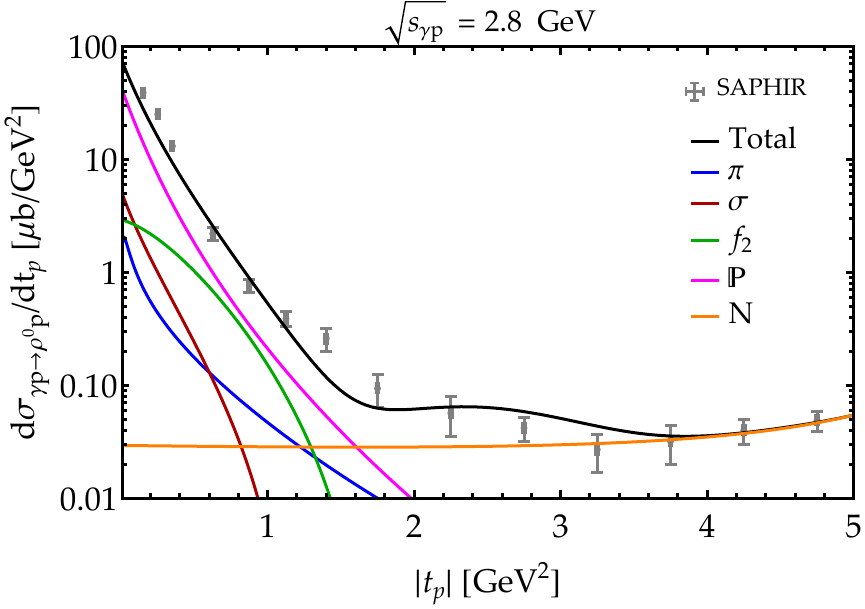}~\includegraphics[width=0.33\linewidth]{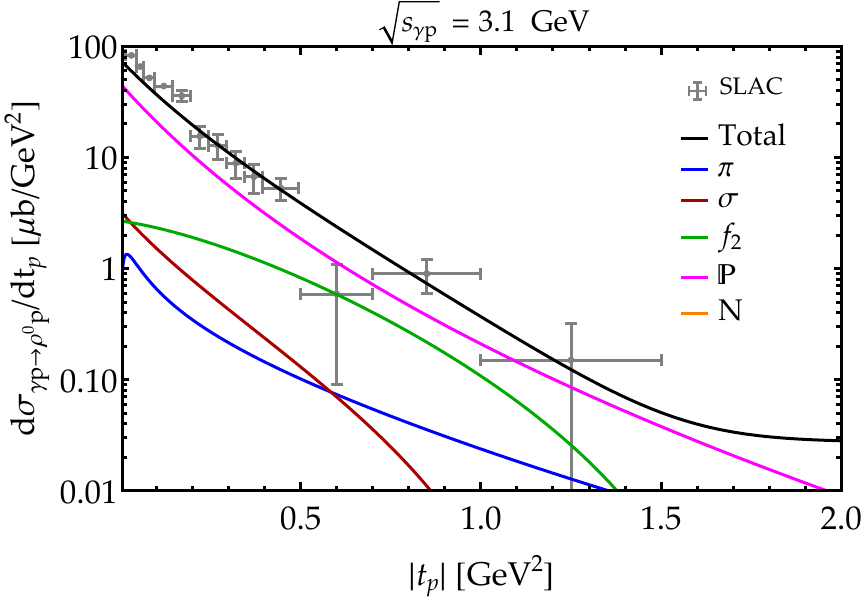}
    \caption{The differential cross-section $d\sigma_{\gamma p \to \rho^{0}p}/dt_{p}$ at the rest frame of the proton target, for various energies of the incoming photon, $E_{\gamma}$.
    The contributions from various intermediate states -- the nucleons $N$, the Pomeron $\mathbb{P}$, the scalar mesons $\sigma$, the pion $\pi$, and the tensor mesons $f_{2}$ -- are shown in colored curves, with the total differential cross-section shown in black.
    The gray brackets denote the experimental measurements with error bars, coming from GlueX~\cite{GlueX:2023fcq}, CLAS~\cite{CLAS:2001zxv}, and SLAC~\cite{Ballam:1972eq}. The individual colored curves represent one phenomenological decomposition of the fitted full amplitude and should not be interpreted as a unique microscopic separation.}
    \label{fig:differential-cross-section-rho}
\end{figure}
The resulting description of the $\rho^{0}$ photoproduction matches the differential cross-section data in the allowed domain $2\,\GeV\lesssim \sqrt{s_{\gamma p}}\lesssim 4\,\GeV$ well, see Fig.~\ref{fig:differential-cross-section-rho}, as well as reproduces the total $\rho^{0}$ photoproduction cross-section within $20\%$ for masses $\sqrt{s_{\gamma p}}>4\,\GeV$.
\subsection{Cross sections and event sampling}
\label{sec:Xsec}

The differential cross section of the process in Eq.~\eqref{eq:process} can be written as
\begin{align}
    \label{eq:d4sigma-generic}
    \frac{d^{4}\sigma_{X}}{ds_{\gamma^{*}p}\,dQ^{2}\,dt_{p}\,ds_{1}}
    =
    \frac{1}{256\pi^{4}}\,\;
    \frac{\overline{|\mathcal{M}_{X}|^{2}}}{
    \lambda(s,m_{e}^{2},m_{p}^{2})\,\;
    \sqrt{\lambda(s_{\gamma^{*}p},-Q^{2},m_{p}^{2})\,(s_{1}-s_{1}^{-})\,(s_{1}^{+}-s_{1})}
    }\,,
\end{align}
Here, $\mathcal{M}_{X}$ is the electroproduction matrix element given by Eq.~\eqref{eq:electroproduction-matrix-element}.
The kinematical variables $\xi = \{s_{\gamma^{*}p},Q^{2},t_{p},s_{1}\}$ are defined in Eq.~\eqref{eq:LorentzInv} and their boundaries $\xi^{\pm}$ are discussed in Appendix~\ref{app:2-to-3}.
Finally, $\lambda(a,b,c) = a^{2}+b^{2}+c^{2}-2ab - 2ac -2bc$.

For the event analysis below, we need to retrieve the full event kinematics, \emph{i.e.}, the 4-momenta of  $e'$, $p'$, $X$, and, in particular, to sample distributions in correlated quantities.
The procedure below, based on Monte Carlo sampling, ensures an accurate sampling of the events, in particular close to the boundaries~\cite{Byckling:1969sx}.

We start by generating $N$ trial points
$\mathcal{U}=\{(s_{\gamma^{*}p},Q^{2},t_{p},s_1)\}$ by sequentially sampling
$s_{\gamma^{*}p}$, $Q^{2}$, $t_{p}$, and $s_1$ within their kinematic ranges, given in Appendix~\ref{app:2-to-3}.
To reduce variance, we use logarithmic proposals for $s_{\gamma^{*}p}$, $Q^2$, and $-t_{p}$, and a flat proposal for $s_1$.
Concretely, we sample
\[
z_s=\ln s_{\gamma^{*}p},\qquad
z_Q=\ln Q^2,\qquad
z_t=\ln(-t_p)
\]
uniformly within their allowed intervals, and then sample $s_1$ uniformly in
$\big[s_1^{-}(s,s_{\gamma^{*}p},Q^2,t_p),\,s_1^{+}(s,s_{\gamma^{*}p},Q^2,t_p)\big]$.
The corresponding proposal densities are therefore
\begin{align}
p_s(s_{\gamma^{*}p})&=
\frac{1}{s_{\gamma^{*}p}\,\ln\!\big(s_{\gamma^{*}p,\max}/s_{\gamma^{*}p,\min}\big)}\,,\nonumber\\
p_Q(Q^2\,|\,s_{\gamma^{*}p})&=
\frac{1}{Q^2\,\ln\!\big(Q^2_{\max}(s,s_{\gamma^{*}p})/Q^2_{\min}(s,s_{\gamma^{*}p})\big)}\,,\nonumber\\
p_t(t_p\,|\,s_{\gamma^{*}p},Q^2)&=
\frac{1}{(-t_p)\,\ln\!\big((-t_p^{-})/(-t_p^{+})\big)}\,,\nonumber\\
p_{s_1}(s_1\,|\,s_{\gamma^{*}p},Q^2,t_p)&=
\frac{1}{s_1^{+}-s_1^{-}}\,,
\end{align}
where $Q^2_{\min,\max}$, $t_p^{\pm}$, and $s_1^{\pm}$ are the exact kinematic boundaries for the previously sampled variables.

Each trial point is assigned the importance-sampling weight
\begin{align}
    \label{eq:weight-event}
    \omega_i(s;s_{\gamma^{*}p},Q^{2},t_{p},s_1)=
    \frac{\displaystyle \frac{d^{4}\sigma}{ds_{\gamma^{*}p}\,dQ^{2}\,dt_{p}\,ds_{1}}}
    {p_s(s_{\gamma^{*}p})\,p_Q(Q^2\,|\,s_{\gamma^{*}p})\,p_t(t_p\,|\,s_{\gamma^{*}p},Q^2)\,p_{s_1}(s_1\,|\,s_{\gamma^{*}p},Q^2,t_p)}\,.
\end{align}
Equivalently,
\begin{align}
    \omega_i
    =
    \frac{d^{4}\sigma}{ds_{\gamma^{*}p}\,dQ^{2}\,dt_{p}\,ds_{1}}\;
    s_{\gamma^{*}p}\ln\!\frac{s_{\gamma^{*}p,\max}}{s_{\gamma^{*}p,\min}}\;
    Q^2\ln\!\frac{Q^2_{\max}}{Q^2_{\min}}\;
    (-t_p)\ln\!\frac{-t_p^{-}}{-t_p^{+}}\;
    (s_1^{+}-s_1^{-})\,,
\end{align}
with all kinematic boundaries evaluated at the sampled point.
The Monte Carlo estimate of the total cross-section is given by
\begin{equation}
    \label{eq:cross-section-MC}
    \sigma_{X}
    =
    \frac{1}{N} \sum_{i=1}^N \omega_i \, .
\end{equation}

\begin{figure}[t!]
\centering
\includegraphics[width=0.5\linewidth]{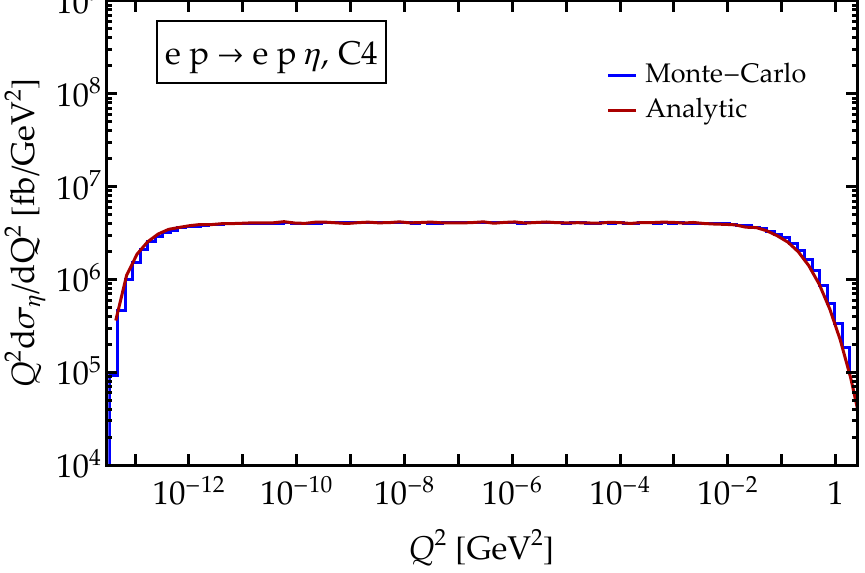}~
\includegraphics[width=0.5\linewidth]{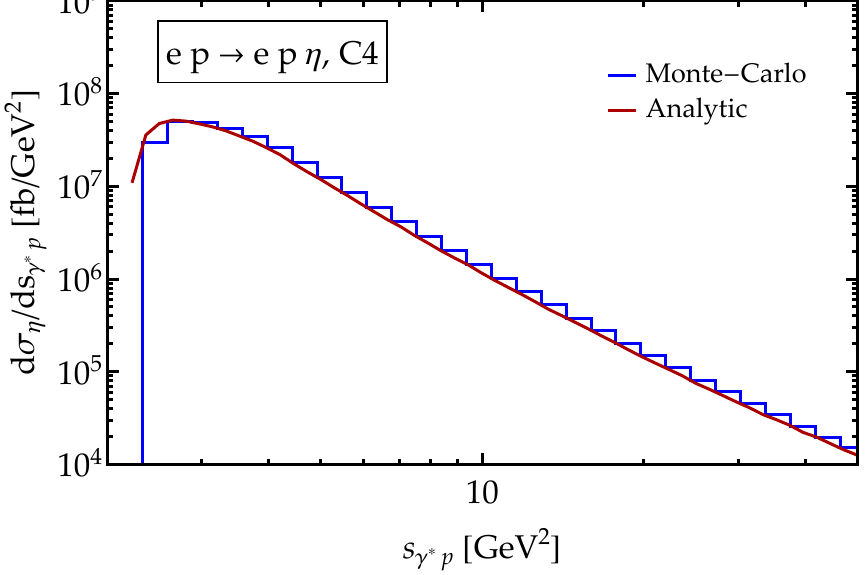}
\caption{Comparison between the analytic (red) and Monte Carlo sampler (blue) implementations of the full $2\to 3$ kinematics of the coherent electroproduction process~\eqref{eq:process}, considering the production of $\eta$ mesons for the setup with energies $E_{p}\times E_{e} = 100~\mathrm{GeV}\times 10~\mathrm{GeV}$. Left: $d\sigma/d\ln(Q^{2})$. Right: $d\sigma/ds_{\gamma^{*}p}$.}
\label{fig:comparison-MC-vs-analytic}
\end{figure}

Finally, for each event, we may sample the full kinematics -- the 4-momenta of the outgoing $e', p'$, and $X$, with optional sampling of $X$ decays.
The 4-momenta are calculated in terms of the Mandelstam invariants using the conservation laws; details are summarized in Appendix~\ref{app:kinematics-from-invariants}.

As a cross-check, we verify that the inclusive cross section of Eq.~\eqref{eq:cross-section-MC} reproduces the direct analytic integration of Eq.~\eqref{eq:d4sigma-generic} at the few-percent level for $m_X\gtrsim 40\,\MeV$, both for total rates and for representative one-dimensional projections, see Fig.~\ref{fig:comparison-MC-vs-analytic}.
For smaller masses, instabilities may appear at extremely small $Q^2$ due to the sharp phase-space boundaries and floating-point precision limitations.

\section{Kinematical distributions at the EIC}
\label{sec:kinematics-domain}

\subsection{Energy configurations and event selections at the EIC}
\label{sec:events}

We consider different beam configurations for the EIC, which span an invariant mass range of $\sqrt{s} = (28 - 140)\,\GeV$~\cite{Accardi:2012qut,AbdulKhalek:2021gbh} and are summarized in Tbl.~\ref{tab:energy-configurations}.
We focus primarily on Configurations 1, 4, and 8, which cover distinct kinematic regimes, and present selected results for the remaining setups.
\begin{table}[t]
\centering
\begin{tabular}{cccc}
\hline\hline
Configuration & $E_p\,[\GeV]$ & $E_e\,[\GeV]$ & $\sqrt{s}\,[\GeV]$ \\
\hline
C1 & 41  & 5  & $ 29$ \\
C2 & 100 & 5  & $ 45$ \\
C3 & 130 & 5  & $ 51$ \\
C4 & 100 & 10 & $ 63$ \\
C5 & 250 & 5  & $ 71$ \\
C6 & 130 & 10 & $72$ \\
C7 & 250 & 10 & $100$ \\
C8 & 275 & 18 & $ 140$ \\
\hline\hline
\end{tabular}
\caption{The considered energy beam configurations at the EIC.}
\label{tab:energy-configurations}
\end{table}
We define the kinematics domain, relevant to all beam configurations, in which the final-state particles $e'$, $X$, and $p'$ in the electroproduction process
\begin{equation}
    e+p\to e'+p'+X\nonumber
\end{equation}
can be reconstructed at the EIC.
Our discussion is phrased in terms of the pseudorapidities $\eta_{e'}$, $\eta_{X}$, $\eta_{p'}$, the scattered proton energy $E_{p'}$, the energy carried by the produced system $E_{X}$, and the squared electron momentum transfer $Q^{2}$.
For detector layouts and acceptance considerations, we follow Ref.~\cite{Balkin:2025rtc} based on Refs.~\cite{ePICDetector,PhysRevC.104.065205,dvcs2025}.

Electron reconstruction is possible in two complementary kinematic regimes.
The first domain corresponds to the central detector, which is conventionally characterized by the fiducial pseudorapidity range $|\eta_{e'}|<4$.
The second regime corresponds to the far-backward detector~(FBD), targeting large pseudorapidities $\eta_{e'}\lesssim -4$.
In the EIC studies, the domain of the $e'$ kinematics $E_{e'},\eta_{e'}$ where FBD has the maximal reconstruction capabilities is often effectively formulated in terms of
\begin{align}
    \label{eq:domain-FBD}
    10^{-3}\,\GeV^{2}<Q^{2}(E_{e'},\eta_{e'})<0.1\,\GeV^{2}\,.
\end{align}
The reconstruction in the range of tiny $Q^{2}<10^{-3}\,\GeV^{2}$ is limited by smearing induced by the beam angular divergence and is rather irreducible. The intermediate region $0.1\,\GeV^{2}\lesssim Q^{2}\lesssim 1\,\GeV^{2}$ is associated with partially uninstrumented directions around the beam-pipe, while $Q^{2}>1\,\GeV^{2}$ typically falls to the central detector acceptance.

However, Eq.~\eqref{eq:domain-FBD} is a decent description of the efficiency coverage only if $Q^{2}(E_{e'},\eta_{e'})$ is dominated by sizable electron energy loss. This is not the case for our signal: as we will see in Fig.~\ref{fig:Eepr-eta_epr}, the electroproduction process~\eqref{eq:process} with hadronically interacting $X$ particle features very small electron fractional energy loss $x \equiv (E_{e}-E_{e'})/E_{e} \ll 0.1$. As a result, $Q^{2}$ is almost uniquely characterized by $\eta_{e'}$.

For this kinematics, the reconstruction capabilities of the current FBD design are limited even if falling into Eq.~\eqref{eq:domain-FBD}; the reconstruction is possible only within a narrow backward pseudorapidity band $-6.5<\eta_{e'}<-5.5$, and outside this band, the scattered electron escapes through uninstrumented apertures.
Moreover, even within $-6.5<\eta_{e'}<-5.5$, the reconstruction efficiency is only $\simeq 0.2$~\cite{AbdulKhalek:2021gbh,Klest:2025fwx}.

This motivates extending the backward instrumentation of the main EIC detector to cover the full range of $\theta_{e'}$, for instance, by adding tracking coverage at larger $\theta_{e'}$. This is possible for the current layout of ePIC.
The $\eta_{e'}$ coverage by FBD may be extended up to $\eta_{e'}<-4.2$, with the domain $-4.2<\eta_{e'}<-4$ being uncovered neither by FBD nor by the central detector because of the beam pipe.
Moreover, for the alternative detector design~\cite{BURKERT2023104032, NADEL-TURONSKI:2024/c}, additional backward coverage could further increase the fraction of events with a reconstructed $e'$ by a factor of a few; operationally, this also corresponds to enlarging the covered $\theta_{e'}$ range (and thereby extending the kinematic reach of the backward region).  Further, we assume that in the domain $Q^{2}(\eta_{e'},E_{e'})>10^{-3}\,\GeV^{2}$ and $\eta_{e'}<-4.2$, electrons may be reconstructed by the FBD with unit efficiency.

The domain where the final state $X$ may be reconstructed depends on its properties.
For the central detectors, $|\eta|<4$, both charged and neutral particles may be reconstructed.
This is also possible in the partially instrumented region $4.5<\eta_{X}<6$, while in $\eta > 6$, only neutral particles and heavy charged states such as protons may be reliably reconstructed.
Conservatively, we focus solely on the fully instrumented region $|\eta_{X}|<4$.
Additionally, we require $X$ to carry a non-negligible fraction of the proton energy $E_{X}>0.1E_{p}$, so that $X$ (or its decay products) can be reconstructed with adequate resolution.
In principle, for $X$ decaying to charged final states, we can also use other kinematical regions to improve the calibrations.

For the scattered proton, we restrict to the far-forward reconstruction domain $\eta_{p'}>4.5$ and the energy $E_{p'}>0.5E_{p}$, where the proton reconstruction efficiency is maximal~\cite{Jentsch:2023krakow,Pitt:2024utg}.
As we discussed above, for the hadronically coupled $X$ states considered below, the electron typically retains nearly its beam energy, $E_{e'}\simeq E_{e}$, and the energy transfer is dominantly carried by the proton leg.
In this limit, one may use $E_{X}\simeq E_{p}-E_{p'}$ as a good approximation, and the requirement that $X$ carries at least a fixed fraction of $E_{p}$ can be equivalently enforced as an upper bound on $E_{p'}<0.9E_{p}$.

The resulting fiducial reconstruction domain is summarized by the following compact selection sets:
\begin{align}
    \label{eq:selection-electron}
    &
    |\eta_{e'}|<4
    \quad\text{or}\quad
    \left[Q^{2}(E_{e'},\eta_{e'})>10^{-3}\,\GeV^{2}
    \quad \text{and} \quad
    \eta_{e'}<-4.2\right]
      \,,
    \\
    \label{eq:selection-proton}
    &\eta_{p'}>4.5\,,
    \qquad
    0.5<E_{p'}/E_{p}<0.9\, ,
    \\
    \label{eq:selection-X}
    &|\eta_{X}|<4\, ,
    \qquad
    E_{X}/E_p>0.1 \, .
\end{align}
Given that $Q^{2}(E_{e'},\eta_{e'})$ is strongly dominated by $\eta_{e'}$, we may approximately translate the FBD acceptance into the following bounds on the $e'$ pseudorapidity (using Eq.~\eqref{eq:Q2-Eepr-etaepr} below):
$\eta_{e'} \in [-5.8, -4.2],[-6.4,-4.2]$ and $[-7.0, -4.2]$ for the EIC beam configurations from Tbl.~\ref{tab:energy-configurations} with $E_{e'}\approx E_{e}=5,10$ and $18\,$GeV, respectively.
We denote the total cross-section by $\sigma_{X,\rm tot}$ and the cross-section calculated under the selection criteria of Eqs.~\eqref{eq:selection-electron}-\eqref{eq:selection-X} by $\sigma_{X,\rm fid}$.

\subsection{Kinematic distributions}
\label{sec:kinematics-correlations}

Let us start with the distributions of the electroproduction events under the selection criteria of Eqs.~\eqref{eq:selection-electron}-\eqref{eq:selection-X} in $s_{\gamma^{*}p}$, the proton momentum transfer $t_{p}$, and the electron momentum transfer $Q^{2}$.
The first two distributions are crucial for assessing the validity of our approach, which requires $\sqrt{s_{\gamma^{*}p}}\gtrsim 2\,\mathrm{GeV}$ and $|t_{p}|\lesssim \text{few GeV}^{2}$. Below $\sqrt{s_{\gamma^{*}p}}\sim 2\,\mathrm{GeV}$, our framework does not incorporate effects from on-shell intermediate baryonic resonances.
The $Q^{2}$ domain is important to understand how the outgoing electron $e'$ is likely to be detected at the EIC - by the FBD, or by the central detectors, and is simultaneously convenient for understanding the applicability of the equivalent photon approximation (see Sec.~\ref{sec:EPA}).

We consider two energy configurations from Tbl.~\ref{tab:energy-configurations}: C1 and C4, which we refer to as the low- and high-energy configurations, respectively.
Next, we will consider the $\pi^{0}$ and $\omega$ particles production, serving as representatives of a ``light'' and a ``heavy'' state $X$, respectively.
The distributions of all other mesons, $\eta,\rho^{0},\dots$, qualitatively belong to the ``heavy'' state as well.
\begin{figure}[t!]
    \centering
    \includegraphics[width=0.33\linewidth]{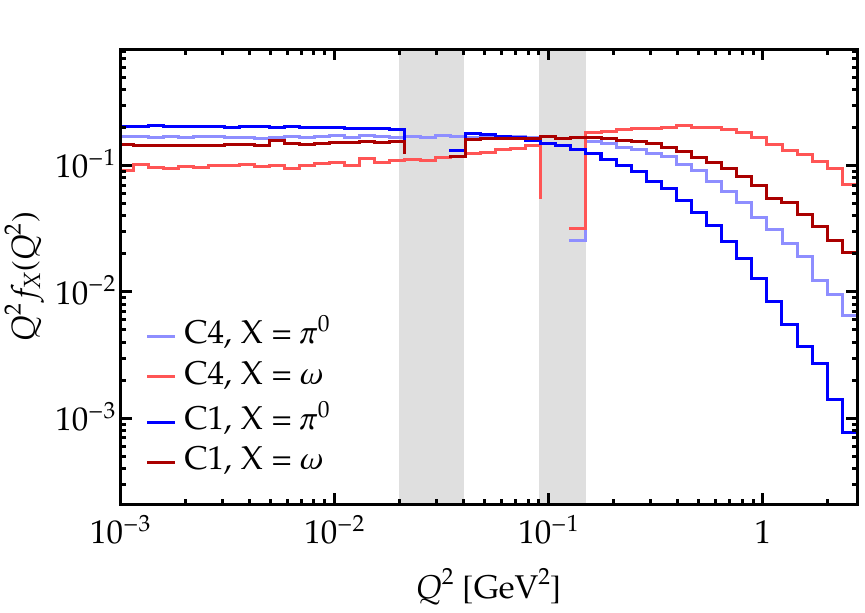}~\includegraphics[width=0.33\linewidth]{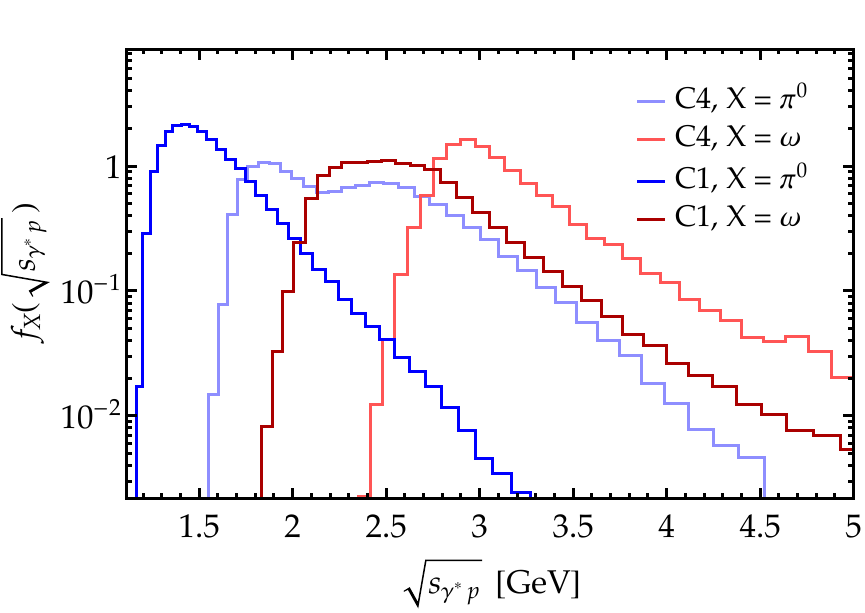}~\includegraphics[width=0.33\linewidth]{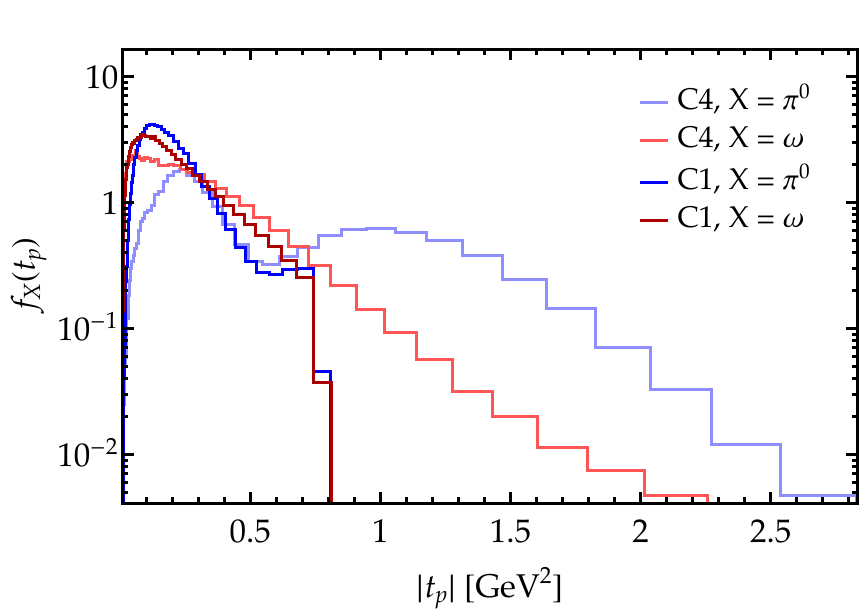}
    \caption{The differential distributions of the $X = \pi^{0},\omega$ electroproduction events under the selection~\eqref{eq:selection-electron}-\eqref{eq:selection-X}. We formulate them in terms of the probability density functions $f_{X}(y) =\frac{1}{\sigma_{X,\text{fid}}}
    \frac{d\sigma_{X,\text{fid}}}{dy}$, where $\sigma_{X,\text{fid}}$ is the cross-section after imposing the cuts. \textit{Left panel}: $Q^{2}f_{X}(Q^{2})$. \textit{Central panel}: $f_{X}(s_{\gamma^{*} p})$. \textit{Right panel}: $f_{X}(t_{p})$. In all plots, the blue (red) lines denote the $\pi^{0}$ ($\omega$) distributions, with the darker (lighter) lines corresponding to the beam energy configuration C1 (C4) from Tbl.~\ref{tab:energy-configurations}. For the $Q^{2}$ distribution, the gaps at $0.02<Q^{2}/\GeV^{2}<0.04$ (for C1) and $0.09<Q^{2}/\GeV^{2}<0.15$ (for C4), shown in gray, arise because of the separation of the FBD and central detectors in the covered pseudorapidity regions: $-4.2<\eta_{e'}<-4$ cannot be covered.}
    \label{fig:distributions-sgammap-Q2-tp}
\end{figure}
The distributions are shown in Fig.~\ref{fig:distributions-sgammap-Q2-tp}, and we summarize them below.

Before imposing the selection (see Fig.~\ref{fig:comparison-EPA-vs-2-3}, left panel), the distribution $Q^{2}d\sigma_{X}/dQ^{2}$ is roughly flat in the domain $Q^{2}_{\text{min}}\ll Q^{2}\lesssim 0.1\,\GeV^{2}$, where $Q^{2}_{\text{min}}$ is the lower kinematic threshold
\begin{align}
    Q^{2}_{\text{min}}
    =
    \frac{(2m_{X}m_{p}+m_{X}^{2})^{2}m_{e}^{2}}{4E_{e}E_{p}(4E_{e}E_{p}-2m_{X}m_{p}-m_{X}^{2})} \, .
\end{align}
This follows from the $1/Q^{2}$ scaling of the differential cross-section, and from the phase space closure near $Q^{2}_{\text{min}}$. Larger $Q^{2}\gtrsim 0.1\,\GeV^{2}$ are suppressed by the form-factor $F_{X}(Q^{2})$ in Eq.~\eqref{eq:photo-to-electro}. However, depending on the EIC energy setup, the event selection prefers larger $Q^{2}$'s unless they are close to the upper kinematic threshold of the electroproduction process, $Q^{2}_{\rm max}$.
For the C1 configuration, the kinematic threshold is $Q_{\rm max}^{2}\simeq (3-4)\,\GeV^{2}$, so the relative fraction of the large-$Q^{2}$ distribution after selection remains heavily suppressed. However, for the C4 configuration, the threshold is far away, and the distribution remains roughly flat even for $Q^{2}\lesssim 2-3\,\GeV^{2}$.

The $t_{p}$ distributions show that most of the events are located in the domain $|t_{p}|\lesssim 2\,\GeV^{2}$ -- within the validity range of our approach.
For the C1, the $t_{p}$  truncation at $|t_{p}| \gtrsim 0.7\,\GeV^{2}$ is caused by the limited transverse momentum of the scattered proton allowed by the $\eta_{p'} > 4.5$ cut at this proton beam energy.
The shape of the distribution for the $\pi^{0}$ events closely follows the shape of the photoproduction cross-section $d\sigma/dt_{p}$ in the domain $\sqrt{s_{\gamma p}} = \text{few GeV}$~\cite{Kashevarov:2017vyl}.

\begin{figure}[t!]
    \centering
    \includegraphics[width=0.5\linewidth]{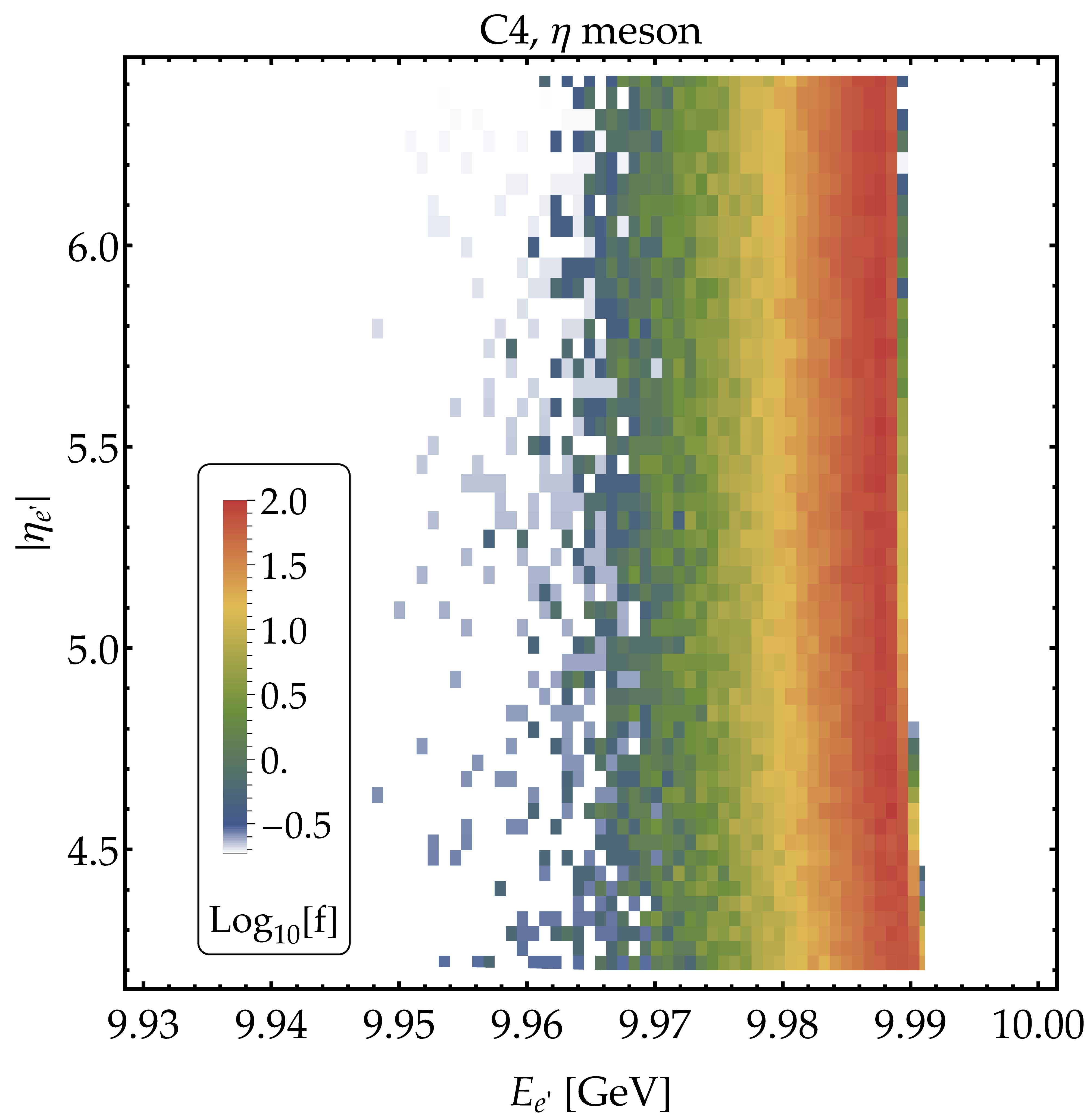}~    \includegraphics[width=0.5\linewidth]{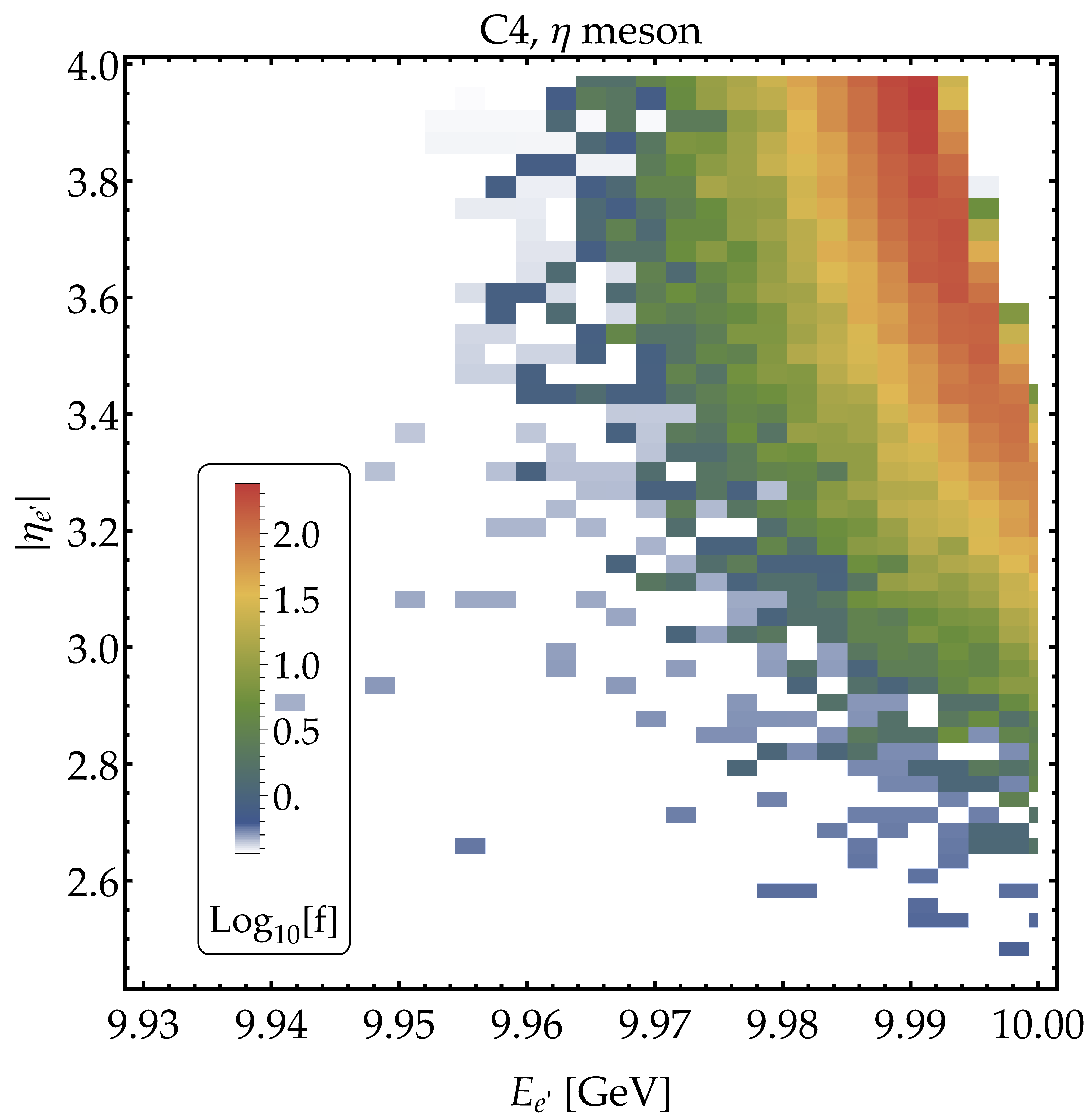}
    \caption{The distribution $f \equiv 1/\sigma_{\eta,{\rm fid}}d^{2}
    \sigma_{\eta,{\rm fid}}/dE_{e'}d\eta_{e'}$ for the events passing the selection~\eqref{eq:selection-electron}-\eqref{eq:selection-X}, for C4.
    The left panel (right) corresponds to the events where $e'$ is within the FBD (central detectors) coverage.}
    \label{fig:Eepr-eta_epr}
\end{figure}

Finally, let us discuss the $s_{\gamma^{*}p}$ distribution.
Apart from the production of the $\pi^{0}$ particle in the C1 configuration, the distribution mainly lives in the $\sqrt{s_{\gamma^{*} p}}>2\,\GeV$ domain.
There are two reasons for this.
First, increasing the $X$ mass shifts the distribution towards higher $\sqrt{s_{\gamma^{*}p}}$, which naturally follows from the naive reaction threshold, $s_{\gamma^{*}p}> (m_{X}+m_{p})^{2}$, mostly small $Q^{2}$, and the fact that the off-shell photon $\gamma^{*}$ tends to be very soft for any $Q^{2}$ (see Fig.~\ref{fig:Eepr-eta_epr}).
Second, for any configuration apart from C1, the $\sqrt{s_{\gamma^{*}p}}$ is distributed well above the naive threshold $(m_{X}+m_{p})$.
In particular, for the neutral pion, most of the events populate the domain above $\sqrt{s_{\gamma^{*}p}}>2\,\GeV$.
This follows from adding the cut $|\eta_{X}|<4$ on top of $\eta_{p'}>4.5, \, E_{p'}<0.9 E_{p}$.
Indeed, near threshold, the sub-process $\gamma^{*}+p\to X+p'$ has very little relative momentum in its center-of-mass frame, so after boosting to the lab, both $X$ and $p'$ are carried to similar forward rapidities.
Requiring simultaneously a very forward proton and a central $X$ therefore selects events sufficiently above threshold, where the final-state recoil is large enough to separate their rapidities.
This effect becomes much weaker when lowering the energy of the incoming proton $E_{p}$.
In that case, the overall longitudinal boost of the $\gamma^{*}p$ system is typically smaller, so even near-threshold configurations can satisfy $|\eta_X|<4$ while keeping $\eta_{p'}>4.5$, and the accepted $s_{\gamma^{*}p}$ spectrum stays much closer to threshold.
In particular, for $\pi^{0}$ production in the C1 configuration, most accepted events remain in the domain $\sqrt{s_{\gamma^{*}p}}<2\,\GeV$.

Therefore, we conclude that our approach breaks down for the C1 configuration, in the case of the production of $\pi^{0}$ mesons and similar particles with mass $m_{X}\lesssim m_{\pi^{0}}$.
Nevertheless, it may still suggest qualitative behavior of the kinematic distributions, and we keep this configuration for completeness.

Next, let us focus on the final electron kinematics.
It is important in the context of a non-uniform $e'$ reconstruction efficiency by the FBD detector, as we highlighted above -- for the marginal portions of the parameter space, its dependence on the kinematics cannot be described solely by $Q^{2}$, being sensitive to both $E_{e'},\theta_{e'}$.
Namely,
\begin{align}
\label{eq:Q2-Eepr-etaepr}
    Q^{2}
    \approx
    2E_{e}E_{e'}(1+\cos(\theta_{e'}))
    \approx E_{e}E_{e'}(\pi-\theta_{e'})^{2}
    \approx \frac{4E_{e}E_{e'}}{1+ e^{-2\eta_{e'}}},
\end{align}
where $\theta_{e'}\,(\eta_{e'})$ is the polar angle\,(pseudorapidity) of the electron $e'$ with respect to the direction of the incoming proton.
In the first equality, we neglected the electron mass, while the second approximate equality holds if the outgoing electrons are nearly collinear to the incoming electron, $\theta_{e'}\to \pi$.
In Fig.~\ref{fig:Eepr-eta_epr}, we plot the $\eta$ meson production event density distributions in the $\eta_{e'}-E_{e'}$ plane, for the C4 setup;
the case of other mesons and setups is qualitatively similar.
We will consider the two domains of the outgoing electron kinematics according to the selection~\eqref{eq:selection-electron}: the FBD domain $Q^{2}>10^{-3}\,\GeV^{2}$, $\eta_{e'}<-4.2$ (the left panel), and $|\eta_{e'}|<4$, which belongs to the central detectors coverage (the right panel).
We see that the majority of the events have tiny electron energy recoil $(E_{e}-E_{e'})/E_{e} < 2$--$3\%$, independent of the range of $Q^{2}$. These two plots, together with Eq.~\eqref{eq:Q2-Eepr-etaepr}, suggest that there is a clear correlation between $Q^{2}$ and $\eta_{e'}$ for our electroproduction events.

As we discussed in Sec.~\ref{sec:events}, this tiny electron energy recoil and, correspondingly, large $\eta_{e'}$ prevent such events from efficient reconstruction by the current FBD setup (that may only cover $-6.5<\eta_{e'}<-5.5$ and reconstruct there the $e'$ particles with $E_{e'} \approx E_{e}$ with the efficiency $\simeq 0.2$), motivating a potential future upgrade to accommodate these measurements.

\begin{figure}
    \centering
    \includegraphics[width=0.5\linewidth]{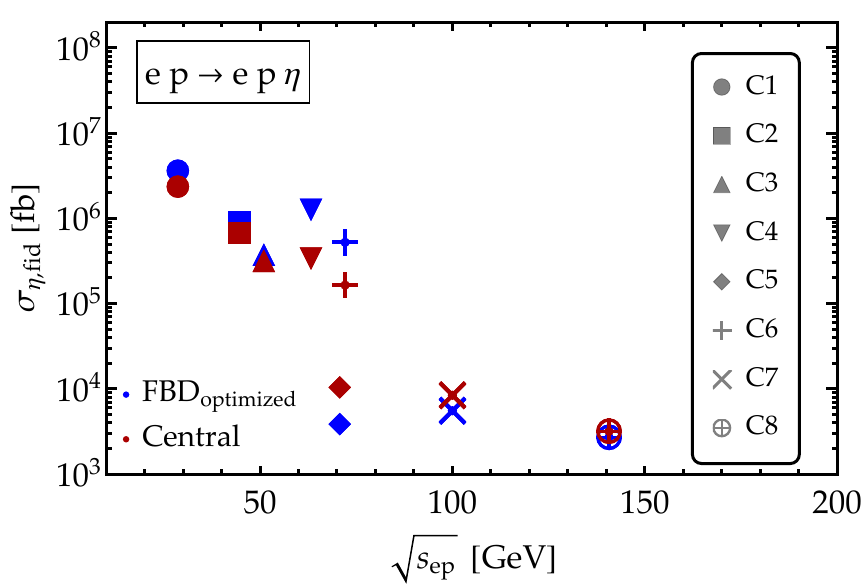}~\includegraphics[width=0.5\linewidth]{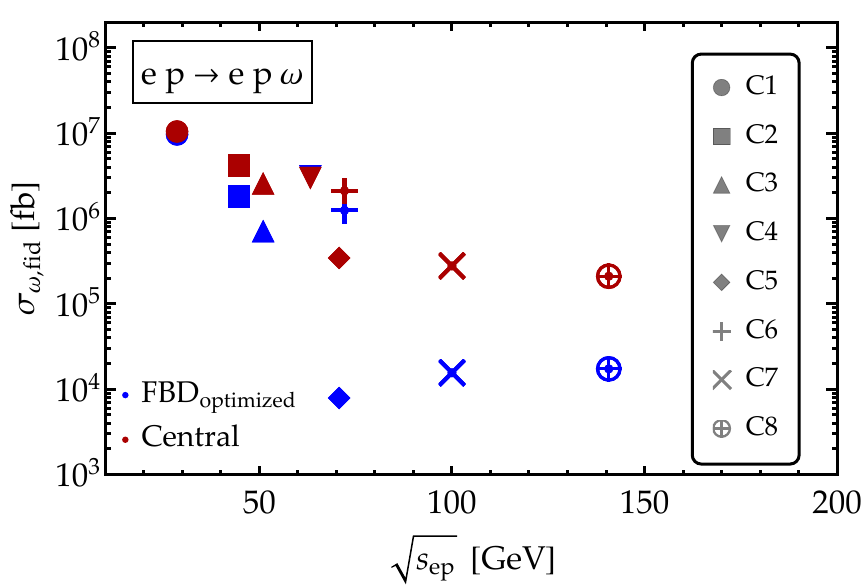}
    \caption{The electroproduction cross-sections after the selection in Eqs.~\eqref{eq:selection-electron}-\eqref{eq:selection-X}, for the energy beam configurations C1--C8 in Tbl.~\ref{tab:energy-configurations}. Two final states are considered: $\eta$ mesons (left) and $\omega$ mesons (right).
    The blue markers show the selection requiring the $e'$ reconstruction by EIC's Far-Backward Detector (FBD), whereas the red markers consider reconstruction by central detectors. The clarification ``optimized'' $\text{FBD}_{\text{optimized}}$ refers to our assumption that the FBD has $\mathcal{O}(1)$ $e'$ kinematics reconstruction efficiency in the full domain $Q^{2}>10^{-3}\,\GeV^{2}$, $\eta_{e'}<-4.2$ of Eq.~\eqref{eq:selection-electron} (see text for details).
    }
    \label{fig:MPE-cross-sections}
\end{figure}

Next, we analyze the behavior of the cross-section $\sigma_{X,\rm fid}$ obtained by imposing the selections of Eqs.~\eqref{eq:selection-electron}-\eqref{eq:selection-X}.
We assume that independently of the energy beam configuration, the reconstruction efficiency of the events falling under the selection is unity.
In Fig.~\ref{fig:MPE-cross-sections}, we report the cross-sections of the events with the production of $\eta$ and $\omega$ mesons falling under the selection, separating the events with the electrons $e'$ that may be reconstructed by the FBD and the central detector.
All the beam energy configurations listed in Tbl.~\ref{tab:energy-configurations} are used to study the efficiency of the search scheme.
The selection is less efficient for high-energy configurations such as C6-C8, mainly due to the $|\eta_{X}|<4$ cut.
This is because for the configurations with very energetic protons with $\eta_{p'}>4.5$, the kinematics mostly prefers far-forward $X$ particles.
For the same reason, the fiducial cross-section is maximized for C1, which is the least energetic case.
A significant decrease in the signal yield with increasing proton energy may be compensated by including the partially instrumented $4.5<\eta_X<6$ domain, but this requires a full detector simulation.
In addition, for the configurations C1-C4 and C6, the fraction of events with the $e'$ to be detected with the central detector is roughly comparable to the FBD-detected $e'$, and also becomes significantly larger for C5, C7, and C8.
\begin{figure}
    \centering
    \includegraphics[width=0.5\linewidth]{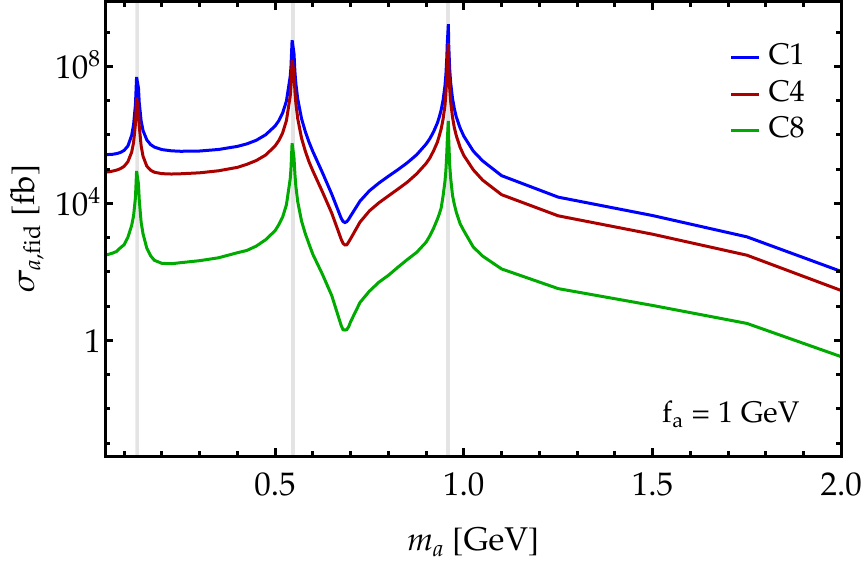}
    \caption{The production cross-section of ALPs, $\sigma_{a,{\rm fid}}$, after imposing the selection~\eqref{eq:selection-electron}-\eqref{eq:selection-X}, for the C1, C4, and C8 configurations from Tbl.~\ref{tab:energy-configurations}. For $m_a\gtrsim1\,\GeV$, the displayed curves are conservative lower estimates: mixing with omitted heavier pseudoscalar excitations may increase the rates by an order of magnitude or more.}
    \label{fig:sigma-ALP}
\end{figure}
Finally, Fig.~\ref{fig:sigma-ALP} shows the production cross-section of the gluonic ALPs after imposing the selection~\eqref{eq:selection-electron}-\eqref{eq:selection-X}, for the C1, C4, and C8 configurations. The results are in qualitative agreement with the production of the pseudoscalar mesons from Fig.~\ref{fig:MPE-cross-sections}. Note that the cross-section looks smooth at the transition $m_{a}\simeq 1.2\,\GeV$ between the small-mass and large-mass ALP generators in Eqs.~\eqref{eq:Ta-gluon-light},~\eqref{eq:Ta-gluon-heavy}; the discontinuity is at the level of $\simeq 10\%$.
\section{Comparison to the equivalent photon approximation}
\label{sec:EPA}

In this section,  we cross-check our full $2\to3$ framework against the equivalent photon approximation~(EPA)~\cite{Budnev:1975poe}.
The EPA expresses the electroproduction cross section Eq.~\eqref{eq:d4sigma-generic} in a flux-factorized form, where the electron line is encoded into an effective virtual-photon flux and the hadronic dynamics is approximated by the corresponding photoproduction subprocess, supplemented by a model for the finite-$Q^{2}$ dependence.

Our comparison is organized around two complementary kinematic regimes. First, we consider the region of small photon virtualities, $Q^2 \ll 1\,\GeV^2$, in which the exchanged photon is quasi-real and the EPA is expected to provide an accurate description of inclusive observables, such as total cross sections and single-differential distributions. In this regime, agreement with the EPA serves as a nontrivial validation of our full $2\to3$ implementation against a simpler and well-established approximation. Second, we consider the region of larger virtualities, $Q^2 \gtrsim 1\,\GeV^2$, where the basic assumptions underlying the EPA are no longer expected to hold at the level of inclusive rates, and deviations from the full $2\to3$ result are therefore anticipated. In addition, we study kinematic correlations to assess the limitations of the EPA beyond inclusive quantities. As we will show, even at very small $Q^2$, where the EPA reproduces inclusive observables well, it can still fail to describe the detailed event-by-event kinematics accurately.

\subsection{EPA formalism}
\label{sec:EPA-description}

Within the EPA approach, the estimate of the cross section of $e^-p\to e^- p X$ can be written as
\begin{align}
    \label{eq:EPA}
    \sigma_{X}^{\rm EPA}
    \simeq
    \int_{x_{\rm min}}^{x_{\rm max}} dx\int_{Q^2_{\rm min}}^{Q^2_{\rm max}} dQ^2\
    \mathrm{f}_{\gamma/e}(x,Q^2)\, \sigma_{\gamma^* p\to p X}(x,Q^{2})\,.
\end{align}
Here, $x\equiv E_\gamma/E_e$ is the energy fraction of the virtual photon, and $\mathrm{f}_{\gamma/e}(x,Q^2)$ is the photon flux from the electron beam~\cite{Budnev:1975poe}
\begin{align}
    \label{eq:EPA-flux}
    \mathrm{f}_{\gamma/e}(x,Q^2)
    =
    \frac{\alpha}{2\pi}\frac{1}{Q^2}
    \left[\frac{1+(1-x)^2}{x}-\frac{2(1-x)}{x}\frac{Q^2_{\rm min}(x)}{Q^2}\right]\,.
\end{align}
The  $\gamma^* p\to p X$ cross section can be approximated as~\cite{H1:1999pji,Lomnitz:2018juf}
\begin{align}
    \label{eq:sigma-2-2-EPA}
    \sigma_{\gamma^* p\to p X}(x,Q^{2})
    \approx
    \sigma_{\gamma p\to p X}(x,Q^{2} = 0)\times F_{X,\text{EPA}}(Q^2)\,.
\end{align}
Here, $\sigma_{\gamma p\to p X}(x,Q^{2} = 0)$ is the cross section with an on-shell photon.
$F_{X,\text{EPA}}$ is a form-factor cumulatively capturing finite $Q^2$ effects.
We adopt $F_{X,\text{EPA}}(Q^{2}) = |F_{X}(Q^{2})|^{2}$, where the form-factors $F_{X}(Q^{2})$ are defined in Eq.~\eqref{eq:photo-to-electro} and are given by Eq.~\eqref{eq:photo-to-electro-pseudoscalars} (\eqref{eq:photo-to-electro-vectors}) for the production of pseudoscalar (vector) particles.

The $x$ and $Q^{2}$ ranges in Eq.~\eqref{eq:EPA} are given by
\begin{align}
    \label{eq:integration-domain-x}
    &x_{\rm min}
    =
    \frac{(m_{X} + m_p)^2-m_p^2}{4 E_e E_p}\, ,
    \qquad
    x_{\rm max}
    =
    1-\frac{m_e}{2E_e}\, , \\
    \label{eq:integration-domain-Q2}
    &Q^2_{\rm min}(x)
    =
    \frac{x^2m_e^2}{1-x}\, ,
    \qquad\quad\quad \quad \ \ \
    Q^2_{\rm max}(x) = 4 E_e^2(1-x)\,,
\end{align}
where the $Q^2$ range can be obtained when the recoil electron moves along with and opposite to the incoming electron.
Motivated by the event selection from Eq.~\eqref{eq:selection-electron}, we introduce the sub-domain $Q^{2} \in (Q^{2}_{1},Q^{2}_{2})$, \emph{i.e.},
\begin{align}
    \label{eq:Q1Q2range}
    Q^{2}_{\text{min}}(x)<Q_{1}^{2}<Q^{2}<Q_{2}^{2}<Q^{2}_{\text{max}}(x)\,.
\end{align}
The ordering of integration in Eq.~\eqref{eq:EPA} gives us the cross-section $d\sigma/dx$ after integrating over $Q^{2}$. To get $d\sigma/dQ^{2}$, we may change the ordering by integrating first over the variable $x$ and obtain the differential cross section $d\sigma/dQ^2$, where now
\begin{align}
    \label{eq:EPA-domain-Q2}
    &\text{max}\left[Q^{2}_{1},\bar{Q}^{2}_{\text{min}}\right]<Q^{2}<\text{min}\left[Q^{2}_{2},\bar{Q}^{2}_{\text{max}}\right]\,, \\  &\bar{Q}^{2}_{\text{min}} = \frac{(2m_{X}m_{p}+m_{X}^{2})^{2}m_{e}^{2}}{4E_{e}E_{p}(4E_{e}E_{p}-2m_{X}m_{p}-m_{X}^{2})}\,, \quad \bar{Q}^{2}_{\text{max}} = \frac{E_{e}}{E_{p}}\left(4E_{e}E_{p}-2m_{X}m_{p}-m_{X}^{2}\right)\,,\\
    &\bar{x}_{\rm min}
    =
    \frac{(m_{X}+m_p)^2-m_p^2}{4 E_e E_p}\,,
    \quad
    \bar{x}_{\rm max}
    =
    {\rm min}\left[\frac{Q^2}{2m_e^2}\left(\sqrt{1+\frac{4m_e^2}{Q^2}}-1\right),1-\frac{Q^2}{4E^2_e}\right]\,.
\end{align}

Equation~\eqref{eq:sigma-2-2-EPA} embodies two distinct approximations.
First, the virtual-photon subprocess is approximated by the corresponding on-shell photoproduction cross section multiplied by a simple $Q^2$-dependent factor.
Second, the exchanged photon is treated within the EPA kinematics as emitted collinearly with the incoming electron, so that its transverse momentum is neglected and its momentum is characterized essentially by the longitudinal energy fraction $x$.

\subsection{Comparison between the EPA and the $2 \to 3$ kinematics}
\label{sec:EPA-comparison}

Now, let us proceed to a detailed comparison between the EPA and $2\to 3$ results.
We compare
\begin{itemize}
\item[--] The differential cross-sections $d\sigma/dQ^{2}$ and $d\sigma/dy$, where
    \begin{align}
        \label{eq:y}
        y \equiv (p_{p}\cdot p_{\gamma^{*}})/(p_{p}\cdot p_{e})
        = \frac{s_{\gamma^{*}p} - m_{p}^{2}+Q^{2}}{s-m_{p}^{2}-m_{e}^{2}}
    \end{align}
is the standard inelasticity variable.
\item[--] The integrated cross-section, with two specific $Q^2$ ranges in Eq.~\eqref{eq:EPA-domain-Q2}:
    \begin{equation}
    \text{total}: \ (Q^{2}_{1},Q^{2}_{2}) = (\bar{Q}^{2}_{\text{min}}, \ \bar{Q}^{2}_{\text{max}}) \quad \text{and} \quad \text{window}: \ (Q^{2}_{1},Q^{2}_{2}) = (1\,\GeV^2, \ \bar{Q}^{2}_{\rm max})\,.
    \label{eq:QiQii}
    \end{equation}
The second choice corresponds to the domain where the outgoing electron is typically within the range of the central EIC detectors. At the same time, it provides a direct test of the roughness of the approximation in Eq.~\eqref{eq:sigma-2-2-EPA}, since it emphasizes the larger-$Q^2$ region where the factorized on-shell treatment is expected to become less accurate.
\end{itemize}
This comparison serves a double purpose.
First, it validates the correctness of our full $2\to 3$ calculation for inclusive rates in the kinematic domain where the simpler EPA approach is expected to work very well.
Second, it reveals the limitations of the EPA that originate from the on-shellness and collinearity approximations encoded in Eq.~\eqref{eq:sigma-2-2-EPA}.
\begin{figure}[t!]
    \centering
    \includegraphics[width=0.5\linewidth]{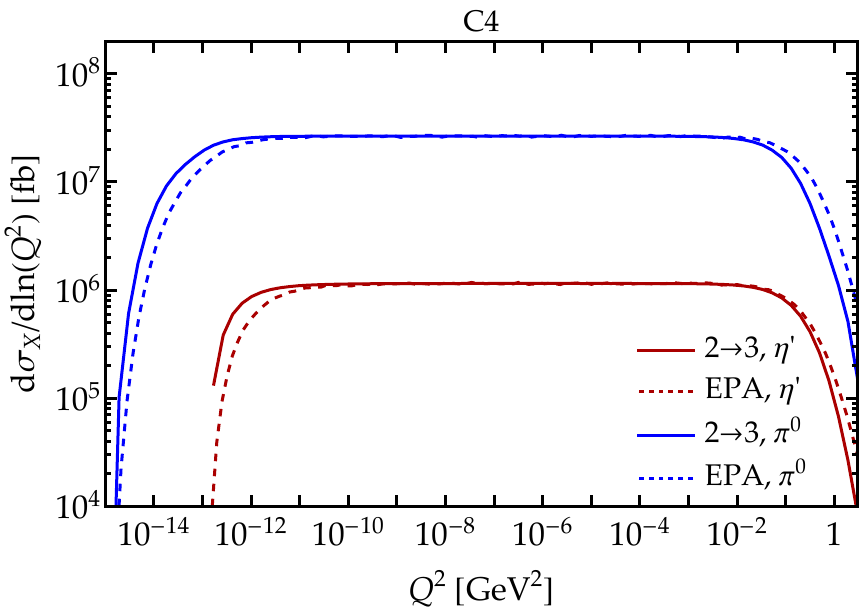}~\includegraphics[width=0.5\linewidth]{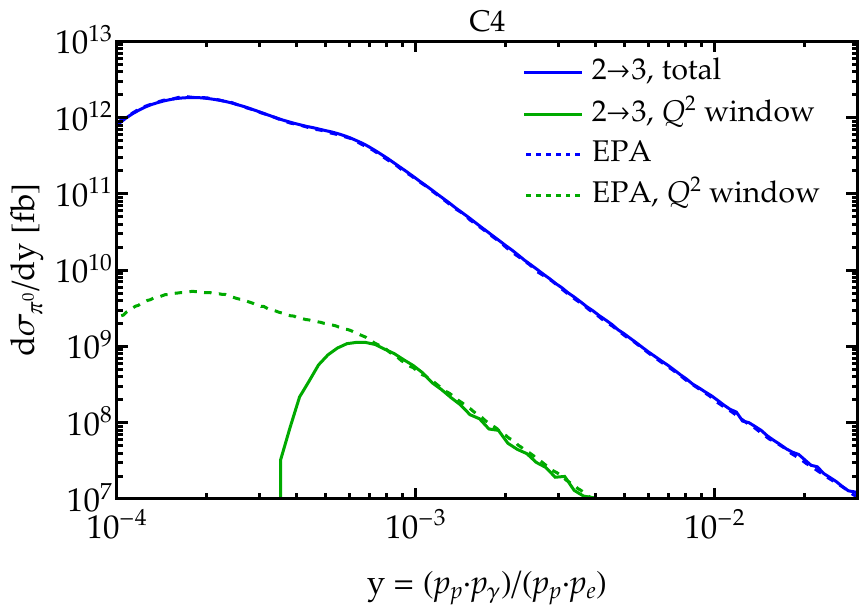}
    \caption{The differential cross-sections for the coherent production process~\eqref{eq:process}, as obtained by the $2\to 3$ approach from Sec.~\ref{sec:Xsec} (solid lines) and the EPA approach from Eq.~\eqref{eq:EPA} (dashed lines), considering the setup C8 from Tbl.~\ref{tab:energy-configurations}. \textit{Left panel}: the differential distribution in $d\sigma_{X}/d\ln(Q^{2})$, for the production of $X = \pi^{0}$ (blue) and $X = \eta'$ (red) mesons. \textit{Right panel}: the differential distribution $d\sigma_{\pi^{0}}/dy$ in the quantity $y$ from Eq.~\eqref{eq:y}. The blue (green) lines show the integration over the domain of $Q^{2}$ given by the first (second) set in Eq.~\eqref{eq:QiQii}.
    For other mesons and energy configurations, the comparison is very similar.}
    \label{fig:comparison-EPA-vs-2-3}
\end{figure}
Let us start by comparing the $Q^{2}$ differential cross-section, which we present in the form
\begin{align}
    \label{eq:sigma-diff-Q2}
    \frac{d\sigma}{d\ln(Q^{2})}
    =
    Q^{2}\frac{d\sigma}{dQ^{2}}\,.
\end{align}
The comparison of the $2\to3$ and EPA results for Eq.~\eqref{eq:sigma-diff-Q2} is shown in Fig.~\ref{fig:comparison-EPA-vs-2-3} (left panel).
Away from the kinematic boundaries, the two approaches are in good agreement and reproduce the approximately constant behavior of $d\sigma/d\ln(Q^{2})$.
Near the lower kinematic thresholds and for $Q^{2}\gtrsim 10^{-2}\,\GeV^{2}$, however, the EPA prediction becomes less accurate.
This is due both to its approximate treatment of the exact $2\to 3$ phase-space boundaries and to the simplified description of the virtual-photon subprocess in Eq.~\eqref{eq:sigma-2-2-EPA}.
In particular, once the photon virtuality is no longer very small, the use of an on-shell $\gamma p\to pX$ cross section multiplied by a simple factor $F_{X,\mathrm{EPA}}(Q^2)$ becomes too crude: in the full $2\to3$ amplitude, the dependence on $Q^2$ is correlated with other invariants, such as $t_p$ and $s_{\gamma^{*}p}$, through propagators, form factors, and scalar-product combinations, and therefore cannot, in general, be reduced to a universal multiplicative correction.

For the $d\sigma/dy$ distribution, in the EPA treatment underlying Eq.~\eqref{eq:sigma-2-2-EPA}, the exchanged photon is approximated as collinear with the incoming electron, $p_{\gamma^{*}}^{\mu} \approx (E_{\gamma},0,0,E_{\gamma})$, and therefore
\begin{align}
    y
    \approx
    2\frac{E_{p}E_{\gamma}+p_{p}E_{\gamma}}{s-m_{p}^{2}-m_{e}^{2}}
    \approx
    x \, ,
\end{align}
where we use $E_{p}\gg m_{p}, E_{e}\gg m_{e}$.
In the case of the full $2\to 3$ calculation, we use the exact definition of Eq.~\eqref{eq:y}.
The resulting comparison is shown in Fig.~\ref{fig:comparison-EPA-vs-2-3} (right panel) for the $\pi^{0}$ case, where we consider two $Q^{2}$ windows, as outlined above.
The differential cross-sections are very similar if considering the whole kinematically allowed $Q^{2}$ domain, but become qualitatively different if only large $Q^{2}>1\,\GeV^{2}$ are considered. In particular, the EPA predicts a distribution extending down to tiny $y$s, while the $2\to 3$ approach shows a sharp cut at higher $y$s.

\begin{table}[t]
    \centering
    \begin{tabular}{cccccc}
     \hline\hline
     Particle $X$ & Configuration & $\sigma_{X,2\to 3}$ [fb] & $\sigma_{X, \rm EPA}$ [fb] & $\sigma_{X,2\to 3}^{Q^{2}}$ [fb] & $\sigma_{X, \rm EPA}^{Q^{2}}$ [fb] \\
     \hline
$\pi^{0}$ &
\makecell{C8 \\ C4 \\ C1 } &
\makecell{$8.5\times 10^{8}$\\ $7.6\times 10^{8}$\\ $6.8\times 10^{8}$} &
\makecell{$8.4\times 10^{8}$\\ $7.5\times 10^{8}$\\ $6.6\times 10^{8}$} &
\makecell{$7.4\times 10^{5}$\\ $7.3\times 10^{5}$\\ $7.3\times 10^{5}$} &
\makecell{$2.1\times 10^{6}$\\ $2.1\times 10^{6}$\\ $2.0\times 10^{6}$}
\\
\hline
$\eta$ &
\makecell{C8 \\ C4 \\ C1 } &
\makecell{$1.3\times 10^{8}$\\ $1.1\times 10^{8}$\\ $1.0\times 10^{8}$} &
\makecell{$1.2\times 10^{8}$\\ $1.1\times 10^{8}$\\ $9.7\times 10^{7}$} &
\makecell{$1.4\times 10^{5}$\\ $1.4\times 10^{5}$\\ $1.4\times 10^{5}$} &
\makecell{$3.2\times 10^{5}$\\ $3.2\times 10^{5}$\\ $3.2\times 10^{5}$}
\\
\hline
$\eta'$ &
\makecell{C8 \\ C4 \\ C1 } &
\makecell{$3.4\times 10^{7}$\\ $3.0\times 10^{7}$\\ $2.7\times 10^{7}$} &
\makecell{$3.3\times 10^{7}$\\ $3.0\times 10^{7}$\\ $2.6\times 10^{7}$} &
\makecell{$4.3\times 10^{4}$\\ $4.3\times 10^{4}$\\ $4.3\times 10^{4}$} &
\makecell{$8.9\times 10^{4}$\\ $8.9\times 10^{4}$\\ $8.9\times 10^{4}$}
\\
\hline\hline
\end{tabular}
\caption{The integrated production cross-section of the pseudoscalar mesons, $X=\pi^{0}, \eta, \eta'$, as predicted by the full $2\to 3$ calculations of the process in Eq.~\eqref{eq:process} and by the ``naive'' EPA of Eq.~\eqref{eq:EPA}. Three energy configurations of the proton-electron beams from Tbl.~\ref{tab:energy-configurations} are considered: C8, C4, and C1. The total cross section $\sigma_{X}$ is given together with the cross section integrated over the region $Q^2>1~{\rm GeV}^2$ (denoted by $\sigma_{X}^{Q^{2}}$).
    }
\label{tab:cross-section-all-Q2-total}
\end{table}
Tbl.~\ref{tab:cross-section-all-Q2-total} compares the total and $Q^{2}$-windowed cross-sections for the three beam configurations C8, C4, and C1, where the fiducial region is defined through the $Q^2$ domain of Eq.~\eqref{eq:QiQii}.
Since the differential cross-section falls rapidly for $Q^2 \gtrsim 0.1\,\GeV^2$, the total cross-section is dominated by the low-$Q^2$ region, in which the EPA closely reproduces the full $2\to3$ result.
Accordingly, the total cross-sections obtained in the EPA and in the exact calculation are in very good agreement for all energy configurations considered.
The situation is different for the $Q^{2}$-windowed cross-section, where the agreement is visibly worse, in line with the behavior already seen in Fig.~\ref{fig:comparison-EPA-vs-2-3}.
The deviation is typically by a factor of 2-3, depending on the final-state particle $X$, and shows only a very weak dependence on the beam-energy configuration.

The deviations for the $Q^{2}>1\,\GeV^{2}$ domain come precisely from the region in which the on-shell approximation of Eq.~\eqref{eq:sigma-2-2-EPA} starts breaking down:
a simple factorized ansatz of the form $\sigma_{\gamma p\to pX}(x,Q^2=0)\,F_{X,\mathrm{EPA}}(Q^2)$ cannot reproduce the correlated dependence of the full $2\to3$ matrix element on $Q^2$, $t_p$, and $s_{\gamma^{*}p}$.

\section{Conclusions}
\label{sec:conclusions}

In this work, we estimated the exclusive electroproduction of bosons in the $e p \to e p X$ processes (see Eq.~\eqref{eq:process}), where $X$ can be a SM meson or a new physics state.\footnote{The full framework developed in this work, including the amplitudes, cross-section implementation, and event-generation tools used in our analysis, will be made publicly available in conjunction with the publication.}
This process has dual importance for the upcoming EIC.
First, it tests exclusive production in a new kinematical region, where hadronic physics plays a crucial role.
Second, exclusive channels are clean and have small backgrounds, thus they can be used to search for rare decays or productions of SM particles or new particles.

Our approach treats the process directly at the level of the exact $2\to 3$ phase space, retaining the full event kinematics and the correlations among the invariants relevant for experimental analyses at the EIC.
The hadronic currents are constructed from phenomenological amplitudes calibrated to available photoproduction data and generalized to finite virtuality through a practical electroproduction ansatz.
By construction, the amplitudes satisfy the Ward identities for both on-shell and off-shell photons.

On the theory side, we have assembled a common description for pseudoscalar and vector final states, including $\pi^{0}$, $\eta^{(\prime)}$, $\rho^{0}$, $\omega$, and $\phi$, and extended it to hadronically coupled ALPs and new vector mediators through mixing with the corresponding light-meson families.
This yields a compact and modular framework in which the hadronic assumptions remain explicit, can be confronted with existing data, and can be systematically refined as new measurements become available.
A central outcome of the study is that a full $2\to 3$ treatment is both feasible and useful for exclusive-production studies at the EIC, especially in situations where the signal selection depends on correlated cuts on the proton, the electron, and the produced state $X$.

Further, we analyzed the events' kinematics in the domain where both the outgoing $e'$, $p'$ particles may be reconstructed at the EIC, and found that the events concentrate in a characteristic kinematic regime.
For most channels and beam configurations, the selected events populate predominantly small photon virtualities, forward proton kinematics, and momentum transfers $|t_p| \lesssim \text{few GeV}^2$, while the $\gamma^{*}p$ invariant mass is typically above the baryon-resonance region, $\sqrt{s_{\gamma^{*}p}} \gtrsim 2\,\GeV$.
An important exception is the lightest pseudoscalar channel, $\pi^0$, in the lowest-energy EIC setup, where the accepted events can remain closer to threshold, and the sensitivity to the resonance region becomes more pronounced.
These observations clarify the regime in which the present framework is expected to be quantitatively reliable, and also identify the corners of phase space where future improvements are most needed.

The event analysis further highlights several practical implications for EIC searches based on the missing-proton-energy signature.
We find that the final-state proton typically remains very forward and carries most of the beam energy, while the produced state $X$ takes a moderate but experimentally relevant fraction of the proton-beam energy.
At the same time, the scattered electron usually retains nearly all of its initial energy, implying a strong preference for tiny electron recoil and a marked correlation between $Q^2$, $E_{e'}$, and $\eta_{e'}$.
This makes the far-backward electron acceptance especially important for maximizing the usable event sample, and suggests that improved backward instrumentation could significantly enhance the sensitivity of such searches.
We also find that lower- and intermediate-energy beam configurations tend to give the largest fiducial signal yields under conservative central-detector requirements on $X$.
In contrast, very high-energy proton beams more often push the produced state into the far-forward region.

We have also benchmarked the framework against the equivalent photon approximation (EPA).
For total rates and selected single-differential distributions in the near-real regime, when the virtuality of the intermediate photon is $Q^{2}\ll 1\,\GeV^{2}$, the agreement is very good, providing a useful validation of the full implementation. However, once $Q^{2}\gtrsim 1\,\GeV^{2}$, the EPA fails to reproduce both the integrated rates and the differential distributions. This confirms that the exact $2\to 3$ treatment is necessary for accurate signal studies.

Overall, the framework developed here provides a practical baseline for exclusive electroproduction studies of both SM mesons and new bosonic states at the EIC.
Its main advantages are that it preserves the full event kinematics, enforces electromagnetic consistency at the amplitude level, and remains flexible enough to incorporate improved hadronic input as it becomes available.
Future EIC measurements of exclusive electroproduction, in both visible and invisible channels, will provide a direct opportunity to test and refine this description further.

\begin{acknowledgments}
We are grateful to Craig Roberts for useful discussions.
The research of RB is supported in part by the U.S. Department of Energy grant number DE-SC0010107. The work of HL and AJ is supported by the U.S. Department of Energy under Grant Contract DE-SC0012704.
TC and YS are supported by the ISF (grant No. 597/24) and by BSF (grant No.
2024091).
YS thanks CERN-TH for the scientific associateship.
ST is supported by the Swiss National Science Foundation - project n. PZ00P2\_222350, and acknowledges the CERN TH Department for hospitality while this research was being carried out. This project has also received funding from the European Union’s Horizon Europe research and innovation programme under the Marie Skłodowska-Curie Staff Exchange grant agreement No 101086085 – ASYMMETRY.
\end{acknowledgments}
\appendix
\onecolumngrid

\section*{Appendix}

\setcounter{equation}{0}
\setcounter{table}{0}
\makeatletter

\section{Couplings of pseudoscalar mesons to nucleons}
\label{app:couplings}

In this appendix, we will derive the coupling of pseudoscalars to nucleons.
Let us start with the baryon-baryon-pseudoscalar interaction~\cite{Borasoy:1999nd,Bruns:2019fwi,Blinov:2021say}
\begin{align}
    \label{eq:BBP_lagrangian}
    \cL_{BBP}
    &=
    \frac{D_s}{2}\Tr\left(\bar{B}\gamma_\mu\gamma_5B\right)\Tr\left(u^\mu\right)+
    \frac{D}{2}\Tr\left(\bar{B}\gamma_\mu\gamma_5\left\{ u^\mu,B \right\}\right)+
\frac{F}{2}\Tr\left(\bar{B}\gamma_\mu\gamma_5\left[ u^\mu,B \right]\right)
\,,
\end{align}
where the trace $\Tr$ acts in the $SU(3)$ flavor space, $D \approx 0.8$, $F \approx 0.46$, $D_s \approx -0.41$ are phenomenological parameters, $B$ is the matrix of the nucleon octet,
\begin{align}
    B
    &\equiv
    \begin{pmatrix}
    \frac{1}{\sqrt{2}}\Sigma^0+\frac{1}{\sqrt{6}}\Lambda & \Sigma^+ & p\\
    \Sigma^- & -\frac{1}{\sqrt{2}}\Sigma^0+\frac{1}{\sqrt{6}}\Lambda & n\\
    \Xi^- & \Xi^0 & -\frac{2}{\sqrt{6}}\Lambda
    \end{pmatrix}\,,
\end{align}
and $u^\mu$ is the so-called vielbein.
In the leading order in $f_{\pi}/f_{a}$, it is
\begin{align}
    u^\mu
    &\equiv
    \frac{2}{f_\pi}
    \left(\partial^\mu P+\frac{f_\pi}{f_a}c_q \partial^\mu a\right),
\end{align}
where $P$ is the pseudoscalar nonet matrix
\begin{align}
    \label{eq:P_matrix}
    P
    &\equiv
    \frac{1}{\sqrt{2}}
    \begin{pmatrix}
    \frac{1}{\sqrt{2}}\pi^0+\frac{1}{\sqrt{3}}\eta+\frac{1}{\sqrt{6}}\eta' & \pi^+ & K^+\\
    \pi^- & -\frac{1}{\sqrt{2}}\pi^0+\frac{1}{\sqrt{3}}\eta+\frac{1}{\sqrt{6}}\eta' & K^0\\
    K^- & \overline{K}^0 & -\frac{1}{\sqrt{3}}\eta+\frac{2}{\sqrt{6}}\eta'
    \end{pmatrix}\,,
\end{align}
and $\eta^{(\prime)}$ are mass states.

Focusing on the proton and neutron, Eq.~\eqref{eq:BBP_lagrangian} leads to the interaction terms
\begin{align}
\cL_{N}
&=
\sum_{N = n,p}g_{NX}\ \overline{N}\gamma^{\mu}\gamma_{5}N \,
\partial_{\mu} X \, .
\end{align}
The nucleon-pseudoscalar couplings $g_{NX}$ are given by (see also Eq.~(B.39) of~\cite{Ovchynnikov:2025gpx})
\begin{align}\label{eq:BBP_vertex}
g_{NP}
&=
\frac{1}{f_\pi}\Tr\left(T_{P} T_{N}\right)\,,
\end{align}
where $T_N$ is $T_{p}\equiv \mathrm{diag}(D_s+D+F,D_s,D_s+D-F)$ or $T_{n}\equiv \mathrm{diag}(D_s,D_s+D+F,D_s+D-F)$ for the proton and neutron respectively, $T_{P}$ is the generator associated with the meson $P$ (given by Eq.~\eqref{eq:TP}).
In particular, we find
\begin{align}
    \label{eq:Coupl-ppi0}
    g_{p\pi^0}
    &=
    -g_{n\pi^0}
    =
    \frac{D+F}{2f_\pi}
    \approx
    6.8\, \GeV^{-1} \,, \\
    \label{eq:Coupl-peta}
    g_{p\eta}
    &=
    g_{n\eta}
    =
    \frac{2F+D_s}{\sqrt{6}f_\pi}
    \approx
    2.2\,\GeV^{-1}\,, \\
    \label{eq:Coupl-petapr}
    g_{p\eta'}
    &=
    g_{n\eta'}
    =
    \frac{3D-F+4D_s}{\sqrt{12}f_\pi}
    \approx
    0.93\ \GeV^{-1}\,.
\end{align}
%

\section{Couplings of ALPs}
\label{app:couplings-X}

In this appendix, we provide the explicit expressions for the couplings between ALPs and mesons.

The relevant Lagrangian is
\begin{align}
\cL_{aNN}
&=
\sum_{N = n,p}g_{Na}\ \overline{N}\gamma^{\mu}\gamma_{5}N \,
\partial_{\mu} a \, ,
    \\
    \cL_{V\gamma a}
    &=
    -\frac{g_{V\gamma a}}{2}\,a\,F^{\mu\nu}\tilde V_{\mu\nu}\,.
\end{align}
%

%

The nucleon couplings follow from Eq.~\eqref{eq:BBP_lagrangian}:
\begin{align}
g_{Na}
&=
\frac{1}{f_\pi}\Tr\left(\widetilde{T}_{a} T_{N}\right)=\sum_{P = \pi^{0},\eta,\eta'} \kappa_{aP}g_{PN}\,.
\end{align}
Likewise, the vector couplings are given by:
\begin{align}
g_{V\gamma a}
&=
\sum_{P = \pi^{0},\eta,\eta'} \kappa_{aP}g_{V\gamma P}\,.
\end{align}
%

Here, $\widetilde T_{a}$ is the ALP generator given by Eqs.~\eqref{eq:Ta-gluon-light},~\eqref{eq:Ta-gluon-heavy} with $\kappa_{aP}\equiv2\Tr\left[\widetilde{T}_aT_P\right]$,  $g_{PN}$ are the couplings of pseudoscalar mesons to nucleons from Eq.~\eqref{eq:Coupl-pP} and $g_{V\gamma P}$ their coupling to vector mesons from Tbl.~\ref{tab:Coupl-VPgamma}.

Concretely, we obtain the following gluonic-ALP couplings for $m_a\lesssim 4\pi f_\pi$:
\begin{multline}
    \label{eq:Coupl-pa}
    g_{pa}(m_a)\big|_{m_a<1.2\,\GeV}
    =
    \frac{m_0^2}{9f_a}\bigg[
    \frac{2F+D_s}{m_\eta^2-m_a^2}
    +
    \frac{2(3D-F+4D_s)}{m_{\eta'}^2-m_a^2}
    \\ +
    \frac{\delta_I m_\pi^2(D+F)}{m_\pi^2-m_a^2}
    \left(
    \frac{1}{m_\eta^2-m_a^2}
    +
    \frac{2}{m_{\eta'}^2-m_a^2}
    \right)\bigg]\,.
\end{multline}
The neutron coupling differs only in the sign of the isospin-breaking term.
For the radiative couplings, we similarly find
\begin{align}
    \label{eq:Coupl-Pgammaa}
    g_{\rho^{0}\gamma a}(m_a)\big|_{m_a<1.2\,\GeV}
    &=
    \frac{egm_0^2}{36\pi^2 f_a}\left[
    \frac{3}{m_\eta^2-m_a^2}
    +
    \frac{6}{m_{\eta'}^2-m_a^2}
    +
    \frac{\delta_I m_\pi^2}{m_\pi^2-m_a^2}
    \left(
    \frac{1}{m_\eta^2-m_a^2}
    +
    \frac{2}{m_{\eta'}^2-m_a^2}
    \right)\right]\,,
    \nonumber\\
    g_{\omega\gamma a}(m_a)\big|_{m_a<1.2\,\GeV}
    &=
    \frac{egm_0^2}{36\pi^2 f_a}\left[
    \frac{1}{m_\eta^2-m_a^2}
    +
    \frac{2}{m_{\eta'}^2-m_a^2}
    +
    \frac{\delta_I m_\pi^2}{m_\pi^2-m_a^2}
    \left(
    \frac{3}{m_\eta^2-m_a^2}
    +
    \frac{6}{m_{\eta'}^2-m_a^2}
    \right)\right]\,,
    \nonumber\\
    g_{\phi\gamma a}(m_a)\big|_{m_a<1.2\,\GeV}
    &=
    \frac{egm_0^2}{36\pi^2 f_a}\left[
    \frac{\sqrt{2}}{m_\eta^2-m_a^2}
    -
    \frac{4\sqrt{2}}{m_{\eta'}^2-m_a^2}
    \right]\,.
\end{align}
For the domain $m_{a} > 1.2\,\GeV$, the couplings read
\begin{align}
\nonumber
    &g_{pa}(m_{a})\big|_{m_{a}>1.2\,\GeV}
    =
    g_{na}(m_{a})\big|_{m_{a}>1.2\,\GeV}
    =
    -\frac{\sqrt{2/3}\,\alpha_s(m_a)}{f_a}(3D_s+2D)\cF(m_{a})\,{\rm max}\left[\frac{\beta_{\cF}}{m_{a}},1\right]\,,
\\ \nonumber
    &g_{\rho^{0}\gamma a}(m_{a})\big|_{m_{a}>1.2\,\GeV}
    =
    -\frac{eg\sqrt{3}}{2\sqrt{2}\pi^2 f_a}\,\alpha_s(m_a)\cF(m_{a})\,,
    \\ \nonumber
    &g_{\omega\gamma a}(m_{a})\big|_{m_{a}>1.2\,\GeV}
    =
    -\frac{eg}{2\sqrt{6}\pi^2 f_a}\,\alpha_s(m_a)\cF(m_{a})\,,
    \\ \label{Coupl--1}
    &g_{\phi\gamma a}(m_{a})\big|_{m_{a}>1.2\,\GeV}
    =
    \frac{eg}{2\sqrt{3}\pi^2 f_a}\,\alpha_s(m_a)\cF(m_{a})\,,
\end{align}
where $\beta_{\cF} =1.4\,\GeV$, and $\cF(m_{a})$ is defined in Eq.~\eqref{eq:cF}.

\section{$2\to 3$ process in terms of Mandelstam invariants}
\label{app:2-to-3}

In this section, we describe the calculation of the $2\to 3$ scattering cross-section in terms of Mandelstam invariants.

\subsection{$2\to 3$ process: generic cross section and phase space}

Consider the exclusive electroproduction process
\begin{align}\label{eq:2to3-generic-process}
e(p_{e})+p(p_{p})\to e'(p_{e'})+p'(p_{p'})+X(p_{X})\,,
\end{align}
We define the invariant mass of the colliding $ep$ pair,
\begin{align}
s
&\equiv
(p_{e}+p_{p})^2\,.
\end{align}
Then, the differential cross section is
\begin{align}\label{eq:dsigma-generic-master}
d\sigma_{X}
&=
\frac{\overline{|\mathcal{M}|^2}}{2\sqrt{\lambda(s,m_{e}^{2},m_{p}^{2})}}\; d\Phi_3\,,
\end{align}
where $\overline{|\mathcal{M}|^2}$ denotes the spin-averaged squared matrix element, $\lambda$ is the K\"all\'en function,
\begin{align}\label{eq:kallen-def}
\lambda(x,y,z)
&\equiv
x^{2}+y^{2}+z^{2}-2(xy+xz+yz)\,,
\end{align}
and the three-body phase space is
\begin{align}\label{eq:Phi3-9D}
d\Phi_3
&=
(2\pi)^4\,\delta^{(4)}\!\Big(p_{e}+p_{p}-p_{e'}-p_{p'}-p_{X}\Big)
\prod_{i=\{e',p',X\}}\frac{d^3\mathbf{p}_i}{(2\pi)^3\,2E_i}\,.
\end{align}
%

\subsection{Reduction to four invariants}

Following Ref.~\cite{Byckling:1971vca}, we introduce the following set of kinematic invariants:
\begin{align}\label{eq:Mandelstam-invariants}
s_{1}
&\equiv
(p_{e'}+p_X)^{2}\,,
\qquad
s_{\gamma^{*}p}
\equiv
(p_{p'}+p_{X})^{2}
=
(p_{p}+p_{e}-p_{e'})^{2}\,,
\nonumber\\
Q^{2}
&\equiv
- (p_{e}-p_{e'})^{2}\,,
\qquad
t_{p}
\equiv
(p_{p}-p_{p'})^{2}\,.
\end{align}
After integrating out the trivial angular variables, the phase-space measure reduces to
\begin{align}\label{eq:PS-measure-generic}
d\Phi_3
&=
\frac{\pi}{16\,\sqrt{\lambda(s,m_{e}^{2},m_{p}^{2})}}\;
\frac{4\; ds_{\gamma^{*}p}\,dQ^{2}\,dt_{p}\,ds_{1}}
{\sqrt{\,\lambda(s_{\gamma^{*}p},-Q^{2},m_{p}^{2})\,(s_{1}-s_{1}^{-})\,(s_{1}^{+}-s_{1})\,}}\,,
\end{align}
so that
\begin{align}\label{eq:differential-cross-section-generic-app}
d\sigma
&=
\frac{d^{4}\sigma}{ds_{\gamma^{*}p}\,dQ^{2}\,dt_{p}\,ds_{1}}\;
ds_{\gamma^{*}p}\,dQ^{2}\,dt_{p}\,ds_{1}\,,
\end{align}
with the fully differential cross section
\begin{align}\label{eq:d4sigma-generic-app}
\frac{d^{4}\sigma}{ds_{\gamma^{*}p}\,dQ^{2}\,dt_{p}\,ds_{1}}
&=
\frac{1}{256\pi^{4}}\;
\frac{\overline{|\mathcal{M}|^{2}}}{
\lambda(s,m_{e}^{2},m_{p}^{2})\;
\sqrt{\lambda(s_{\gamma^{*}p},-Q^{2},m_{p}^{2})\,(s_{1}-s_{1}^{-})\,(s_{1}^{+}-s_{1})}
}\,.
\end{align}

The endpoints $s_{1}^{\pm}$ are determined by Gram-determinant constraints and can be
written compactly as
\begin{multline}\label{eq:s1pm-compact}
s_{1}^{\pm}(s\,,s_{\gamma^{*}p}\,,Q^{2}\,,t_{p})
=
s + m_{p}^{2}
\\-
\frac{
\Delta \mp 2\sqrt{\,G\!\left(s\,, -Q^{2}\,, s_{\gamma^{*}p};\,m_{e}^{2}\,,m_{p}^{2}\,,m_{e}^{2}\right)\;
                G\!\left(s_{\gamma^{*}p}\,,t_{p}\,,m_{p}^{2};\, -Q^{2}\,,m_{p}^{2}\,,m_{X}^{2}\right)}
}{
\lambda(s_{\gamma^{*}p}\,, -Q^{2}\,, m_{p}^{2})
}\,,
\end{multline}
where
\begin{align}\label{eq:G-def}
G(x\,,y\,,z;\,u\,,v\,,w)
&\equiv
-\frac{1}{2}
\det\!\begin{pmatrix}
0 & 1 & 1 & 1 & 1\\
1 & 0 & v & x & z\\
1 & v & 0 & u & y\\
1 & x & u & 0 & w\\
1 & z & y & w & 0
\end{pmatrix}\,,
\end{align}
and
\begin{align}\label{eq:Delta-def}
\Delta
&\equiv
\det\!\begin{pmatrix}
2m_{p}^{2} & s_{\gamma^{*}p}+Q^{2}+m_{p}^{2} & m_{p}^{2}+m_{p}^{2}-t_{p}\\
s_{\gamma^{*}p}+Q^{2}+m_{p}^{2} & 2s_{\gamma^{*}p} & s_{\gamma^{*}p}-m_{X}^{2}+m_{p}^{2}\\
s-m_{e}^{2}+m_{p}^{2} & s+s_{\gamma^{*}p}-m_{e}^{2} & 0
\end{pmatrix}\,.
\end{align}
%

\subsection{Kinematic domain and useful scalar products}

The physically allowed domain in $(s_{\gamma^{*}p},Q^{2},t_{p},s_{1})$ at fixed $s$ is:
\begin{align}
&s_{\gamma^{*}p}^{-}(s)
\le
s_{\gamma^{*}p}
\le
s_{\gamma^{*}p}^{+}(s)\,, \\
&Q^{2}_{-}(s\,,s_{\gamma^{*}p})
\le
Q^{2}
\le
Q^{2}_{+}(s\,,s_{\gamma^{*}p})\,,
\\
&t_{p}^{-}(s\,,s_{\gamma^{*}p}\,,Q^{2})
\le
t_{p}
\le
t_{p}^{+}(s\,,s_{\gamma^{*}p}\,,Q^{2})\,,
\\
&s_{1}^{-}(s\,,s_{\gamma^{*}p}\,,Q^{2}\,,t_{p})
\le
s_{1}
\le
s_{1}^{+}(s\,,s_{\gamma^{*}p}\,,Q^{2}\,,t_{p})\,,
\end{align}
where:
\begin{align}
\label{eq:s2_bounds}
&s_{\gamma^{*}p}^{-}(s)
=
(m_{p}+m_{X})^{2}\,,
\qquad
s_{\gamma^{*}p}^{+}(s)
=
(\sqrt{s} - m_{e})^{2}\,,
\\
\label{eq:t1_bounds}
&Q^{2}_{\pm}(s\,,s_{\gamma^{*}p})
=
-2m_{e}^{2}
+\frac{(s+m_{e}^{2}-m_{p}^{2})(s-s_{\gamma^{*}p}+m_{e}^{2})
\pm \sqrt{\lambda(s\,,m_{e}^{2}\,,m_{p}^{2})\,\lambda(s\,,s_{\gamma^{*}p}\,,m_{e}^{2})}}{2s}\,,
\\
\label{eq:t2_bounds}
&t_{p}^{\pm}(s\,,s_{\gamma^{*}p}\,,Q^{2})
=
2m_{p}^{2}
\\ &\qquad\qquad\qquad-\frac{(s_{\gamma^{*}p}+m_{p}^{2}+Q^{2})(s_{\gamma^{*}p}+m_{p}^{2}-m_{X}^{2})
\mp \sqrt{\lambda(s_{\gamma^{*}p}\,,m_{p}^{2}\,,-Q^{2})\,\lambda(s_{\gamma^{*}p}\,,m_{p}^{2}\,,m_{X}^{2})}}
{2s_{\gamma^{*}p}}\,.
\end{align}
Here, $Q^{2}_{-}$ and $Q^{2}_{+}$ correspond to the minimal and maximal allowed values of $Q^{2}$, respectively.

We close the discussion of the $2\to 3$ phase space by expressing various scalar products for our electroproduction process~\eqref{eq:process}:
\begin{align}\label{eq:scalar-products-generic}
2\,p_{e}\!\cdot p_{p}
&=
s - m_{p}^{2} - m_{e}^{2},
&
2\,p_{e}\!\cdot p_{e'}
&=
2m_{e}^{2}+Q^{2},
\nonumber\\
2\,p_{p}\!\cdot p_{p'}
&=
2m_{p}^{2}-t_{p},
&
2\,p_{e'}\!\cdot p_{X}
&=
s_{1}-m_{e}^{2}-m_{X}^{2},
\nonumber\\
2\,p_{X}\!\cdot p_{p'}
&=
s_{\gamma^{*}p}-m_{X}^{2}-m_{p}^{2},
&
2\,p_{e'}\!\cdot p_{p'}
&=
s-s_{1}-s_{\gamma^{*}p}+m_{X}^{2},
\nonumber\\
2\,p_{e}\!\cdot p_{X}
&=
s_{1}-Q^{2}-t_{p}-m_{e}^{2},
&
2\,p_{p}\!\cdot p_{X}
&=
s_{\gamma^{*}p}+Q^{2}+t_{p}-m_{p}^{2},
\nonumber\\
2\,p_{e}\!\cdot p_{p'}
&=
s-s_{1}+t_{p}-m_{p}^{2},
&
2\,p_{p}\!\cdot p_{e'}
&=
s-s_{\gamma^{*}p}-Q^{2}-m_{e}^{2}.
\end{align}
%
\section{Alternative description of the electroproduction of $\pi^{0}$}
\label{app:pi0-production-kaskulov}

Depending on the beam energy configuration at the EIC, the electroproduction of light pseudoscalar mesons such as $\pi^{0}$ may be mostly sensitive to the domain $\sqrt{s_{\gamma^{*} p}} \lesssim 2\,\GeV$ (see Fig.~\ref{fig:distributions-sgammap-Q2-tp}), where some of the nucleon excitations in the bremsstrahlung in the diagrams (b), (c) of Fig.~\ref{fig:diagrams-pseudoscalar} may be on-shell.
Hence, the description of Ref.~\cite{Kashevarov:2017vyl} that we adopt for our study and that is accurately fitted to the photoproduction data in the range $\sqrt{s_{\gamma p}}\gtrsim 3\,\GeV$ may break down.

To better understand the systematics stemming from the extrapolation of the $\pi^{0}$ photoproduction in the domain $\sqrt{s_{\gamma^{*}p}} = (2-3)\GeV$, in this section, we incorporate the alternative approach of Refs.~\cite{Kaskulov:2008ej,Kaskulov:2011wd,Vrancx:2013fra}.
It describes the photoproduction via the $t$-channel exchanges by vector mesons $V = \rho,\omega$ and $s/u$-exchanges by baryonic resonances $B$ (the diagrams (b), (c) in Fig.~\ref{fig:diagrams-pseudoscalar}).
In these works, the photoproduction input was tested against $d\sigma_{\gamma p\to \pi^{0}p}/dt_{p}$ data at different photon energies from SLAC and DESY~\cite{Anderson:1971xh,Braunschweig:1970dp}.

Based on Fig.~1 of Ref.~\cite{Kaskulov:2011wd}, we conclude that the approach seems to overestimate the photoproduction cross-section at low $\sqrt{s_{\gamma p}}\lesssim 3.5\,\GeV$, and we therefore expect that it will deliver systematically larger cross-sections.

The matrix element $\cM^{\nu}_{P}$ in Eq.~\eqref{eq:electroproduction-matrix-element} is given by
\begin{align}
    \label{eq:matrix-element-pseudoscalar-meson-1}
    \cM^{\nu}_{P}
    &=
    \cM^{\nu}_{P,V}+\cM^{\nu}_{P,B}\,,
\end{align}
where the vector\,(nucleon) contribution is denoted by $\cM^{\nu}_{P,V}\,(\cM^{\nu}_{P,B})$.
Within the description of Ref.~\cite{Kaskulov:2011wd}, the objects $\cM^{\nu}_{P,V},\cM^{\nu}_{P,B}$ already include the finite-$Q^{2}$ form-factors.
Therefore, in Eq.~\eqref{eq:photo-to-electro} relating the matrix element~\eqref{eq:matrix-element-pseudoscalar-meson-1}, we use $F_{X}(Q^{2}) \equiv 1$.

The contribution from the exchange of vector mesons is given by
\begin{align}
    \label{eq:matrix-element-vector}
    \cM^{\nu}_{P,V}
    &=
    \sum_{V = \rho,\omega}
    g_{V\gamma P}\, g_{V pp}\,F_{V\gamma P}(Q^{2})
    \varepsilon^{\nu\kappa\alpha\beta} p_{\gamma^{*},\kappa}
    (p_{p}-p_{p'})_{\alpha}
    \nonumber\\
    &\quad\times
    \bar{u}_{s'}(p_{p'})
    \left[  (1 + \kappa_{V}) \gamma_{\beta}
    - \frac{\kappa_{V}}{2m_p} (p_{p}+p_{p'})_{\beta} \right]
    u_s(p_{p}) \times \cR_{V}(t_{p},s_{\gamma^{*}p})\,,
\end{align}
where the couplings $g_{V\gamma \pi^{0}}$, $g_{Vpp}$, and $\kappa_V$ are all taken as in Ref.~\cite{Kaskulov:2011wd} and are somewhat different from those used in Ref.~\cite{Kashevarov:2017vyl}.
$F_{V\gamma P}=\left(1+Q^{2}/m_{V}^{2}\right)^{-1}$ is the effective form-factor suppressing the large-$Q^{2}$ contribution to the process. Finally, $\cR_{V}(t_{p},s_{\gamma^{*}p})$ corresponds to the Reggeized propagator, accounting for the family of $V$ resonances with fixed quantum numbers~\cite{Kaskulov:2010kf}:
\begin{align}
    \label{eq:regge-pion-vector}
    \cR_{V}(t_{p},s_{\gamma^{*}p})
    &=
    - \alpha'_{V}\left[\frac{1 - e^{-i\pi\alpha_{V}(t_{p})}}{2}\right]
    \Gamma[1-\alpha_{V}(t_{p})]
    (\alpha'_{V}s_{\gamma^{*} p})^{\alpha_{V}(t_{p})-1}\,,
\end{align}
where $\Gamma$ is the Gamma function, and $\alpha_{V}(t_{p})$ are Regge trajectories for vector mesons~\cite{Donnachie:1992ny,Kaskulov:2011wd}:
\begin{align}
    \label{eq:production-vector-couplings-1}
    \alpha_{\rho}(t_{p})
    &=
    0.53+\alpha'_{\rho} t_{p}\,,
    \qquad\quad
    \alpha_{\omega}(t_{p})
    =
    0.40 + \alpha'_{\omega} t_{p}\,,
\end{align}
with $\alpha'_{\rho}=\alpha'_{\omega}=0.85\,\GeV^{-2}$, again slightly different from those used in Ref.~\cite{Kashevarov:2017rmk}.

Next, the contribution of the nucleon resonances to the $\pi^{0}$ electroproduction in the domain $\sqrt{s_{\gamma^{*}p}}>2\,\GeV$, described in Refs.~\cite{Kaskulov:2008ej,Kaskulov:2010kf,Kaskulov:2011wd}, is
\begin{align}
    \label{eq:matrix-element-nucleon}
    \cM^{\nu}_{\pi^{0},N}
    &=
    ig_{\pi^{0} NN}e \,\bar{u}_{s'}(p_{p'})
    \bigg[
    G_{s_{\gamma^{*} p}}(Q^{2},s_{\gamma^{*} p},t_{p})
    \gamma_5\frac{\slashed{p}_{p}+\slashed{p}_{\gamma^{*}} + m_p}
    {s_{\gamma^{*} p}-m_{p}^2}\,\gamma^{\nu}
    \nonumber\\
    &\qquad
    + G_{u}(Q^{2},u,t_{p})\gamma^{\nu}\,
    \frac{\slashed{p}_{p'}-\slashed{p}_{\gamma^{*}} + m_{p}}{u-m_p^2}\gamma_5
    + \text{GI}^{\nu} \bigg] u_s(p_{p})\,,
\end{align}
with $u \equiv (p_{p'}-p_{\gamma^{*}})^{2}$ and $g_{\pi^{0} NN}= 2m_{p}g_{\pi^{0}p} \approx 12.8 $ is the pion-nucleon coupling, which is deduced from Eq.~\eqref{eq:Coupl-ppi0}.
The $s_{\gamma^{*}p}$- and $u$-channel transition form-factors are
\begin{align}\label{eq:F1generic}
G_{s_{\gamma^{*}p}(u)}(Q^{2},s_{\gamma^{*}p}(u),t_{p})
&=
\cG_{s_{\gamma^{*}p}(u)}(Q^{2},s_{\gamma^{*}p}(u))\,(t_{p}-m_\pi^{2})\,
\cR_{N,s_{\gamma^{*}p}(u)}\!\left(\alpha(t_{p})\right)\,.
\end{align}
Here, $\cG_{s_{\gamma^{*}p}(u)}$ are effective form-factors, incorporating the tower of baryonic resonances with higher spins $S>1/2$ via the Bloom-Gilman duality connection~\cite{Bloom:1970xb,Bloom:1971ye},
\begin{align}\label{eq:F1_s_res}
\cG_{s_{\gamma^{*}p}}(Q^{2},s_{\gamma^{*}p})
&=
\frac{
s_{\gamma^{*}p}\,\ln\!\left[\frac{\xi Q^{2}}{m_p^{2}}+1\right]\,
\frac{(2\xi Q^{2}+s_{\gamma^{*}p})}{(\xi Q^{2})^{2}}
-\frac{s_{\gamma^{*}p}\,\big(\xi Q^{2}+s_{\gamma^{*}p}\big)}
{\xi Q^{2}\,\big(\xi Q^{2}+m_p^{2}\big)}
+\ln\!\left(\frac{s_{\gamma^{*}p}-m_p^{2}}{m_p^{2}}\right)-i\pi}
{\left(\frac{\xi Q^{2}}{s_{\gamma^{*}p}}+1\right)^{2}
\left(\frac{s_{\gamma^{*}p}^{2}+2s_{\gamma^{*}p}m_p^{2}}{2m_p^{4}}
+\ln\!\left(\frac{s_{\gamma^{*}p}-m_p^{2}}{m_p^{2}}\right)-i\pi \right)}\,,
\end{align}
and
\begin{align}
    \label{eq:F1_u_res}
    \cG_{u}(Q^{2},u)
    &=
    \frac{
    u\,\ln\!\left[\frac{\xi Q^{2}}{m_p^{2}}+1\right]\,
    \frac{(2\xi Q^{2}+u)}{(\xi Q^{2})^{2}}
    -\frac{u\,\big(\xi Q^{2}+u\big)}{\xi Q^{2}\,\big(\xi Q^{2}+m_p^{2}\big)}
    +\ln\!\left(\frac{m_p^{2}-u}{m_p^{2}}\right)}
    {\left(\frac{\xi Q^{2}}{u}+1\right)^{2}
    \left(
    \frac{u^{2}+2u m_p^{2}}{2m_p^{4}}
    +\ln\!\left(\frac{m_p^{2}-u}{m_p^{2}}\right) \right)}\,,
\end{align}
with the cutoff fixed to $\xi=0.4$~\cite{Kaskulov:2010kf}, while the Regge trajectories read
\begin{align}
    \label{eq:Regge-pion-nucleon}
    \cR_{N,u}\!\left(t_{p},s_{\gamma^{*}p}\right)
    &=
    -\alpha'\,\Gamma[-\alpha(t_{p})]\,
    e^{\alpha(t_{p})\,\ln(\alpha' s_{\gamma^{*}p})}\,,
    \qquad
    \cR_{N,s_{\gamma^{*}p}}\!\left(t_{p},s_{\gamma^{*}p}\right)
    =
    e^{-i\pi \alpha(t_{p})}\cR_{N,u}\!
    \left(t_{p},s_{\gamma^{*}p}\right)\,,
\end{align}
with
\begin{align}
    \label{eq:alpha_def}
    \alpha(t_{p})
    &=
    \alpha'\,(t_{p}-m_\pi^{2})\,,
    \qquad
    \alpha'
    =
    \frac{0.74\,\GeV^{-2}}{1+2.4\,\frac{Q^{2}}{s_{\gamma^{*}p}}}\,.
\end{align}
Finally, the term $\text{GI}^{\nu}$ must be generally introduced in the case $G_{s_{\gamma^{*}p}} \neq G_{u}$ (in particular in the limit $Q^{2}\to 0$) to ensure gauge invariance.
The explicit form we utilize is the contact term
\begin{align}
    \label{eq:pole-free-cancellation}
    \text{GI}^{\nu}
    &=
    -\left[\frac{(2p_{p}+p_{\gamma^{*}})^{\nu}}{s_{\gamma^{*}p}-m_{p}^{2}}(G_{s_{\gamma^{*}p}}-\hat{G}) + \frac{(2p_{p'}-p_{\gamma^{*}})^{\nu}}{u-m_{p}^{2}}(G_{u}-\hat{G})\right]\gamma_{5}\,,
\end{align}
that simultaneously ensures the Ward identity and the correct soft-photon limit of the differential cross-section~\cite{Davidson:2001rk}.
The function $\hat{\mathcal{G}}$ is fixed to be
\begin{align}
    \label{eq:Fhat-minimal}
    \hat{G}
    &=
    G_{u}+G_{s_{\gamma^{*}p}}-G_{u}G_{s_{\gamma^{*}p}}\,.
\end{align}
The residual degree of freedom -- the arbitrariness left by the residual condition -- is minimized by calibrating on the data.
Details are summarized in Appendix~\ref{app:gauge-dependence}.

\begin{figure}[t!]
    \centering
    \includegraphics[width=0.33\linewidth]{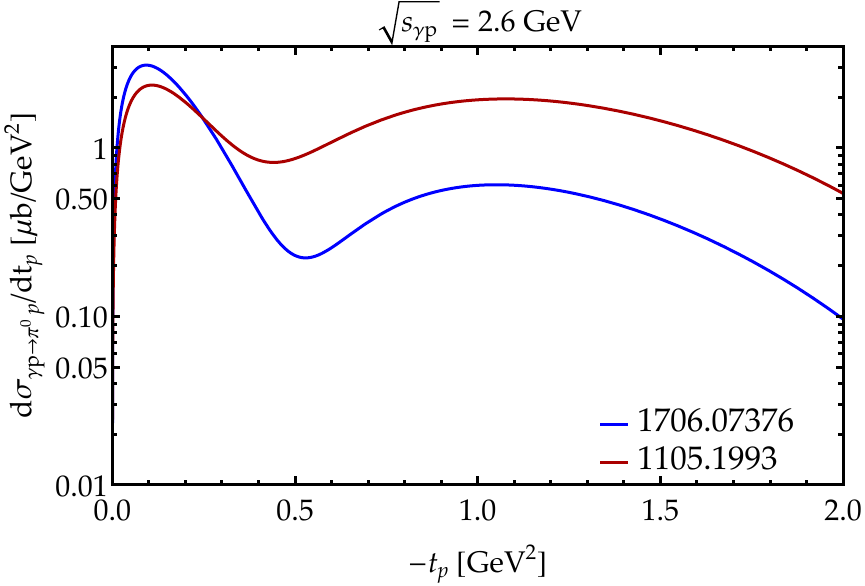}~\includegraphics[width=0.33\linewidth]{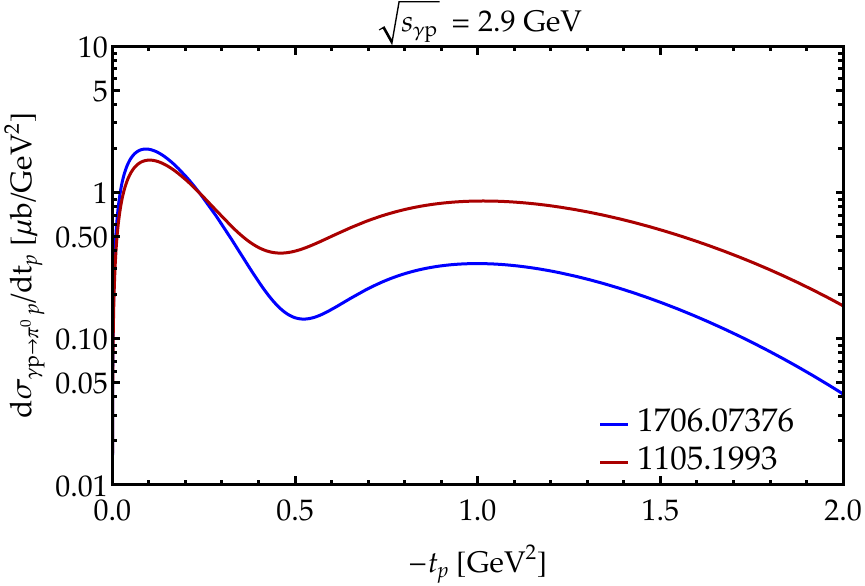}~\includegraphics[width=0.33\linewidth]{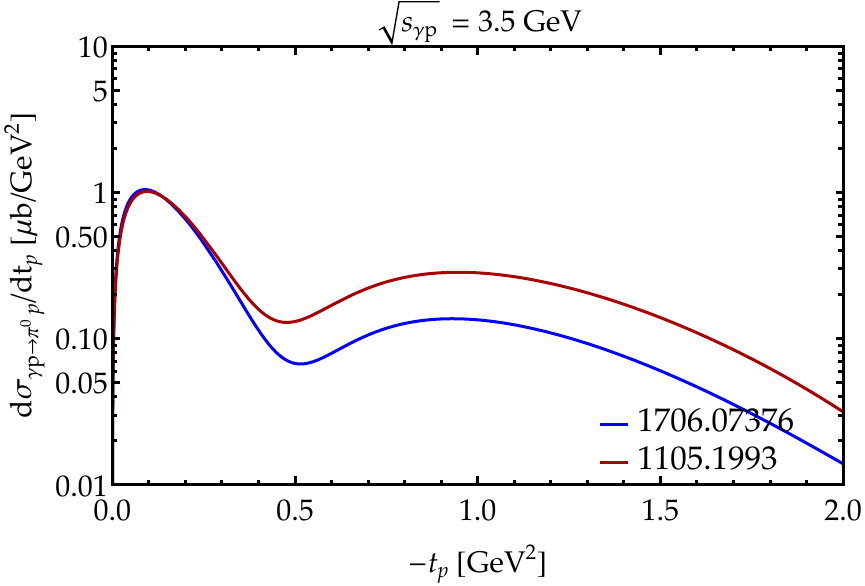}
    \caption{Comparison of the differential cross-sections of the pion photoproduction, as obtained following the approaches of Ref.~\cite{Kashevarov:2017rmk} (blue) and Ref.~\cite{Kaskulov:2011wd} (red), for various invariant masses of the $\gamma p$ system. See text for details.}
    \label{fig:comparison-kaskulov-vs-kashevarov}
\end{figure}

We compare the differential cross-sections $d\sigma_{\gamma p\to \pi^{0}p}/dt_{p}$ for various invariant masses $\sqrt{s_{\gamma p}} = 2.6, 2.9, 3.5\,\GeV$ (corresponding to $E_{\gamma} = 3,\,4,\,6\,\GeV$) in Fig.~\ref{fig:comparison-kaskulov-vs-kashevarov}. The last two configurations have data from DESY~\cite{Braunschweig:1968pra} to which the description of Ref.~\cite{Kashevarov:2017vyl} fits well, while the approach of Ref.~\cite{Kaskulov:2011wd} significantly overestimates the production for $|t_{p}|\gtrsim 0.5\,\GeV^{2}$. Moreover, as the invariant mass decreases, the discrepancy worsens. This trend explains the difference by a factor of $\simeq 5$ in $\sigma_{\rm fid}$ as obtained using these two approaches.

\subsection{Ensuring gauge invariance of the nucleon summand}
\label{app:gauge-dependence}

Let us discuss $\text{GI}_{\mu}$ entering Eq.~\eqref{eq:matrix-element-nucleon} in more detail.
To restore Ward identities, Refs.~\cite{Kaskulov:2008ej,Kaskulov:2011wd} utilized the pole term
\begin{align}\label{eq:pole-like-cancellation}
    \text{GI}_{\mu}
    &=
    -\gamma_{5}(\mathcal{G}_{s_{\gamma^{*} p}}-\mathcal{G}_{u})\frac{p_{\gamma^{*},\mu}}{-Q^{2}},
    \qquad
    p_{\gamma^{*}} \equiv p_{e}-p_{e'},
\end{align}
and then analyzed the $\gamma p \to \pi^{0}p$ and $\gamma^{*}p\to \pi^{0}p$ scatterings. However, we cannot use it for our calculations of the meson production in the far-forward direction, because it induces incorrect scaling of the differential cross-section with $Q^{2}$.

The issue follows from the fact that apart from violating the Ward identities, ``dressing'' the propagators in Eq.~\eqref{eq:matrix-element-nucleon} (\emph{i.e.}, introducing $\mathcal{G}_{u/s_{\gamma^{*}p}}\neq 1$) leads to artifacts in the analytic structure of the matrix element. Namely, in the correct theory, the Born squared matrix element ($\mathcal{G} = 1$) behaves as $|\mathcal{M}|^{2}\propto 1/Q^{4}$, but the integration over $s_{1}$ removes one power of $Q^{2}$, giving the standard scaling $d\sigma/dQ^{2}ds_{\gamma^{*}p}dt_{p} \propto 1/Q^{2}$ (and the same scaling for $d\sigma/dQ^{2}$, after integrating over $s_{\gamma^{*}p}$ and $t_{p}$). This is aligned with soft photon theorems~\cite{Low:1958sn,Burnett:1967km}, according to which the transverse part of the hadronic current must scale as $J_{N}\propto p_{\gamma^{*}}$ in the soft-photon limit, which automatically cancels one power of the squared propagator $1/Q^{4}$ after integrating over the phase space.

The ``dressed'' propagators generically break this cancellation.
Indeed, as long as $\mathcal{G}_{s_{\gamma^{*} p}}\neq \mathcal{G}_u$, the $s_{\gamma^{*}p}$- and $u$-channel pieces in~\eqref{eq:matrix-element-nucleon} modify the nucleon-pole residues: they are unequal.
As a result, the transverse part of the hadronic current no longer scales as $J_{N}\propto p_{\gamma^{*}}$ in the soft limit, and the cancellation does not occur.

The pole-like term~\eqref{eq:pole-like-cancellation}, being
proportional to $p_{\gamma^{*}}$ and carrying no $s/u$ denominators, cannot repair those residues: it enforces $p_{\gamma^{*}} \cdot J=0$ but vanishes identically in the physical matrix element after contracting with the electron current or the photon polarization, leaving the $s/u$-channel residues (and, generically, crossing properties) different from their Born values.

\begin{figure}[t]
\centering
\includegraphics[width=0.5\linewidth]{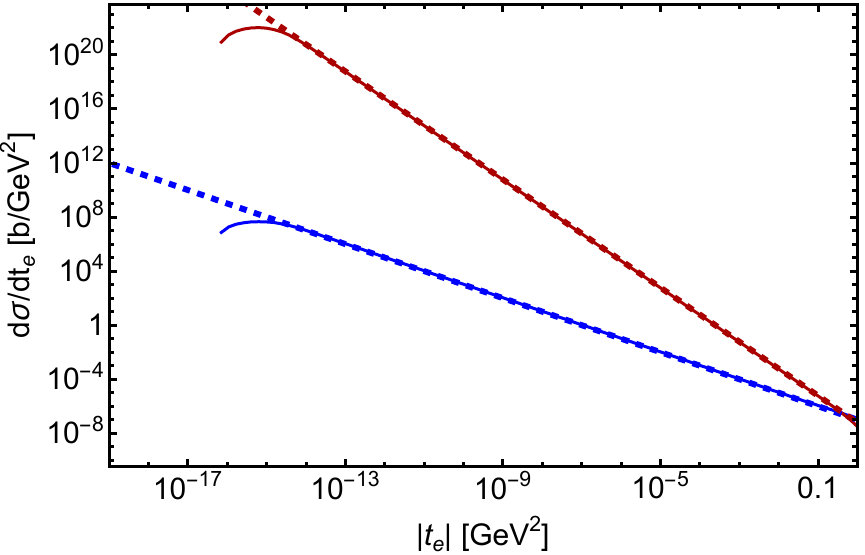}
\caption{The differential cross-section $d\sigma/dt_{e}$ with $t_{e} = -Q^{2}$ of the neutral pion production computed using two different descriptions of the term $\text{GI}_{\mu}$ from Eq.~\eqref{eq:matrix-element-nucleon}:
the ``seagull'' term~\eqref{eq:pole-like-cancellation} (the solid red line) and the pole-free term~\eqref{eq:pole-free-cancellation} (the solid blue line), with $\hat{\mathcal{G}}$ fixed by Eq.~\eqref{eq:Fhat-minimal}.
The seagull term does not fix the unphysical scaling of the hadronic current with $Q^{2}$, which leads to the $\propto 1/Q^{4}$ behavior (the dashed red line), whereas the pole-free term leads to the correct scaling $\propto 1/Q^{2}$ (the dashed blue line).}
\label{fig:cross-section-unphysical}
\end{figure}

Explicitly, we have ensured that, while the matrix element~\eqref{eq:matrix-element-pseudoscalar-meson} with Eq.~\eqref{eq:pole-like-cancellation} satisfies the Ward identities, the differential cross-section $d\sigma/dQ^{2}$ has the unphysical behavior with $Q^{2}$.
Namely, the small-$Q^{2}$ limit of the nucleon contribution to the differential cross-section has the leading contribution
\begin{align}
    \label{eq:bad-gauge-fixing-limit}
    Q^{2}\frac{d\sigma_{\pi^{0}}}{dQ^{2}ds_{\gamma^{*}p}dt_{p}}
    &\to
    \alpha_{\text{EM}}^{2}(2m_{p}g_{\pi p})^{2}\frac{t_{p}|\mathcal{G}_{s_{\gamma^{*}p}}-\mathcal{G}_{u}|^{2}(s-s_{\gamma^{*}p})}{4\pi Q^{2} (s-m_{p}^{2})(s_{\gamma^{*}p}-m_{p}^{2})^{3}} + \mathcal{O}(Q^{0})\,.
\end{align}
Crucially, the $1/Q^{2}$ piece must vanish, as after the integration, it gives the scaling $d\sigma/dQ^{2}\propto 1/Q^{4}$. However, for tiny $Q^{2}$, the form-factors $\mathcal{G}_{s_{\gamma^{*}p}}$ behave as
\begin{align}
|\mathcal{G}_{u/s_{\gamma^{*}p}}| \approx \exp[i\pi\alpha_{u/s_{\gamma^{*}p}}]\times (c+d_{u/s_{\gamma^{*}p}}\times Q^{2})
\end{align}
(recall Eqs.~\eqref{eq:F1_s_res},~\eqref{eq:F1_u_res}). Given that $\alpha_{u}\neq \alpha_{s_{\gamma^{*}p}}$, the difference $\mathcal{G}_{s_{\gamma^{*}p}}-\mathcal{G}_{u}$ does not vanish for finite $Q^{2}$. The impact of this $1/Q^{4}$ term is shown in Fig.~\ref{fig:cross-section-unphysical}, being an order of magnitude enhancement.

Instead of the term~\eqref{eq:pole-like-cancellation}, we consider the pole-free construction~\eqref{eq:pole-free-cancellation}, where $\hat{\mathcal{G}}$ is an arbitrary scalar function that must be pole-free and ensure that the residues of the propagators in the limit $s_{\gamma^{*}p}\to m_{p}^{2}$, $u\to m_{p}^{2}$ still match the ``bare'' theory $\mathcal{G}_{u}=\mathcal{G}_{s_{\gamma^{*} p}} = 1$.

The minimal choice is given by Eq.~\eqref{eq:Fhat-minimal}, $\hat{\mathcal{G}} =\mathcal{G}_{s_{\gamma^{*}p}}+\mathcal{G}_{u} -\mathcal{G}_{s_{\gamma^{*}p}}\mathcal{G}_{u}$.
However, the choice is not unique; for instance, adding the structure of the type $\lambda (1-\mathcal{G}_{s_{\gamma^{*}p}})(1-\mathcal{G}_{u})$ has no impact on the residues.
Fortunately, the term~\eqref{eq:pole-free-cancellation} contains the transverse part and hence explicitly enters the observables, such as the $\gamma+p\to \pi^{0}+p$ cross-section.
As a result, the ambiguity may be fixed by comparing the prediction of Eqs.~\eqref{eq:matrix-element-nucleon},~\eqref{eq:pole-free-cancellation} with the data on the $\gamma p \to \pi^{0}p$ scattering (see Fig.~1 of Ref.~\cite{Kaskulov:2011wd}).
We found that the minimal operator~\eqref{eq:Fhat-minimal} provides a very good agreement with the data, whereas sizable modifications significantly distort the shape $d\sigma_{\gamma p\to p\pi^{0}}/dt_{p}$.
Further, we will use the minimal choice~\eqref{eq:Fhat-minimal} for $\hat{\mathcal{G}}$.

We have checked that with this prescription, not only the matrix element satisfies the Ward identities, but also $d\sigma/dQ^{2}\propto Q^{-2}$ (see Fig.~\ref{fig:cross-section-unphysical}).

\section{Fit for the $g_{\rho\sigma\gamma}$ coupling}
\label{app:g-rho-sigma-gamma}

In this section, we derive a conservative upper limit on the $\rho\sigma\gamma$ coupling $g_{\rho\sigma\gamma}$ (defined in Eq.~\eqref{eq:Lagr-Vsigmagamma}) from the process $\rho^{0}\to\pi^{0}\pi^{0}\gamma$, to which the $\sigma$ contributes via $\rho^{0}\to \sigma^{*}\gamma \to \pi^{0}\pi^{0}\gamma$.
The experimentally measured partial width is
$[\Gamma(\rho^{0}\to\pi^{0}\pi^{0}\gamma)]_{\text{exp}} \approx 6.8\pm 1.1\,\text{keV}$~\cite{ParticleDataGroup:2024cfk}.
Within the Hidden Local Symmetry implementation of vector-meson dominance~\cite{Fujiwara:1984mp}, this decay also receives the contribution $\rho^{0}\to\omega^{*}\pi^{0}\to\pi^{0}\pi^{0}\gamma$, at the level of $\approx 1\,\text{keV}$. It is comparable to the experimental uncertainty and does not saturate the total rate. We therefore obtain a conservative bound by attributing the full measured rate to the $\sigma$-mediated channel and requiring
$[\Gamma(\rho^{0}\to\pi^{0}\pi^{0}\gamma)]_{\sigma} \le [\Gamma(\rho^{0}\to\pi^{0}\pi^{0}\gamma)]_{\text{exp}}$ (optionally at $1\sigma$),
which yields an upper limit on $g_{\rho\sigma\gamma}$.

To describe the $\sigma$ contribution, we will model it as a Breit-Wigner resonance, with mass $m_{\sigma} = 500\text{  MeV}$ and width $\Gamma_{\sigma} = 100$--$800\text{ MeV}$ as reported in PDG~\cite{ParticleDataGroup:2024cfk}. The choice of the $\sigma$ mass follows Refs.~\cite{Yu:2016zut,Yu:2017vvp}, which adopted it to build the Regge trajectory of the scalar-meson family for the photoproduction of vector mesons, and which we use in our approach.

We start with the matrix element
\begin{align}
    \cM(\rho^{0}\to 2\pi^{0}\gamma)
    &=
    2e g_{\rho\sigma\gamma}\,
    \big((p_{\gamma}\!\cdot p_{V})(\epsilon_{\gamma}\!\cdot \epsilon^{*}_{V})
    -(p_{\gamma}\!\cdot \epsilon^{*}_{V})(p_{V}\!\cdot \epsilon_{\gamma})\big)\,
    \frac{g_{\sigma \pi\pi}(p_{\pi,1}\!\cdot p_{\pi,2})}{s_{\pi\pi}-m_{\sigma}^{2}+i\Gamma_{\sigma}(s_{\pi\pi})\sqrt{s_{\pi\pi}}}\,
     \, ,
\end{align}
where $2g_{\sigma \pi\pi}(p_{\pi,1}\!\cdot p_{\pi,2})$ is the derivative $\sigma\pi\pi$ vertex~\cite{Fariborz:1999gr}, $s_{\pi\pi} = (p_{\pi,1}+p_{\pi,2})^{2}$, and $\Gamma_{\sigma}(s)\propto \sqrt{s-4m_{\pi}^{2}}(s-2m_{\pi}^{2})^{2}/s$ encodes the invariant-mass scaling of the $\sigma$ width for this vertex.
Next, we express $g_{\sigma\pi\pi}$ in terms of the total width $\Gamma_{\sigma}\approx 3\Gamma(\sigma\to 2\pi^{0})$ as a function of $m_{\sigma}$.
Then, we fix the value of $g_{\rho\sigma\gamma}$ from the measured width $\Gamma(\rho^{0}\to 2\pi^{0}\gamma)$.

Finally, to estimate the uncertainty, we scanned over a range of widths $\Gamma_{\sigma} = (100-800)\,\MeV$; we obtain the allowed range $|g_{\rho\sigma\gamma}| = (1.0-1.4)\,\GeV^{-1}$.

\section{Decays of $f_{2}$ meson}
\label{app:f2-decays}

In this section, we provide a description of the decays of an $f_{2}$ meson into various final states.
Further, we assume an effective interaction of the form
\begin{align}
    \label{eq:f2-interaction-generic}
    \cL_{\rm int}
    &=
    g_T\,f_{2,\mu\nu}\,T^{\mu\nu},
\end{align}
where $T^{\mu\nu}$ is the hadronic stress-energy tensor.
For a given final state $X$, the decay amplitude is proportional to the matrix element $\langle X|T^{\mu\nu}(0)|0\rangle$, and we allow for channel-dependent effective couplings $g_{T,X}$.

We are interested in the decays $f_{2}\to 2\pi$, $f_{2}\to 4\pi$, and $f_{2}\to V\gamma$, with $V = \rho^{0},\omega,\phi$.
The widths of the first two processes are reported in PDG~\cite{ParticleDataGroup:2024cfk}, with the central values $\Gamma(f_{2}\to \pi\pi) \approx 157\,\MeV$ and $\Gamma(f_{2}\to 2\pi^{+}2\pi^{-})\approx 5.2\,\MeV$.
The widths $f_{2}\to \omega/\phi\gamma$ are not reported, but there are some theoretical calculations based on the quark model~\cite{Ishida:1988uw}: $\Gamma(f_{2}\to \phi \gamma)\approx 1.3\,\keV$, $\Gamma(f_{2}\to \omega\gamma)\approx 27\,\keV$.

The parts of the stress-energy tensor relevant for these decays are:
\begin{align}
    \label{eq:Tmunu-Vgamma}
    T_{\mu\nu}^{\pi\pi}
    &=
    \partial_{\mu}\pi^{(+}\partial_{\nu}\pi^{-)} + \partial_{\mu}\pi^{0}\partial_{\nu}\pi^{0}
    -g_{\mu\nu}\left(\partial_{\alpha}\pi^{+}\partial^{\alpha}\pi^{-} -m_{\pi}^{2}\pi^{+}\pi^{-}+\frac{1}{2}(\partial_{\alpha}\pi^{0})^{2}-\frac{m_{\pi}^{2}}{2}(\pi^{0})^{2}\right),
    \nonumber\\
    T_{\mu\nu}^{VV}
    &=
    -V_{\mu\alpha}V_{\nu}^{\ \alpha}+m_{V}^{2}V_{\mu}V_{\nu}
    - g_{\mu\nu}\left(-\frac{1}{4}V_{\alpha\beta}V^{\alpha\beta}+\frac{m_{V}^{2}}{2}V_{\alpha}V^{\alpha}\right),
    \nonumber\\
    T_{\mu\nu}^{V\gamma}
    &=
    -2c_{V\gamma}V_{\mu\alpha}F_{\nu}^{\ \alpha} + g_{\mu\nu}\frac{c_{V\gamma}}{2}F_{\alpha\beta}V^{\alpha\beta},
\end{align}
where $V_{\mu\nu} = \partial_{\mu}V_{\nu}-\partial_{\nu}V_{\mu}$, $\partial_{\mu}x^{(a}\partial_{\nu}x^{b)} \equiv \partial_{\mu}x^{a}\partial_{\nu}x^{b}+\partial_{\mu}x^{b}\partial_{\nu}x^{a}$, and the first summand in Eq.~\eqref{eq:Tmunu-Vgamma} is automatically symmetrized after contracting with the symmetric $f_{2}$ field (either on- or off-shell).
The terms $\propto g_{\mu\nu}$ vanish after contracting with on-shell $f_{2}$ polarization, but we keep them for generality. Finally,
\begin{align}
    c_{\rho \gamma}
    &=
    \frac{e}{g}\,,
    \quad
    c_{\omega \gamma}
    =
    \frac{e}{3g}\,,
    \quad
    c_{\phi \gamma}
    =
    -\frac{\sqrt{2}e}{3g}\,,
\end{align}
with $g = m_{\rho}/(\sqrt{2}f_{\pi})\approx 5.9$, are given by the mixing of the photon with the vector meson $V$.
The corresponding vertices are:
\begin{align}
    \label{eq:f2pipi}
    \Gamma_{f_{2}\pi^{+}\pi^{-}}^{\mu\nu}
    &=
    ig_{T,\pi\pi}\left( p_{1}^{\mu}p_{2}^{\nu}+p_{1}^{\nu}p_{2}^{\mu} - g^{\mu\nu}((p_{1}\cdot p_{2})-m_{\pi}^{2})\right),
    \\
    \label{eq:f2VV}
    \Gamma_{f_{2}VV}^{\mu\nu,\alpha\beta}
    &=
    ig_{T,VV}\left( \tilde{\Gamma}^{\mu\nu,\alpha\beta}_{f_{2}VV}(p_{1},p_{2},m_{V}) + \tilde{\Gamma}^{\mu\nu,\beta\alpha}_{f_{2}VV}(p_{2},p_{1},m_{V})\right),
    \\
    \Gamma_{f_{2}V\gamma}^{\mu\nu,\alpha\beta}
    &=
    2\,ig_{T,V\gamma}c_{V\gamma}\left( \tilde{\Gamma}^{\mu\nu,\alpha\beta}_{f_{2}VV}(p_{V},p_{\gamma},0)\right),
\end{align}
where
\begin{align}
    \tilde{\Gamma}^{\mu\nu,\alpha\beta}_{f_{2}VV}(p_1,p_2,m_V)
    &=
    \big(p_1\!\cdot\! p_2 + m_V^{2}\big)\, g^{\mu\alpha} g^{\nu\beta}
    + p_1^{\mu} p_2^{\nu}\, g^{\alpha\beta}
    - p_1^{\mu} p_2^{\alpha}\, g^{\nu\beta}
    - p_1^{\beta} p_2^{\nu}\, g^{\mu\alpha}
    \nonumber\\
    &\quad
    -\frac{1}{2}\, g^{\mu\nu}\left[
    \big(p_1\!\cdot\! p_2 + m_V^{2}\big)\, g^{\alpha\beta}
    - p_1^{\beta} p_2^{\alpha} \right].
\end{align}
In Eq.~\eqref{eq:f2pipi}, the momenta $p_{1},p_{2}$ denote the momenta of the $\pi^{+},\pi^{-}$ mesons, while in Eq.~\eqref{eq:f2VV}, they denote the momenta of the two $V$ mesons.

Within vector meson dominance, the couplings $g_{T,V\gamma}$ are related to the couplings $g_{T,VV}$ by $g_{T,V\gamma} = g_{T,VV}$.
If assuming the universality of the interaction~\eqref{eq:f2-interaction-generic} with hadrons, we would also have
\begin{align}
    \label{eq:f2-universality}
    g_{T,\pi\pi}
    &=
    g_{T,\rho\rho}=g_{T,\omega\omega}=g_{T,\phi\phi}\,.
\end{align}
However, the coupling may not be universal.
For instance, embedding $f_{2}$ into the $SU(3)$ representation of the tensor meson nonet, one would associate it with the generator $T_{f_{2}} \propto \text{diag}(1,1,0)$~\cite{Guo:2011ir}, which means that it does not prefer the $s\bar{s}$ admixtures such as $\phi$.

Using the polarization sum rule for the $f_{2}$ meson, Eq.~\eqref{eq:f2-polarization-sum-rule}, we find the decay widths for $f_{2}\to 2\pi$ and $f_{2}\to V\gamma$:
\begin{align}
    \label{eq:width-f2-to-2-pi}
    \Gamma(f_{2}\to \pi\pi)
    &=
    \frac{3}{2}\Gamma(f_{2}\to \pi^{+}\pi^{-})
    =
    g_{T,\pi\pi}^{2}\,\frac{\bigl(m_{f_{2}}^{2}-4m_{\pi}^{2}\bigr)^{5/2}}
    {320\,\pi\,m_{f_{2}}^{2}}
    \approx
    157\,\MeV\,\left(\frac{g_{T,\pi\pi}}{9.3\,\GeV^{-1}}\right)^{2}\,,\\
    \label{eq:width-f2-to-V-gamma}
    \Gamma(f_{2}\to V\gamma)
    &=
    g_{T,V\gamma}^{2}c_{V\gamma}^{2}\,\frac{\bigl(m_{f_{2}}^{2}-m_{V}^{2}\bigr)^{3}\,
    \bigl(6m_{f_{2}}^{4}+3m_{f_{2}}^{2}m_{V}^{2}+m_{V}^{4}\bigr)}
    {120\pi\,m_{f_{2}}^{7}} \,.
\end{align}
Next, we calculate the width $f_{2}\to 2\rho^{0*}\to 2\pi^{+}2\pi^{-}$ using two approaches: the narrow width approximation $f_{2}\to \rho^{0}\pi^{+}\pi^{-}$ and the full 4-body calculation, using Ref.~\cite{Ovchynnikov:2025gpx}.
The amplitude of the process reads
\begin{align}
    \cM_{f_{2}\to 2\pi^{+}2\pi^{-}}
    &=
    \cM_{1}+\cM_{2},
\end{align}
where
\begin{align}
    \cM_{1}
    &=
    \epsilon_{\mu\nu}^{f_{2}}\Gamma_{f_{2}\rho\rho}^{\mu\nu,\alpha\beta}D^{(\rho)}_{\alpha\alpha_{1}}(p_{1}+p_{2})D^{(\rho)}_{\beta\beta_{1}}(p_{3}+p_{4})\Gamma_{\rho \pi\pi}^{\alpha_{1}}(p_{1},p_{2})\Gamma_{\rho \pi\pi}^{\beta_{1}}(p_{3},p_{4}),
\end{align}
and
$\mathcal{M}_{2} = \mathcal{M}_{1}\big|_{p_{4}\leftrightarrow p_{2}}$, $\Gamma_{\rho \pi\pi}^{\gamma}(p_{1},p_{2}) = ig(p_{1}-p_{2})^{\gamma}$. $D^{(\rho)}_{\beta\beta_{1}}(p)$ is the $\rho$ meson propagator
\begin{align}
    D^{(\rho)}_{\beta\beta_{1}}(p)
    &=
    \frac{-g_{\beta\beta_{1}}+\frac{p_{\beta}p_{\beta_{1}}}{m_{\rho}^{2}}}{p^{2}-m_{\rho}^{2}+i\tilde{\Gamma}_{\rho}(p^{2})m_{\rho}},
\end{align}
with $\tilde{\Gamma}_{\rho}(p^{2}) = \Gamma_{\rho}\times \frac{m_{\rho}}{\sqrt{p^{2}}}\left(\frac{s-4m_{\pi}^{2}}{m_{\rho}^{2}-4m_{\pi}^{2}} \right)^{\frac{3}{2}}$ is the modulation of $\rho$ decay width as a function of the transferred momentum~\cite{Guo:2011ir}.
We found
\begin{align}
    \label{eq:width-f2-to-4-pi}
    2\times\Gamma(f_{2}\to \rho^{0}\pi^{+}\pi^{-})
    &\approx
    \Gamma(f_{2} \to 2\rho^{0*}\to 2\pi^{+}2\pi^{-})
    \approx
    5.2\,\MeV\,\left(\frac{g_{T,\rho \rho}}{6.1\,\GeV^{-1}}\right)^{2}\,,
\end{align}
where the first equality holds with $\cO(10\%)$ accuracy, and a factor of 2 comes from the fact that both $\rho^{0*}$ in the 3-body process may be on-shell.
The values of the couplings $g_{T,\phi\gamma}$, $g_{T,\omega\gamma}$ needed to reproduce the quark-model prediction for $f_{2}\to \phi/\omega \gamma$ are $g_{T,\omega \gamma} \approx 4.5\,\GeV^{-1}$, $g_{T,\phi \gamma} \approx 0.7\,\GeV^{-1}$.
These values are perfectly consistent with the couplings quoted in Refs.~\cite{Yu:2016zut,Yu:2017vvp}, once one accounts for the different normalization convention used there for the $f_{2}V\gamma$ vertex; in our notation, their coupling corresponds to $c_{V\gamma}g_{T,V\gamma}/2$.

As we see from Eqs.~\eqref{eq:width-f2-to-2-pi},~\eqref{eq:width-f2-to-4-pi}, the universality model~\eqref{eq:f2-universality} would predict overly large width $\Gamma(f_{2}\to 4\pi)$ -- by a factor of 2 larger than the experimentally observed value.
On the other hand, if we assume the universality in the couplings $g_{T,VV}$ and fix $g_{T,VV}$ from the requirement that the width~\eqref{eq:width-f2-to-4-pi} saturates the observed $f_{2}\to2\pi^{+}2\pi^{-}$ width, it would predict $g_{T,\omega \gamma}(g_{T,\phi \gamma})$ larger by a factor of $1.4$\,($4.3$) than if these couplings had been obtained from the quark-model prediction.
These results do not contradict the observations, though.

\section{Full event kinematics from Mandelstam invariants}
\label{app:kinematics-from-invariants}

\subsection{Reconstructing energies and polar angles from invariants}

Given a sampled point $(s_1,s_{\gamma^{*}p},Q^{2},t_{p})$, we reconstruct the event kinematics in the $ep$ c.m.\ frame.
The final state energies follow from standard three-body kinematics:
\begin{align}\label{eq:cm_energies_momenta}
E_{e'}^{*}
&=
\frac{s+m_e^{2}-s_{\gamma^{*}p}}{2\sqrt{s}},
\qquad
p_{e'}^{*}
=
\frac{\sqrt{\lambda(s,m_e^{2},s_{\gamma^{*}p})}}{2\sqrt{s}},
\nonumber\\
E_{p'}^{*}
&=
\frac{s+m_p^{2}-s_{1}}{2\sqrt{s}},
\qquad
p_{p'}^{*}
=
\frac{\sqrt{\lambda(s,m_p^{2},s_{1})}}{2\sqrt{s}},
\nonumber\\
E_{X}^{*}
&=
\frac{s+m_X^{2}-s_{3}}{2\sqrt{s}},
\qquad
p_{X}^{*}
=
\frac{\sqrt{\lambda(s,m_X^{2},s_{3})}}{2\sqrt{s}},
\end{align}
with $s_3 = (p_{e'}+p_{p'})^{2} = s + m_{e}^{2}+m_{X}^{2}+m_{p}^{2} - s_{1}-s_{\gamma^{*}p}$.

We choose the $z$ axis along the incoming proton in the c.m.\ frame and define
\begin{align}
E_p^{*}
&=
\frac{s+m_p^2-m_e^2}{2\sqrt{s}},
\qquad
k^{*}
=
\frac{\sqrt{\lambda(s,m_e^2,m_p^2)}}{2\sqrt{s}}\,.
\end{align}
The c.m.\ polar angles $\theta_i^{*}$ follow from
\begin{align}\label{eq:cos_theta_star_master}
\cos\theta_i^{*}
&=
\frac{E_p^{*}E_i^{*}-p_p\!\cdot p_i}{k^{*}\,p_i^{*}}
\qquad (i=e',X,p')\,,
\end{align}
using
\begin{align}\label{eq:pp_dot_products_for_angles}
2\,p_p\!\cdot p_{p'}
&=
2m_p^2-t_{p}\,,
\qquad
2\,p_p\!\cdot p_{e'}
=
s-s_{\gamma^{*}p}-Q^{2}-m_e^2\,,
\qquad
2\,p_p\!\cdot p_X
=
s_{\gamma^{*}p}+Q^{2}+t_{p}-m_p^2\,.
\end{align}

Boosting to the laboratory frame is a longitudinal boost along the proton beam with
\begin{align}
\gamma
&=
\frac{E_e+E_p}{\sqrt{s}},
\qquad
\beta
=
\frac{\sqrt{E_p^2-m_p^2}-\sqrt{E_e^2-m_e^2}}{E_e+E_p}.
\end{align}
For each final-state particle $i$,
\begin{align}\label{eq:boost_longitudinal}
E_i
&=
\gamma\big(E_i^{*}+\beta\,p_i^{*}\cos\theta_i^{*}\big)\,,
\qquad
p_{z,i}
=
\gamma\big(p_i^{*}\cos\theta_i^{*}+\beta\,E_i^{*}\big)\,,
\qquad
p_{T,i}
=
p_i^{*}\sin\theta_i^{*}\,,
\end{align}
so that $\tan\theta_i=p_{T,i}/p_{z,i}$.

\subsection{Azimuthal angles and full four-momenta}

The invariants fix all dot products among external momenta; in a collinear-beam setup, the remaining continuous freedom is a global rotation about the beam axis. We sample this freedom by drawing one azimuth uniformly and determining the other two from transverse-momentum closure.

Define the transverse-momentum magnitudes
\begin{align}
a
&\equiv
p_{T,X}=p_X\sin\theta_X\,,
\qquad
b
\equiv
p_{T,e'}=p_{e'}\sin\theta_{e'}\,,
\qquad
c
\equiv
p_{T,p'}=p_{p'}\sin\theta_{p'}\,.
\end{align}
We draw a global azimuth $\phi_X\sim \mathrm{Unif}[-\pi,\pi]$ and define $\phi_T\equiv \phi_X+\pi$, so that
$\mathbf{p}_{T,T}\equiv-\mathbf{p}_{T,X}$ points at azimuth $\phi_T$. Transverse momentum conservation requires
\begin{align}\label{eq:pT_closure_vector}
\mathbf{p}_{T,e'}+\mathbf{p}_{T,p'}=\mathbf{p}_{T,T}\,.
\end{align}
In the transverse plane, the vectors form a triangle with side lengths $(a,b,c)$, hence the relative angle $\Delta$ between $\mathbf{p}_{T,e'}$ and $\mathbf{p}_{T,p'}$ is fixed by
\begin{align}
\cos\Delta
&=
\frac{a^2-b^2-c^2}{2bc},
\qquad
\Delta
=
\arccos(\cos\Delta).
\end{align}
There are two mirror solutions; we sample them with $\sigma=\pm1$ (equiprobably) and set
\begin{align}\label{eq:phi_solutions}
\alpha
&=
\operatorname{atan2}\!\Big(\sigma\,c\sin\Delta,\; b+c\cos\Delta\Big)\,,
\qquad
\phi_{e'}
=
\phi_T+\alpha\,,
\qquad
\phi_{p'}
=
\phi_T+\alpha-\sigma\,\Delta\,,
\end{align}
which enforces Eq.~\eqref{eq:pT_closure_vector}.

Finally, the lab-frame four-momenta are assembled as
\begin{align}\label{eq:full_4momenta}
p_i^\mu
&=
\big(E_i,\; p_i\sin\theta_i\cos\phi_i,\; p_i\sin\theta_i\sin\phi_i,\; p_i\cos\theta_i\big)\,,
\qquad
p_i
=
\sqrt{E_i^2-m_i^2}\,,
\end{align}
for $i\in\{e',p',X\}$. Energy and momentum conservation are then satisfied by construction.

\bibliographystyle{JHEP}
\bibliography{main}

\end{document}